\documentclass[11pt]{article}
\usepackage{amsmath,amssymb}
\usepackage[nosort]{cite}
\usepackage[margin=1in]{geometry}
\usepackage{graphicx}
\makeatletter \@addtoreset{equation}{section} \makeatother

\makeatletter
\newcommand\blfootnote{\xdef\@thefnmark{}\@footnotetext}
\makeatother

\newcommand{\fft}[2]{{\frac{#1}{#2}}}
\newcommand{\ft}[2]{{\textstyle\frac{#1}{#2}}}
\newcommand{\beq}{\begin{equation}}
\newcommand{\eeq}{\end{equation}}
\newcommand{\bea}{\begin{eqnarray}}
\newcommand{\eea}{\end{eqnarray}}
\newcommand{\nn}{\nonumber}
\newcommand{\mut}{\tilde{\mu}}
\newcommand{\half}{\frac12}
\newcommand{\zb}{\bar{z}}
\newcommand{\Tr}{\mbox{Tr}}
\newcommand{\psit}{\tilde{\psi}}
\begin{document}
%%%%%%%%%%%%%%%%%%%%%%%%%%%%%%%%%%%%%%%%

\begin{titlepage}

\hbox to \hsize{{\tt arXiv:0704.2233}\hss
\hbox{\hbox{Brown-HET-1480}\quad
\hbox{MCTP-07-15}\quad
\hbox{MIFP-07-13}\quad
\hbox{NSF-KITP-07-58}
}}

\vspace*{.2cm}

\begin{center}

{\Large\bf Bubbling AdS and droplet descriptions of BPS geometries\\[6pt]
in IIB supergravity}

\vspace*{.5cm}

{\bf Bin Chen$^*$, Sera Cremonini$^\dagger$, Aristomenis Donos$^\ddagger$,
Feng-Li Lin$^\sharp$,\\[4pt]
Hai Lin$^\dagger$, James T.~Liu$^{\dagger,\flat}$,
Diana Vaman$^\dagger$, Wen-Yu Wen$^\S$
}

\blfootnote{E-mail:
\vtop{\hbox{\{seracre, hailin, jimliu, dvaman\}@umich.edu,
bchen01@pku.edu.cn, aristomenis\_donos@brown.edu,}
\hbox{linfengli@phy.ntnu.edu.tw, steve.wen@gmail.com}}}

\vspace*{.2cm}
\small{
$^*${\it  School of Physics, Peking University, Beijing 100871, P.R. China}
\vspace*{.2cm}

$^\dagger${\it Michigan Center for Theoretical Physics,
Randall Laboratory of Physics,\\
The University of Michigan, Ann Arbor, MI 48109--1040, USA}
\vspace*{.2cm}

$^\ddagger${\it
Physics Department, Brown University, Providence, RI 02912, USA}
\vspace*{.2cm}

$^\sharp${\it Department of Physics, National Taiwan Normal University,
Taipei City, 116, Taiwan}
\vspace*{.2cm}

$^\flat${\it George P.\ \& Cynthia W.\ Mitchell Institute for Fundamental
Physics,\\
Texas A\&M University, College Station, TX 77843--4242, USA}
\vspace*{.2cm}

$^\S${\it Department of Physics and Center for Theoretical Sciences,\\
National Taiwan University, Taipei 106, Taiwan}
}
\end{center}

%\vspace*{.2cm}

\begin{abstract}
This paper focuses on supergravity duals of BPS states in ${\cal N}=4$
super Yang-Mills.  In order to describe these duals, we begin with a
sequence of breathing mode reductions of IIB supergravity: first on $S^3$,
then $S^3\times S^1$, and finally on $S^3\times S^1\times CP^1$.  We then
follow with a complete supersymmetry analysis, yielding 1/8, 1/4 and
1/2 BPS configurations, respectively (where in the last step we take the
Hopf fibration of $S^3$).  The 1/8 BPS geometries, which have an $S^3$
isometry and are time-fibered over a six-dimensional base, are determined
by solving a non-linear equation for the K\"ahler metric on the base.
Similarly, the 1/4 BPS configurations have an $S^3\times S^1$ isometry
and a four-dimensional base, whose K\"ahler metric obeys another non-linear,
Monge-Amp\`ere type equation.

Despite the non-linearity of the problem, we develop a universal
bubbling AdS description of these geometries by focusing  on
the boundary conditions which ensure their regularity.  In the 1/8 BPS case,
we find that the $S^3$ cycle shrinks to zero size on a five-dimensional
locus inside the six-dimensional base.  Enforcing regularity of the full
solution requires that the interior of a smooth, generally disconnected
five-dimensional surface be removed from the base.  The AdS$_5\times S^5$
ground state corresponds to excising the interior of an $S^5$, while the
1/8 BPS excitations correspond to deformations (including topology change)
of the $S^5$ and/or the excision of additional droplets from the base.
In the case of 1/4 BPS configurations, by enforcing regularity conditions,
we identify three-dimensional surfaces inside the four-dimensional base
which separate the regions where the $S^3$ shrinks to zero size from
those where the $S^1$ shrinks.

We discuss a large class of examples to show the emergence of a universal
bubbling AdS picture for all 1/2, 1/4 and 1/8 BPS geometries.

\end{abstract}

\end{titlepage}
%%%%%%%%%%%%%%%%%%%%%%%%%%%%%%%%%%%%%%%%

\setcounter{tocdepth}{2}
\tableofcontents

%%%%%%%%%%%%%%%%%%%%%%%%%%%%%%%%%%%%%%%%
\section{Introduction}
%%%%%%%%%%%%%%%%%%%%%%%%%%%%%%%%%%%%%%%%

In its most straightforward incarnation, AdS/CFT duality is a relation
between $\mathcal N=4$ super-Yang Mills theory and IIB string theory
on AdS$_5\times S^5$.  This system has been extensively studied, and
recently there has been much progress in the study of various sectors
of this correspondence.  In general, some of the best understood
aspects of this duality naturally arise through the use of supersymmetry.
A particularly striking example of this was realized in a remarkable
paper by Lin, Lunin and Maldacena (LLM) \cite{Lin:2004nb}, which constructed
explicit regular 1/2 BPS states in IIB supergravity and demonstrated
their relation to the free fermion picture of the corresponding
1/2 BPS sector of the $\mathcal N=4$ super-Yang Mills theory
\cite{Corley:2001zk,Berenstein:2004kk}.

Based on the correspondence with chiral primaries satisfying $\Delta=J$,
LLM examined all regular 1/2 BPS states with $SO(4)\times SO(4)$ isometry
in IIB supergravity with only the metric and self-dual five-form turned on.  Because of this $S^3\times S^3$ isometry, explicit
construction of such 1/2 BPS `bubbling AdS' configurations may be simplified
by working in an effective four-dimensional theory of the form
\begin{equation}
e^{-1}\mathcal L_4=e^{3H}[R+\ft{15}2\partial H^2-\ft32\partial G^2
-\ft14e^{-3(H+G)}F_{\mu\nu}^2+12e^{-H}\cosh G].
\label{eq:llmlag}
\end{equation}
The four-dimensional metric, two scalars $H$ and $G$, and the 2-form field strength $F_{\mu\nu}$ are related to their ten-dimensional counterparts
according to \cite{Lin:2004nb,Liu:2004ru,Liu:2004hy}
\begin{eqnarray}
ds_{10}^2&=&g_{\mu\nu}dx^\mu dx^\nu+e^H(e^Gd\Omega_3^2+e^{-G}d\widetilde
\Omega_3^2),\nonumber\\
F_{(5)}&=&(1+*_{10})F_{(2)}\wedge\Omega_3.
\label{eq:llmans}
\end{eqnarray}
Since the supersymmetric bubbling configurations preserve a time-like
Killing vector $\partial/\partial t$, the construction further simplifies
into a three dimensional one.  The result is that all such 1/2 BPS
states are describable in terms of a single harmonic function
$Z=\fft12\tanh G$ satisfying the linear equation \cite{Lin:2004nb}
\begin{equation}
\left(\partial_1^2+\partial_2^2+y\partial_y\fft1y\partial_y\right)
Z(x_1,x_2,y)=0.
\label{eq:bubble}
\end{equation}
The resulting ten-dimensional metric is then of the form
\begin{equation}
ds_{10}^2=-h^{-2}(dt+\omega)^2+h^2(dx_1^2+dx_2^2+dy^2)
+y(e^Gd\Omega_3^2+e^{-G}d\widetilde\Omega_3^2)
\label{eq:llmmet}
\end{equation}
where $h^{-2}=2y\cosh G$.

The bubbling picture arises through the observation that regularity of
the metric (\ref{eq:llmmet}) demands that only one of the three-spheres
collapses (in an appropriate manner) as $y\to0$.  The necessary boundary
conditions are then simply
\begin{equation}
Z(x_1,x_2,y=0)=\pm\ft12.
\label{eq:llmbc}
\end{equation}
These boundary conditions allow the $y=0$ boundary plane to be identified
with the fermion droplet phase-space plane \cite{Lin:2004nb}, and the
complete form of $Z$ may then be obtained through an appropriate Green's
function solution to (\ref{eq:bubble}).  In fact, a key feature of this
1/2 BPS bubbling AdS$_5\times S^5$ construction is precisely the linearity
of the governing equation (\ref{eq:bubble}).  This linearity is natural
from the free fermion picture on the gauge theory side of the duality,
and at first sight may be thought of as a consequence of the BPS
({\it i.e.}~no force) condition.  However, this is not necessarily the
case, as for example 1/2 BPS configurations in 11-dimensional supergravity
with $SO(3)\times SO(6)$ isometry are described by a Toda-type equation,
which is non-linear \cite{Lin:2004nb}.  Nevertheless, even in this case,
the bubbling picture survives in terms of boundary conditions corresponding
to either the $S^2$ or $S^5$ shrinking on the $y=0$ boundary plane.

Given the elegant bubbling description for 1/2 BPS configurations in both
the gauge theory and string theory side of the AdS/CFT correspondence, it
is natural to extend the above LLM investigation to both 1/4 BPS
\cite{Ryzhov:2001bp,D'Hoker:2001bq,Dolan:2002zh,D'Hoker:2003vf,Dolan:2007rq}
and 1/8 BPS \cite{Mikhailov:2000ya,Biswas:2006tj,Mandal:2006tk} configurations.
While there are several possibilities for obtaining reduced supersymmetry,
we are primarily interested in backgrounds with multiple commuting $R$-charges
turned on.  For $\mathcal N=4$ super-Yang Mills, as well as the dual
description of IIB on AdS$_5\times S^5$, the relevant supergroup is
$PSU(2,2|4)$, which admits the bosonic subgroup $SO(2,4)\times SO(6)$.
On the gravity side, states may be labeled by $(\Delta, S_1, S_2)$ for
energy and spin in AdS$_5$ and $(J_1, J_2, J_3)$ for angular momentum
on $S^5$.  Focusing on the chiral primaries, we take $s$-wave states
in AdS$_5$ satisfying $\Delta=J_1+J_2+J_3$.  Given that the BPS
condition takes the form
\begin{equation}
\Delta\ge\pm gS_1\pm gS_2\pm J_1\pm J_2\pm J_3
\end{equation}
(with an even number of minus signs, and with $g$ the inverse radius of
AdS$_5$),
we see that the generic state with three non-vanishing $R$-charges preserves
1/8 of the supersymmetries.  When $J_3=0$, the eigenvalues of the
Bogomol'nyi matrix pair up, and we are left with a 1/4 BPS state.  Finally,
when $J_2=J_3=0$, the system reduces to the familiar 1/2 BPS case.

When gravitational backreaction is taken into account, the turning on of
$J_1$, $J_2$ and $J_3$ in succession breaks the isometries of the five-sphere
from $SO(6)$ to $SO(4)$, $SO(2)$ and finally the identity.  Combining this
with the unbroken $SO(4)$ isometry of $s$-wave states in AdS$_5$, the
natural family of backgrounds we are interested in takes the form
\begin{equation}
\begin{tabular}{lll}
supersymmetries&chiral primary&isometry\\
\hline
1/2 BPS&$\Delta=J_1$&$S^3\times S^3$\\
1/4 BPS&$\Delta=J_1+J_2$&$S^3\times S^1$\\
1/8 BPS&$\Delta=J_1+J_2+J_3$&$S^3$\\
\end{tabular}
\end{equation}
In this paper, our main  interest is the supergravity description
of such backgrounds.  The 1/2 BPS case was of course the subject of LLM
\cite{Lin:2004nb} and related investigations.  The other two cases have
generally received less attention.  However, the invariant tensor analysis
of \cite{Tod:1983pm,Tod:1995jf,Gauntlett:2002sc,Gauntlett:2002nw} has
recently been applied towards the construction of supergravity backgrounds
corresponding to these two cases.  Backgrounds with $S^3\times S^1$
isometry were initially examined in \cite{Donos:2006iy}, and subsequent
gauging of the $U(1)$ isometry was considered in \cite{Donos:2006ms}.
In addition, solutions preserving an $S^3$ isometry (corresponding to the
1/8 BPS case) may be obtained by double analytic continuation of the
AdS$_3$ solutions investigated in \cite{Kim:2005ez}, as it was later done in 
\cite{Gauntlett:2006ns}.  (Note that 1/4 BPS
and 1/8 BPS solutions of a different nature were also investigated in
\cite{Liu:2004ru} and \cite{Gava:2006pu}, respectively.)

In both cases of $S^3$ isometry \cite{Kim:2005ez} and
$S^3\times S^1$ isometry \cite{Donos:2006iy,Donos:2006ms}, the
invariant tensor analysis and resulting description of the backgrounds
are essentially complete.  However, unlike for LLM geometries, in these
cases the supersymmetry analysis is not particularly constructive.
For example, it was found in \cite{Kim:2005ez} that 1/8 BPS configurations
with an $S^3$ isometry may be written using a metric of the form
\begin{equation}
ds_{10}^2=-e^{2\alpha}(dt+\omega)^2+e^{-2\alpha}h_{ij}dx^idx^j+e^{2\alpha}
d\Omega_3^2,
\label{eq:exksix}
\end{equation}
where $h_{ij}$ is a K\"ahler metric of complex dimension three.  In the end,
the invariant tensor analysis does not provide an actual procedure for 
obtaining this metric short of solving a non-linear equation on its curvature \cite{Kim:2005ez}
\begin{equation}
\square_6R=-R_{ij}R^{ij}+\ft12R^2.
\label{eq:exkim}
\end{equation}
Similarly, the 1/4 BPS analysis of \cite{Donos:2006iy,Donos:2006ms} leads
to a non-linear equation of Monge-Amp\`ere type related to the properties of
the K\"ahler metric on a base of complex dimension two.

Although the presence of such non-linear equations complicates the
analysis of 1/4 and 1/8 BPS states, it is nevertheless possible to
develop a robust picture of bubbling AdS even without complete
knowledge of the supergravity solution.  The main point here is
that the supergravity backgrounds are determined not only by the
imposition of local conditions such as (\ref{eq:exkim}), but also
by the boundary conditions.  In particular, turning back to the
LLM case, we recall that the droplet picture really originates
from the LLM boundary conditions (\ref{eq:llmbc}) imposed to
ensure regularity of the geometry and not directly from the
harmonic function equation (\ref{eq:bubble}).  The LLM boundary
conditions $Z(x_1,x_2,0)=\pm1/2$ ensure that the metric remains
smooth wherever either of the $S^3$ factors collapses to zero
size. Likewise, 1/4 BPS configurations preserving an $S^3\times
S^1$ isometry have potential singularities in the metric whenever
either the $S^3$ or $S^1$ collapses.  Avoiding such singularities
then demands similar boundary conditions: $Z(x_i,y=0)=\pm1/2$,
where this time $i=1,\ldots,4$ and
\begin{equation}
ds_{10}^2=-h^{-2}(dt+\omega)^2+y^{-1}e^{-G}h_{ij}dx^idx^j+h^2dy^2
+y(e^Gd\Omega_3^2+e^{-G}(d\psi+\mathcal A)^2).
\end{equation}
Note that $h^{-2}=2y\cosh G$ is unchanged from the LLM case.  What is
different, however, is that now the metric $h_{ij}$ (as well as the function
$G$) appears rather complicated, and does not admit an easy construction.

The bubbling AdS description of 1/8 BPS configurations is particularly
interesting in that it constitutes the most general case of turning on
all three commuting $R$-charges.  Since the 1/8 BPS metric, given in
(\ref{eq:exksix}), does not involve a $y$ coordinate, there is no 1/8 BPS
equivalent of an LLM $y=0$ phase-space plane.  Nevertheless, the K\"ahler
base can be given in terms of six real coordinates, $x_i$, $i=1,\ldots,6$.
As highlighted in \cite{Berenstein:2007wz}, it is natural to associate
these coordinates with the six real adjoint scalars of the dual $\mathcal
N=4$ super-Yang Mills theory.  In this picture, the eigenvalue distribution
from the matrix description maps into configurations in $\mathbb R^6$
corresponding to the degeneration locus of the $S^3$ in AdS$_5$.  From
the gravity side, this indicates that the six-dimensional base has regions
removed, with the boundary of such regions dual to the eigenvalue
distribution.  The AdS$_5\times S^5$ `ground state' corresponds to removing
a ball from the center of $\mathbb R^6$, and the addition of dual giant
gravitons corresponds to removing other disconnected regions as well.
Although the six-dimensional metric becomes singular as one approaches the
boundary, it must behave in such a manner that, when combined with the
shrinking $S^3$, the full ten-dimensional metric remains regular.

It is the aim of this paper to elucidate the bubbling picture of both 1/4
and 1/8 BPS configurations that we have sketched above, and to justify the
connection between boundary conditions and droplets in an effective
phase-space description of these geometries.  Before we do so, however, we
present a unified treatment of the invariant tensor analysis for 1/8, 1/4
and 1/2 BPS configurations.  In particular, based on symmetry conditions,
we may start with IIB supergravity with the self-dual five-form active,
and perform a breathing mode reduction to seven dimensions on $S^3$.  This
seven-dimensional system is the natural place to start from when discussing
1/8 BPS configurations.  A further reduction on $S^1$ brings the system
down to six dimensions (and allows a description of 1/4 BPS geometries).
Because of the abelian $U(1)$ isometry, we allow a gauge field to be turned
on in this reduction \cite{Donos:2006ms}.  Finally, we may reduce this
system to four dimensions on $CP^1$.  A generic configuration with $S^3\times
S^1\times CP^1$ isometry will preserve 1/4 of the supersymmetries
\cite{Gava:2006pu}.  However, by making use of the Hopf fibration of $S^3$
as $U(1)$ bundled over $CP^1$, we may recover the round $S^3\times S^3$
background of LLM, thus giving rise to the 1/2 BPS system.

Following the chain of breathing mode reductions and the
supersymmetry analysis, we discuss how the bubbling AdS picture
arises in the 1/4 and 1/8 BPS sectors.  Essentially, this is based
on an investigation of the boundary conditions needed to maintain
a smooth geometry wherever any of the various spheres degenerate
to zero size.  Because of the difficulty in providing a
constructive method for obtaining the full supergravity
backgrounds, we will mainly support our arguments with a set of
examples, which we treat separately for the 1/8 BPS and 1/4 BPS
cases. Readers who wish to skip the details of the breathing mode
reductions and invariant tensor analyses are invited to proceed
directly to Section~\ref{sec:bubble}, where the bubbling AdS
description is taken up.

The main technical results of this paper are presented in the following
two sections.  In Section~\ref{sec:breathing}, we perform a chain of
breathing mode reductions, starting with $S^3$, then adding $S^1$
and finally adding $CP^1$.  This allows us to write down effective
seven, six and four-dimensional theories governing 1/8, 1/4 and 1/2 BPS
configurations, respectively.  The supersymmetry analysis is then taken up
in Section~\ref{sec:susyanal};
%much of the material here is a review
this is intended to give a unified treatment of \cite{Kim:2005ez},
\cite{Donos:2006iy,Donos:2006ms}, and \cite{Lin:2004nb}, for the
1/8, 1/4 and 1/2 BPS cases, respectively, and show how the ansatz
of these three cases are embedded into each other. The remaining
parts of this paper are devoted to the development of the bubbling
AdS description of 1/4 and 1/8 BPS states.  In
Section~\ref{sec:bubble}, we present a brief summary of the
supergravity backgrounds, and then show how the LLM boundary
conditions generalize to provide a uniform droplet picture which
survives the reduction from 1/2 BPS down to 1/4 and 1/8 BPS
configurations.  We then turn to examples of 1/8 BPS geometries in
Section~\ref{EightBPSsection} followed by 1/4 BPS geometries in
Section~\ref{QuarterBPSsection}.  In Section~\ref{regular} we
return to the local conditions on the K\"ahler metric for 1/8 BPS
configurations and investigate in particular the interplay between
boundary conditions and regularity of the metric.  Finally, we
conclude in Section~\ref{sec:conclusion} with a summary of the 1/8
BPS droplet picture and how it also encompasses 1/4 and 1/2 BPS
states as special cases.  Various technical details are relegated
to the appendices.

%%%%%%%%%%%%%%%%%%%%%%%%%%%%%%%%%%%%%%%%
\section{Breathing mode compactifications of IIB supergravity}
\label{sec:breathing}
%%%%%%%%%%%%%%%%%%%%%%%%%%%%%%%%%%%%%%%%

The bosonic fields of IIB supergravity are given by the NSNS fields
$g_{MN}$, $B_{MN}$ and $\phi$ as well as the RR field strengths
$F_{(1)}$, $F_{(3)}$ and $F_{(5)}^+$, while the fermionic fields are
the (complex Weyl) gravitino $\Psi_M$ and dilatino $\lambda$, both
transforming with definite chirality in $D=10$.  Because we are
interested in describing giant graviton configurations, which are
essentially built out of D3-branes, we will only concern ourselves
with the self-dual five-form $F_{(5)}^+$ in addition to the metric.
In this sector, the IIB theory admits a particularly simple bosonic
truncation with equations of motion
\begin{equation}
R_{MN}=\fft1{4\cdot4!}(F^2)_{MN},\qquad
F_{(5)}=*F_{(5)},\qquad dF_{(5)}=0.
\label{eq:iibeom}
\end{equation}
The corresponding Lagrangian is given by
\begin{equation}
e^{-1}\mathcal L_{10}=R-\fft1{4\cdot5!}F_{(5)}^2,
\label{eq:iiblag}
\end{equation}
where self-duality of $F_{(5)}$ is to be imposed only after deriving
the equations of motion.

In the absence of the IIB dilaton/axion and three-form field strengths,
the dilatino transformation becomes trivial.  Thus the only relevant
supersymmetry transformation is that of the gravitino, which becomes
\begin{equation}
\delta\Psi_M=\left[\nabla_M+\fft{i}{16\cdot5!}F_{NPQRS}\Gamma^{NPQRS}
\Gamma_M\right]\epsilon.
\label{eq:iibsusy}
\end{equation}
Note that there is a delicate balance between self-duality of
$F_{(5)}$ and the chirality of the spinor parameter $\epsilon$.
With the natural definition of self-duality $F_{M_1\cdots M_5}=\fft1{5!}
\epsilon_{M_1\cdots M_5}{}^{N_1\cdots N_5}F_{N_1\cdots N_5}$, the spinor
$\epsilon$ satisfies $\Gamma^{11}\epsilon=\epsilon$ where $\Gamma^{11}
=\fft1{10!}\epsilon_{M_1\cdot M_{10}}\Gamma^{M_1\cdots M_{10}}$.

The bubbling configurations that we are interested in always preserve
an $S^3$ in AdS$_5$.  However, the isometries of the $S^5$ are naturally
broken depending on the amount of angular momentum (or $R$-charge)
$(J_1,J_2,J_3)$ turned on.  As in \cite{Lin:2004nb}, for 1/2 BPS
configurations we take $J_2=J_3=0$, and the resulting internal isometry
is that of $S^3$.  For 1/4 BPS configurations \cite{Donos:2006iy}
we have $J_3=0$ and hence $S^1$ isometry.  The generic 1/8 BPS case
has all three angular-momenta non-vanishing, resulting in the loss of
all manifest isometry of the original $S^5$.

It is then clear that, to capture this family of solutions, we ought
to consider breathing mode reductions of (\ref{eq:iiblag}) and
(\ref{eq:iibsusy}) on $S^3$, $S^3\times S^1$ and $S^3\times S^3$,
respectively, for 1/8, 1/4 and 1/2 BPS geometries.  It is natural
to proceed with this reduction in steps, at each stage adding additional
symmetries to the system.  Adding a $U(1)$ isometry to the $S^3$ reduction
is straightforward, and a natural way to obtain $S^3\times S^3$ from
$S^3\times U(1)$ is to use the Hopf fibration of the second $S^3$ as
a $U(1)$ bundle over $CP^1$.  This chain of reductions also provides
a natural way of understanding the embedding of 1/2 BPS configurations
into the 1/4 BPS system, and then finally into the 1/8 BPS case.

We note that Kaluza-Klein sphere reductions have been extensively
studied in the literature.  However, the main feature of the
present set of reductions is the inclusion of breathing (and possibly
squashing) modes \cite{Bremer:1998zp}.  Although these bosonic reductions
are consistent (as any truncation to the singlet sector would be
\cite{Duff:1985jd}), the resulting theory is however not supersymmetric,
as the breathing and squashing modes are in general part of the massive
Kaluza-Klein tower.  Nevertheless, it is still instructive to reduce
the original IIB Killing spinor equation (\ref{eq:iibsusy}) along with
the bosonic sector fields.  In this way, any solution to the reduced Killing
spinor equations may then be lifted to yield a supersymmetric background
of the original IIB theory.  Breathing mode reductions of the supersymmetry
variations were previously investigated in \cite{Liu:2000gk}, and in the
LLM context in \cite{Liu:2004ru,Liu:2004hy}.

%%%%%%%%%%%%%%%%%%%%%%%%%%%%%%%%%%%%%%%%
\subsection{$S^3$ reduction to $D=7$}
\label{sec:d=7}

The first stage of the reduction, corresponding to the generic 1/8 BPS case,
is to highlight the $S^3$ isometry inside AdS$_5$, which we always retain.
We thus take a natural reduction ansatz of the form
\begin{eqnarray}
ds_{10}^2&=&ds_7^2+e^{2\alpha}d\Omega_3^2,\nonumber\\
{}^{10}F_{(5)}&=&F_{(2)}\wedge\omega_3 + \widetilde F_{(5)}.
\label{eq:7ans}
\end{eqnarray}
note that self-duality of ${}^{10}F_{(5)}$ imposes the conditions
\begin{equation}
F_{(2)}=-e^{3\alpha}*_7\widetilde F_{(5)},\qquad
\widetilde F_{(5)}=e^{-3\alpha}*_7F_{(2)}.
\end{equation}

The ten-dimensional Einstein equation in (\ref{eq:iibeom}) reduces to
yield the seven-dimensional Einstein equation
\begin{equation}
R_{\mu\nu}-\ft12g_{\mu\nu}R=3(\partial_\mu\alpha\partial_\nu\alpha
-2g_{\mu\nu}(\partial\alpha)^2+\nabla_\mu\nabla_\nu\alpha-g_{\mu\nu}
\square\alpha)+\ft12e^{-6\alpha}[F^2{}_{\mu\nu}-\ft14g_{\mu\nu}F^2]
+3e^{-2\alpha}
\end{equation}
(in the `string frame'), as well as the scalar equation of motion
\begin{equation}
\partial^\mu(3\alpha)\partial_\mu\alpha+\square\alpha=
-\ft18e^{-6\alpha}F^2+2e^{-2\alpha}.
\end{equation}
In addition, the $F_{(5)}$ Bianchi identity and equation of motion
in (\ref{eq:iibeom}) reduce to their seven-dimensional counterparts
\begin{equation}
dF_{(2)}=0,\qquad d(e^{-3\alpha}*_7F_{(2)})=0.
\end{equation}
The above equations of motion may be obtained from an effective
seven-dimensional Lagrangian
\begin{equation}
e^{-1}\mathcal L_7=e^{3\alpha}[R+6(\partial\alpha)^2
-\ft14e^{-6\alpha}F_{(2)}^2+6e^{-2\alpha}].
\label{eq:7lag}
\end{equation}
The run-away potential term arises because of the curvature of the
reduction $S^3$, and will remain unbalanced until the second $S^3$
is introduced.

\subsubsection{Supersymmetry variations}

In order to study supersymmetric configurations, we must also examine
the reduction of the gravitino variation (\ref{eq:iibsusy}).  In
order to do so, we choose a Dirac decomposition of the form
\begin{equation}
\Gamma_\mu=\gamma_\mu\otimes1\otimes\sigma_1,\qquad
\Gamma_a=1\otimes\sigma_a\otimes\sigma_2.
\end{equation}
Defining the 10-dimensional chirality matrix as
$\Gamma^{11}=\fft1{10!} \epsilon_{M_1\cdots
M_{10}}\Gamma^{M_1\cdots M_{10}}$, we find
$\Gamma^{11}=-1\otimes1\otimes\sigma_3$ where we have taken the
seven-dimensional Dirac matrices to satisfy
$\fft1{7!}\epsilon_{\mu_1\cdots\mu_7}\gamma^{\mu_1\cdots\mu_7}=1$.
In this case, the IIB chirality condition
$\Gamma^{11}\epsilon=\epsilon$ translates into the condition that
$\epsilon$ has negative $\sigma_3$ eigenvalue.  This allows us to
decompose the complex IIB spinor as
${}^{10}\epsilon=\epsilon\otimes
\chi\otimes\genfrac{[}{]}{0pt}{}01$ where $\chi$ is a
two-component spinor on $S^3$ satisfying the Killing spinor
equation
\begin{equation}
\left[\hat\nabla_a+\fft{i\eta}2\hat\sigma_a\right]\chi=0,
\label{s3spinor}
\end{equation}
with $\eta=\pm1$.

Using the above decomposition, the 10-dimensional gravitino variation
(\ref{eq:iibsusy}) decomposes into a seven-dimensional `gravitino'
variation
\begin{equation}
\delta\psi_\mu=\left[\nabla_\mu-\fft{i}{16}e^{-3\alpha}F_{\nu\lambda}
\gamma^{\nu\lambda}\gamma_\mu\right]\epsilon,
\label{eq:7gto}
\end{equation}
as well as a `dilatino' variation
\begin{equation}
\delta\lambda=\left[\gamma^\mu\partial_\mu\alpha
+\fft{i}8e^{-3\alpha}F_{\mu\nu}\gamma^{\mu\nu}-\eta e^{-\alpha}\right]
\epsilon
\label{eq:7dto}
\end{equation}
which arises from the components of (\ref{eq:iibsusy}) living on the
$S^3$.  We emphasize here that these are not necessarily the
transformations of any actual seven-dimensional supersymmetric model,
as we only claim the bosonic sector to form a consistent truncation of
the original IIB theory.  Nevertheless, based on their structure,
it is useful to think of these as would-be gravitino and dilatino
variations.  So long as these two `Killing spinor equations' are
satisfied, we are guaranteed that the lifted solution is a
supersymmetric configuration of the original IIB theory.

%%%%%%%%%%%%%%%%%%%%%%%%%%%%%%%%%%%%%%%%
\subsection{Additional reduction on $U(1)$ to $D=6$}
\label{sec:d=6}

In order to describe 1/4 BPS geometries with $S^3\times S^1$ isometry,
we may further reduce the seven-dimensional system (\ref{eq:7lag}) to
$D=6$ along a $U(1)$ direction.  This follows by a traditional Kaluza-Klein
circle reduction, where we take
\begin{eqnarray}
ds_7^2&=&ds_6^2+e^{2\beta}(d\psi+\mathcal A)^2,\nonumber\\
{}^7F_{(2)}&=&F_{(2)}+d\chi\wedge(d\psi+\mathcal A).
\label{eq:6ans}
\end{eqnarray}
This is the most general ansatz consistent with $U(1)$ isometry, and
includes an axionic scalar $\chi$ which in the original IIB picture
corresponds to five-form flux on $S^3\times S^1$ along with a non-compact
dimension.  For a pure bubbling picture with $S^3$ inside AdS$_5$ and $S^1$
independently inside $S^5$, we would want to set $\chi=0$.  However
doing so at this stage would lead to an inconsistent truncation as
demonstrated below.  We thus prefer to work with the most general $U(1)$
reduction including $\chi$ at this stage.

The resulting six-dimensional Einstein equation is
\begin{eqnarray}
R_{\mu\nu}-\ft12g_{\mu\nu}R&=&\ft14\partial_\mu(3\alpha+\beta)
\partial_\nu(3\alpha+\beta)-\ft58g_{\mu\nu}(\partial(3\alpha+\beta))^2
+\nabla_\mu\nabla_\nu(3\alpha+\beta)-g_{\mu\nu}\square(3\alpha+\beta)
\nonumber\\
&&+\ft34[\partial_\mu(\alpha-\beta)\partial_\nu(\alpha-\beta)-\ft12g_{\mu\nu}
(\partial(\alpha-\beta))^2]+\ft12e^{-6\alpha-2\beta}[\partial_\mu\chi
\partial_\nu\chi-\ft12g_{\mu\nu}(\partial\chi)^2]\nonumber\\
&&+\ft12e^{-6\alpha}[F^2{}_{\mu\nu}-\ft14g_{\mu\nu}F^2]
+\ft12e^{2\beta}[\mathcal F^2{}_{\mu\nu}-\ft14g_{\mu\nu}\mathcal F^2]
+3g_{\mu\nu}e^{-2\alpha}
\end{eqnarray}
and the scalar equations are
\begin{eqnarray}
\partial^\mu(3\alpha+\beta)\partial_\mu\alpha+\square\alpha&=&
-\ft14e^{-6\alpha-2\beta}(\partial\chi)^2-\ft18e^{-6\alpha}F^2+2e^{-2\alpha},
\nonumber\\
\partial^\mu(3\alpha+\beta)\partial_\mu\beta+\square\beta&=&
-\ft14e^{-6\alpha-2\beta}(\partial\chi)^2+\ft18e^{-6\alpha}F^2
+\ft14e^{2\beta}\mathcal F^2,\nonumber\\
\partial^\mu(-3\alpha-\beta)\partial_\mu\chi+\square\chi&=&
\ft12e^{2\beta}F_{\mu\nu}\mathcal F^{\mu\nu}.
\label{eq:6seom}
\end{eqnarray}
In addition, the field strengths satisfy the Bianchi identities and
equations of motion
\begin{eqnarray}
d\mathcal F&=&0,\hphantom{d\chi\wedge\mathcal F}\qquad
d(e^{3\alpha+3\beta}*_6\mathcal F)\,=-e^{-3\alpha+\beta}
*_6F\wedge d\chi,\nonumber\\
dF&=&d\chi\wedge\mathcal F,\hphantom{0}\qquad d(e^{-3\alpha+\beta}*_6F)=0.
\label{eq:6bieom}
\end{eqnarray}
The above equations of motion may be derived from an effective six-dimensional
Lagrangian
\begin{equation}
e^{-1}\mathcal L_6=e^{3\alpha+\beta}[R+\ft34(\partial(3\alpha+\beta))^2
-\ft34(\partial(\alpha-\beta))^2-\ft12e^{-6\alpha-2\beta}(\partial\chi)^2
-\ft14e^{-6\alpha}F_{(2)}^2-\ft14e^{2\beta}\mathcal F_{(2)}^2+6e^{-2\alpha}],
\end{equation}
where $F_{(2)}=dA_{(1)}+\chi\mathcal F_{(2)}$.

Note that if we were to take $\chi=0$, its equation of motion (\ref{eq:6seom})
would demand the constraint $F_{\mu\nu}\mathcal F^{\mu\nu}=0$.  This is
consistent with the independence of the $S^3$ in AdS$_5$ and $S^1$ in $S^5$
sectors, where $F_{(2)}$ lives in AdS$_5$ while $\mathcal F_{(2)}$ lives
in $S^5$.

To make a connection with the 1/4 BPS geometries investigated in
\cite{Donos:2006iy,Donos:2006ms}, we may let
\begin{equation}
\alpha=\ft12(H+G),\qquad\beta=\ft12(H-G).
\label{eq:donosHG}
\end{equation}
This results in a metric reduction of the form
\begin{equation}
ds_{10}^2=ds_6^2+e^H[e^Gd\Omega_3^2+e^{-G}(d\psi+\mathcal A)^2],
\end{equation}
as well as an effective Lagrangian
\begin{equation}
e^{-1}\mathcal L_6=e^{2H+G}[R+\ft34(\partial(2H+G))^2-\ft34(\partial G)^2
-\ft14e^{-3(H+G)}F^2-\ft14e^{H-G}\mathcal F^2+6e^{-(H+G)}].
\label{eq:dlag}
\end{equation}
Note that we have set $\chi=0$.  So, in addition to (\ref{eq:dlag}), we
must also impose the constraint $F_{\mu\nu}\mathcal F^{\mu\nu}=0$ indicated
above.

\subsubsection{Supersymmetry variations}

{}From the seven-dimensional point of view, the supersymmetry conditions
are encoded in the gravitino and dilatino variations
(\ref{eq:7gto}) and (\ref{eq:7dto}).  Given the bosonic reduction
(\ref{eq:6ans}), the supersymmetry variations are easily reduced
along the $U(1)$ fiber to give rise to six-dimensional variations.
In particular, we may use the straightforward relation between six
and seven-dimensional Dirac matrices
\begin{equation}
\gamma_\mu\to\begin{cases}\gamma_\mu&\mu=0,\ldots,5,\\
\gamma^7\equiv\fft1{6!}\epsilon_{\mu_1\cdots\mu_6}\gamma^{\mu_1\cdots
\mu_6}&\mu=6,\end{cases}
\end{equation}
and no additional Dirac decomposition is needed.

With this convention, the two-form field strength reduces according to
\begin{equation}
^7F_{\mu\nu}\gamma^{\mu\nu}=F_{\mu\nu}\gamma^{\mu\nu}+2e^{-\beta}
\gamma^\mu\gamma^7\partial_\mu\chi,
\end{equation}
while the spin connections reduce according to
\begin{equation}
^7\omega^{\alpha\gamma}=\omega^{\alpha\gamma}-\ft12e^\beta
\mathcal F^{\alpha\gamma}e^7,\qquad
{}^7\omega^{\alpha7}=-e^{\mu\,\alpha}\partial_\mu\beta e^7-\ft12e^\beta
\mathcal F^{\alpha\gamma}e^\gamma.
\end{equation}
In order to properly reduce the covariant derivative $^7\nabla_\mu$
appearing in the gravitino variation (\ref{eq:7gto}), we must keep
in mind that Killing spinors $\epsilon$ may in fact be charged along
the $U(1)$ fiber \cite{Liu:2004ru}.  We thus take
\begin{equation}
\partial_\psi\leftrightarrow-\ft{i}2n,
\end{equation}
where $n\in\mathbb Z$, and the sign is chosen for later convenience.
This integral choice of $n$ corresponds to the period of $\psi$ being
$2\pi$.

Putting the above together, we find the six-dimensional `gravitino'
variation
\begin{equation}
\delta\psi_\mu=\left[\nabla_\mu+\fft{in}2\mathcal A_\mu
-\fft{i}{16}e^{-3\alpha}F_{\nu\lambda}\gamma^{\nu\lambda}\gamma_\mu
+\fft14e^\beta\mathcal F_{\mu\nu}\gamma^\nu\gamma^7
+\fft{i}8e^{-3\alpha-\beta}\gamma^\nu\partial_\nu\chi
\gamma_\mu\gamma^7\right]\epsilon,
\label{eq:6gto}
\end{equation}
as well as the two `dilatino' variations
\begin{eqnarray}
\delta\lambda_\alpha&=&\left[\gamma^\mu\partial_\mu\alpha+\fft{i}8
e^{-3\alpha}F_{\mu\nu}\gamma^{\mu\nu}+\fft{i}4e^{-3\alpha-\beta}
\gamma^\mu\partial_\mu\chi\gamma^7-\eta e^{-\alpha}\right]\epsilon,\nonumber\\
\delta\lambda_\beta&=&\left[\gamma^\mu\partial_\mu\beta
-\fft{i}8e^{-3\alpha}F_{\mu\nu}\gamma^{\mu\nu}
-\fft14e^\beta\mathcal F_{\mu\nu}\gamma^{\mu\nu}\gamma^7
+\fft{i}4e^{-3\alpha-\beta}\gamma^\mu\partial_\mu\chi
\gamma^7-ine^{-\beta}\gamma^7\right]\epsilon.
\label{eq:6dto}
\end{eqnarray}
Here $\lambda_\alpha$ is identical to $\lambda$ given in (\ref{eq:7dto}),
while $\lambda_\beta=2\gamma^7\psi_7$.  These variations are for the
general reduction, including the axionic scalar $\chi$.  If desired,
we may truncate to $\chi=0$ and furthermore make the substitution
(\ref{eq:donosHG}) to arrive at the transformations
\cite{Donos:2006iy,Donos:2006ms}
\begin{eqnarray}
\delta\psi_\mu&=&\left[\nabla_\mu+\fft{in}2\mathcal A_\mu
+\fft14e^{\fft12(H-G)}\mathcal F_{\mu\nu}\gamma^\nu\gamma^7
-\fft{i}{16}e^{-\fft32(H+G)}F_{\nu\lambda}\gamma^{\nu\lambda}\gamma_\mu
\right]\epsilon,\nonumber\\
\delta\lambda_H&=&\left[\gamma^\mu\partial_\mu H-\fft14e^{\fft12(H-G)}
\mathcal F_{\mu\nu}\gamma^{\mu\nu}\gamma^7-\eta e^{-\fft12(H+G)}
-ine^{-\fft12(H-G)}\gamma^7\right]\epsilon,\nonumber\\
\delta\lambda_G&=&\left[\gamma^\mu\partial_\mu G+\fft14e^{\fft12(H-G)}
\mathcal F_{\mu\nu}\gamma^{\mu\nu}\gamma^7+\fft{i}4e^{-\fft32(H+G)}
F_{\mu\nu}\gamma^{\mu\nu}-\eta e^{-\fft12(H+G)}+ine^{-\fft12(H-G)}
\gamma^7\right]\epsilon,\nonumber\\
\label{eq:6susy}
\end{eqnarray}
corresponding to the truncated Lagrangian of (\ref{eq:dlag}).

%%%%%%%%%%%%%%%%%%%%%%%%%%%%%%%%%%%%%%%%
\subsection{The final reduction on $CP^1$ to $D=4$}
\label{sec:cp1red}

Noting that $S^3$ can be written as $U(1)$ bundled over $CP^1$, we may
obtain an $S^3\times S^3$ solution by reducing the effective six-dimensional
system to four dimensions on $CP^1$.  This procedure will actually allow
for more general geometries, where the second $S^3$ is squashed along the
$U(1)$ fiber.  The generic $(\mbox{squashed }S^3)\times(\mbox{round }S^3)$
system has $SU(2)\times U(1)\times SO(4)$ isometry, and was investigated
in \cite{Gava:2006pu}.

The $CP^1$ reduction proceeds by taking
\begin{eqnarray}
ds_6^2&=&ds_4^2+e^{2\gamma}ds^2(CP^1),\nonumber\\
{}^6F_{(2)}&=&F_{(2)}+2m\chi J,\nonumber\\
{}^6\mathcal F_{(2)}&=&\mathcal F_{(2)}+2mJ,
\label{eq:4ans}
\end{eqnarray}
where $J_{(2)}$ is the K\"ahler form on $CP^1$.  We take the standard
Einstein metric on $CP^1$ with $\hat R_{ab}=\lambda\hat g_{ab}$.

Although the reduction is straightforward, the intermediate steps are
somewhat tedious.  We end up with a four-dimensional Einstein equation
of the form
\begin{eqnarray}
R_{\mu\nu}-\ft12g_{\mu\nu}R&=&\ft16\partial_\mu(3\alpha+\beta+2\gamma)
\partial_\nu(3\alpha+\beta+2\gamma)-\ft7{12}g_{\mu\nu}
(\partial(3\alpha+\beta+2\gamma))^2\nonumber\\
&&+\nabla_\mu\nabla_\nu(3\alpha+\beta+2\gamma)
-g_{\mu\nu}\square(3\alpha+\beta+2\gamma)\nonumber\\
&&+\ft16[\partial_\mu(3\alpha-\beta-2\gamma)\partial_\nu(3\alpha-\beta-2\gamma)
]-\ft12g_{\mu\nu}(\partial(3\alpha-\beta\--2\gamma))^2]\nonumber\\
&&+\ft23[\partial_\mu(\beta-\gamma)\partial_\nu(\beta-\gamma)
-\ft12g_{\mu\nu}(\partial(\beta-\gamma))^2]
+\ft12e^{-6\alpha-2\beta}[\partial_\mu\chi\partial_\nu\chi-\ft12g_{\mu\nu}
(\partial\chi)^2]\nonumber\\
&&+\ft12e^{-6\alpha}[F^2{}_{\mu\nu}-\ft14g_{\mu\nu}F^2]
+\ft12e^{2\beta}[\mathcal F^2{}_{\mu\nu}-\ft14g_{\mu\nu}\mathcal F^2]
\nonumber\\
&&+g_{\mu\nu}[3e^{-2\alpha}+\lambda e^{-2\gamma}
-m^2e^{2\beta-4\gamma}(1+e^{-6\alpha-2\beta}\chi^2)].
\end{eqnarray}
The three scalars $\alpha$, $\beta$ and $\gamma$ are non-canonically
normalized, while the axionic scalar $\chi$ is canonical.  The four
scalar equations of motion are
\begin{eqnarray}
\partial^\mu(3\alpha+\beta+2\gamma)\partial_\mu\alpha+\square\alpha
&=&-\ft14e^{-6\alpha-2\beta}(\partial\chi)^2-\ft18e^{-6\alpha}F^2
+2e^{-2\alpha}-m^2e^{-6\alpha-4\gamma}\chi^2,\nonumber\\
\partial^\mu(3\alpha+\beta+2\gamma)\partial_\mu\beta+\square\beta
&=&-\ft14e^{-6\alpha-2\beta}(\partial\chi)^2+\ft18e^{-6\alpha}F^2
+\ft14e^{2\beta}\mathcal F^2+m^2e^{-6\alpha-4\gamma}\chi^2\nonumber\\
&&+2m^2e^{2\beta-4\gamma},\nonumber\\
\partial^\mu(3\alpha+\beta+2\gamma)\partial_\mu\gamma+\square\gamma
&=&\ft14e^{-6\alpha-2\beta}(\partial\chi)^2+\ft18e^{-6\alpha}F^2
+\lambda e^{-2\gamma}-m^2e^{-6\alpha-4\gamma}\chi^2\nonumber\\
&&-2m^2e^{2\beta-4\gamma},\nonumber\\
\partial^\mu(-3\alpha-\beta+2\gamma)\partial_\mu\chi+\square\chi
&=&\ft12e^{2\beta}F_{\mu\nu}\mathcal F^{\mu\nu}+4m^2e^{2\beta-4\gamma}\chi,
\end{eqnarray}
while the field strengths satisfy the Bianchi identities and equations
of motion
\begin{eqnarray}
d\mathcal F&=&0,\hphantom{d\chi\wedge\mathcal F}\qquad
d(e^{3\alpha+3\beta+2\gamma}*_4\mathcal F)
=-e^{-3\alpha+\beta+2\gamma}*_4F\wedge d\chi,\nonumber\\
dF&=&d\chi\wedge\mathcal F,\hphantom{0}\qquad
d(e^{-3\alpha+\beta+2\gamma}*_4F)=0.
\end{eqnarray}
The four-dimensional Lagrangian which yields the above equations of
motion is then of the form
\begin{eqnarray}
e^{-1}\mathcal L_4&=&e^{3\alpha+\beta+2\gamma}[R
+\ft56(\partial(3\alpha+\beta+2\gamma))^2
-\ft16(\partial(3\alpha-\beta-2\gamma))^2
-\ft23(\partial(\beta-\gamma))^2\nonumber\\
&&-\ft12e^{-6\alpha-2\beta}(\partial\chi)^2
-\ft14e^{-6\alpha}F_{(2)}^2-\ft14e^{2\beta}\mathcal F_{(2)}^2
+6e^{-2\alpha}+2\lambda e^{-2\gamma}\nonumber\\
&&-2m^2e^{2\beta-4\gamma}(1+e^{-6\alpha-2\beta}\chi^2)].
\label{eq:4lag1}
\end{eqnarray}

Although we have introduced two constant parameters, $m$ [which is related to
the fibration in (\ref{eq:4ans})] and $\lambda$ (which is the curvature
of $CP^1$, $\hat R_{ab}=\lambda\hat g_{ab}$), they may be scaled away by
adjusting the breathing and squashing mode scalars $\beta$ and $\gamma$.
In particular, so long as $\lambda\ne0$ and $m\ne0$, we may set $m=\pm1$,
$\lambda=4$ by shifting the fields according to
\begin{eqnarray}
&&\beta\to\beta+\log(\lambda/4|m|),\qquad
\gamma\to\gamma+\ft12\log(\lambda/4),\nonumber\\
&&\chi\to\fft\lambda{4|m|}\chi,\qquad
\mathcal A_\mu\to\fft{4|m|}\lambda\mathcal A_\mu.
\label{eq:4scale}
\end{eqnarray}
Although this transformation rescales the effective Lagrangian by an overall
constant, this has no effect on the classical equations of motion.  Ignoring
this overall factor, (\ref{eq:4lag1}) takes on the parameter free form
\begin{eqnarray}
e^{-1}\mathcal L_4&=&e^{3\alpha+\beta+2\gamma}[R
+\ft56(\partial(3\alpha+\beta+2\gamma))^2
-\ft16(\partial(3\alpha-\beta-2\gamma))^2
-\ft23(\partial(\beta-\gamma))^2\nonumber\\
&&-\ft12e^{-6\alpha-2\beta}(\partial\chi)^2
-\ft14e^{-6\alpha}F_{(2)}^2-\ft14e^{2\beta}\mathcal F_{(2)}^2
+6e^{-2\alpha}+8 e^{-2\gamma}
-2e^{2\beta-4\gamma}(1+e^{-6\alpha-2\beta}\chi^2)].\nonumber\\
\end{eqnarray}

The above system allows for a general squashed $S^3$ geometry,
and corresponds to the case studied in \cite{Gava:2006pu}.
To obtain a round $S^3\times S^3$ reduction, we may take
\begin{equation}
\gamma=\beta,\qquad\chi=0,\qquad\mathcal F=0,
\label{eq:4round}
\end{equation}
where consistency of setting the scalars $\gamma$ and $\beta$ equal to
each other is ensured by the above choice of $|m|=1$ and $\lambda=4$.
The resulting truncation becomes
\begin{equation}
e^{-1}\mathcal L_4=e^{3(\alpha+\beta)}[R+\ft{15}2(\partial(\alpha+\beta))^2
-\ft32(\partial(\alpha-\beta))^2-\ft14e^{-6\alpha}F^2+6(e^{-2\alpha}+
e^{-2\beta})].
\end{equation}
Defining
\begin{equation}
\alpha=\ft12(H+G),\qquad\beta=\ft12(H-G)
\end{equation}
finally gives (\ref{eq:llmlag}), which was obtained in \cite{Liu:2004ru}
by direct $S^3\times S^3$ reduction of the LLM ansatz (\ref{eq:llmans}).

\subsubsection{Supersymmetry variations}

Turning to the supersymmetry variations, our aim is to reduce the
six-dimensional `gravitino' and `dilatino' variations (\ref{eq:6susy})
on $CP^1$ to four-dimensions.  To do so, we start by introducing a
Dirac decomposition
\begin{equation}
^6\gamma_\mu=\gamma_\mu\otimes1,\qquad
{}^6\gamma_a=\gamma^5\otimes\sigma_a,
\end{equation}
where $a=1,2$ correspond to the two directions on $CP^1$.  Note that
we define $\gamma^5=\fft{i}{4!}\epsilon_{\mu\nu\rho\sigma}
\gamma^{\mu\nu\rho\sigma}$, so that $\gamma^7=\fft1{6!}\epsilon_{\mu_1\cdots
\mu_6}{}^6\gamma^{\mu_1\cdots\mu_6}=\gamma^5\otimes\sigma_3$.

{}From (\ref{eq:4ans}), and the definition of the K\"ahler form, we see
that the two-form field strengths reduce according to
\begin{eqnarray}
^6F_{\mu\nu}\gamma^{\mu\nu}&=&F_{\mu\nu}\gamma^{\mu\nu}+4ime^{-2\gamma}
\chi\sigma_3,\nonumber\\
^6\mathcal F_{\mu\nu}\gamma^{\mu\nu}&=&\mathcal F_{\mu\nu}\gamma^{\mu\nu}
+4ime^{-2\gamma}\sigma_3.
\end{eqnarray}
Inserting this into (\ref{eq:6susy}) gives rise to a straightforward
reduction of the `dilatino' variations
\begin{eqnarray}
\delta\lambda_\alpha&=&\left[\gamma^\mu\partial_\mu\alpha
+\fft{i}8e^{-3\alpha}F_{\mu\nu}\gamma^{\mu\nu}+\fft{i}4e^{-3\alpha-\beta}
\gamma^\mu\partial_\mu\chi\gamma^5\sigma_3-\fft12me^{-3\alpha-2\gamma}\chi
\sigma_3-\eta e^{-\alpha}\right]\epsilon,\nonumber\\
\delta\lambda_\beta&=&\biggl[\gamma^\mu\partial_\mu\beta
-\fft{i}8e^{-3\alpha}F_{\mu\nu}\gamma^{\mu\nu}-\fft14e^\beta\mathcal F_{\mu\nu}
\gamma^{\mu\nu}\gamma^5\sigma_3+\fft{i}4e^{-3\alpha-\beta}\gamma^\mu
\partial_\mu\chi\gamma^5\sigma_3\nonumber\\
&&+\fft12me^{-3\alpha-2\gamma}\chi\sigma_3
-i(me^{\beta-2\gamma}+ne^{-\beta}\sigma_3)\gamma^5\biggr]\epsilon.
\end{eqnarray}
In order to reduce the `gravitino' variation, we use the spin connections
\begin{equation}
^6\omega^{\alpha\beta}=\omega^{\alpha\beta},\qquad
{}^6\omega^{\alpha b}=-e^{\mu\,\alpha}\partial_\mu\gamma e^b,\qquad
{}^6\omega^{ab}=e^{-\gamma}\hat\omega_c^{ab} e^c,
\end{equation}
where $\hat\omega_c^{ab}$ is the spin connection on $CP^1$.  This results
in the four-dimensional `gravitino' variation
\begin{eqnarray}
\delta\psi_\mu&=&\biggl[\nabla_\mu+\fft{in}2\mathcal A_\mu
-\fft{i}{16}e^{-3\alpha}F_{\nu\lambda}\gamma^{\nu\lambda}\gamma_\mu
+\fft14e^\beta\mathcal F_{\mu\nu}\gamma^\nu\gamma^5\sigma_3
+\fft{i}8e^{-3\alpha-\beta}\gamma^\nu\partial_\nu\chi\gamma_\mu\gamma^5
\sigma_3\nonumber\\
&&\qquad+\fft14me^{-3\alpha-2\gamma}\chi\gamma_\mu\sigma_3\biggr]\epsilon,
\end{eqnarray}
as well as the variation on $CP^1$
\begin{eqnarray}
\delta\psi_i&=&\left[\hat\nabla_i+\fft{in}2\mathcal A_i\right]\epsilon
+\fft12e^\gamma\gamma^5\hat\sigma_i\biggl[\gamma^\mu\partial_\mu\gamma
-\fft{i}8e^{-3\alpha}F_{\mu\nu}\gamma^{\mu\nu}-\fft{i}4e^{-3\alpha-\beta}
\gamma^\mu\partial_\mu\chi\gamma^5\sigma_3\nonumber\\
&&\qquad-\fft12me^{-3\alpha-2\gamma}\chi\sigma_3+ime^{\beta-2\gamma}\gamma^5
\biggr]\epsilon.
\label{eq:4cp1var}
\end{eqnarray}

At this stage, there are several ways to proceed.  Since we are interested
in writing the squashed $S^3$ as $U(1)$ bundled over $CP^1$, we assume
from now on that both $\lambda$ and $m$ are non-vanishing.  In this case,
the scaling of (\ref{eq:4scale}) allows us to set $\lambda=4$ and
$m=\hat\eta$, where $\hat\eta=\pm1$ is a choice of sign.
Two-component Killing spinors $\hat\epsilon$ on the
squashed sphere can then be taken to either satisfy
\begin{equation}
\left[\hat\nabla_i+\fft{in}2\mathcal A_i\right]\hat\epsilon=0,
\qquad n\ne0,
\label{eq:4kse1}
\end{equation}
or
\begin{equation}
\left[\hat\nabla_i+\frac{i\hat\eta}{2}\hat\sigma_i\right]\hat\epsilon=0,
\qquad n=0. \label{eq:4kse2}
\end{equation}
This second possibility corresponds to ordinary Killing spinors on
$CP^1$.  However, the sign in the Killing spinor equation
(\ref{eq:4kse2}) is not arbitrary, but rather is fixed to ensure that these
Killing spinors descend properly from those on the squashed $S^3$.
At this point a note is in order concerning the $\hat\eta$, which is
the sign of $m$.  From (\ref{eq:4ans}), we my infer that
changing the sign of $\hat\eta$ corresponds to changing the sign of
the gauge bundle on the $U(1)$ fiber, which in term corresponds to
orientation reversal on the squashed $S^3$.  In general, orientation
issues may be rather subtle in squashed sphere compactifications, with
only one choice of sign yielding a supersymmetric configuration
\cite{Duff:1983nu,Duff:1986hr}.  It is for this reason that we have
kept $\hat\eta$ as a parameter.  Nevertheless, it is important to keep in mind
that $\hat\eta$ is a parameter specifying the bosonic field
configuration, and that changing the sign of $\hat\eta$ (flipping the
orientation) in principle changes the solution.  For this reason,
$\hat\eta$ ought to be thought of as a fixed constant, unlike the
Killing spinor sign parameters $\eta$ and $\widetilde\eta$ (defined below),
which may be chosen freely.

For the first case ($n\ne0$), the Killing spinors are charged along the
$U(1)$ fiber, but are (gauge) covariantly constant on $CP^1$.  Integrability
of (\ref{eq:4kse1}) shows that the $U(1)$ charge is given by $n=\pm2$,
with corresponding projection condition
\begin{equation}
\sigma_3\hat\epsilon=\widetilde\eta\hat\epsilon,
\label{eq:4proj1}
\end{equation}
where $\widetilde\eta=\pm1$.  The sign in the projection is correlated
with the $U(1)$ charge according to $n=-2\hat\eta\widetilde\eta$.
Taking these various signs into account, we end up with the `gravitino'
and `dilatino' variations
\begin{eqnarray}
\delta\psi_\mu&=&\biggl[\nabla_\mu-i\hat\eta\widetilde\eta\mathcal A_\mu
-\fft{i}{16}e^{-3\alpha}F_{\nu\lambda}\gamma^{\nu\lambda}\gamma_\mu
+\fft14\widetilde\eta e^\beta\mathcal F_{\mu\nu}\gamma^\nu\gamma^5
+\fft{i}8\widetilde\eta e^{-3\alpha-\beta}\gamma^\nu\partial_\nu\chi
\gamma_\mu\gamma^5\nonumber\\
&&\qquad+\fft14\hat\eta\widetilde\eta e^{-3\alpha-2\gamma}
\chi\gamma_\mu\biggr]\epsilon,\nonumber\\
\delta\lambda_\alpha&=&\left[\gamma^\mu\partial_\mu\alpha+\fft{i}8e^{-3\alpha}
F_{\mu\nu}\gamma^{\mu\nu}+\fft{i}4\widetilde\eta e^{-3\alpha-\beta}\gamma^\mu
\partial_\mu\chi\gamma^5-\fft12\hat\eta\widetilde\eta e^{-3\alpha-2\gamma}\chi
-\eta e^{-\alpha}\right]\epsilon,\nonumber\\
\delta\lambda_\beta&=&\biggl[\gamma^\mu\partial_\mu\beta-\fft{i}8e^{-3\alpha}
F_{\mu\nu}\gamma^{\mu\nu}-\fft14\widetilde\eta e^\beta\mathcal F_{\mu\nu}
\gamma^{\mu\nu}\gamma^5+\fft{i}4\widetilde\eta e^{-3\alpha-\beta}\gamma^\mu
\partial_\mu\chi\gamma^5+\fft12\hat\eta\widetilde\eta e^{-3\alpha-2\gamma}\chi
\nonumber\\
&&\qquad+i\hat\eta(2e^{-\beta}-e^{\beta-2\gamma})\gamma^5\biggr]
\epsilon,\nonumber\\
\delta\lambda_\gamma&=&\left[\gamma^\mu\partial_\mu\gamma-\fft{i}8e^{-3\alpha}
F_{\mu\nu}\gamma^{\mu\nu}-\fft{i}4\widetilde\eta e^{-3\alpha-\beta}\gamma^\mu
\partial_\mu\chi\gamma^5-\fft12\hat\eta\widetilde\eta e^{-3\alpha-2\gamma}\chi
+i\hat\eta e^{\beta-2\gamma}\gamma^5\right]\epsilon.
\label{eq:4var1}
\end{eqnarray}
Note that $\delta\lambda_\gamma$ is obtained from the gravitino variation
$\delta\psi_i$ on $CP^1$.  Because of the projection (\ref{eq:4proj1}),
a complete set of Killing spinors is obtained only after taking into account
both signs of $\widetilde\eta$.

For the second case ($n=0$), the Killing spinors are uncharged along the $U(1)$
fiber.  In this case, we end up with the variations
\begin{eqnarray}
\delta\psi_\mu&=&\biggl[\nabla_\mu
-\fft{i}{16}e^{-3\alpha}F_{\nu\lambda}\gamma^{\nu\lambda}\gamma_\mu
+\fft14\widetilde\eta e^\beta\mathcal F_{\mu\nu}\gamma^\nu\gamma^5
+\fft{i}8\widetilde\eta e^{-3\alpha-\beta}\gamma^\nu\partial_\nu\chi
\gamma_\mu\gamma^5
+\fft14\hat\eta\widetilde\eta e^{-3\alpha-2\gamma}
\chi\gamma_\mu\biggr]\epsilon,\nonumber\\
\delta\lambda_\alpha&=&\left[\gamma^\mu\partial_\mu\alpha+\fft{i}8e^{-3\alpha}
F_{\mu\nu}\gamma^{\mu\nu}+\fft{i}4\widetilde\eta e^{-3\alpha-\beta}\gamma^\mu
\partial_\mu\chi\gamma^5-\fft12\hat\eta\widetilde\eta e^{-3\alpha-2\gamma}\chi
-\eta e^{-\alpha}\right]\epsilon,\nonumber\\
\delta\lambda_\beta&=&\biggl[\gamma^\mu\partial_\mu\beta-\fft{i}8e^{-3\alpha}
F_{\mu\nu}\gamma^{\mu\nu}-\fft14\widetilde\eta e^\beta\mathcal F_{\mu\nu}
\gamma^{\mu\nu}\gamma^5+\fft{i}4\widetilde\eta e^{-3\alpha-\beta}\gamma^\mu
\partial_\mu\chi\gamma^5+\fft12\hat\eta\widetilde\eta e^{-3\alpha-2\gamma}\chi
\nonumber\\
&&\qquad-i\hat\eta e^{\beta-2\gamma}\gamma^5\biggr]
\epsilon,\nonumber\\
\delta\lambda_\gamma&=&\left[\gamma^\mu\partial_\mu\gamma-\fft{i}8e^{-3\alpha}
F_{\mu\nu}\gamma^{\mu\nu}-\fft{i}4\widetilde\eta e^{-3\alpha-\beta}\gamma^\mu
\partial_\mu\chi\gamma^5-\fft12\hat\eta\widetilde\eta e^{-3\alpha-2\gamma}\chi
-i\hat\eta(2e^{-\gamma}-e^{\beta-2\gamma})\gamma^5\right]\epsilon,\nonumber\\
\label{eq:4var2}
\end{eqnarray}
where $\delta\lambda_\gamma$ was obtained by substituting (\ref{eq:4kse2})
into (\ref{eq:4cp1var}).  Although no $\sigma_3$ projection is involved
in this case, it is nevertheless still convenient to break up the
Killing spinor expressions into definite $\sigma_3$ eigenvalues
corresponding to (\ref{eq:4proj1}).  In addition to the lack of
gauge connection $\mathcal A_\mu$ in the `gravitino' variation, these
expressions differ from those in the first case, (\ref{eq:4var1}), in
the `superpotential' gradient terms in the $\lambda_\beta$ and $\lambda_\gamma$
variations.  Note that, in both cases, the orientation sign $\hat\eta$
may be removed by taking $\chi\to\hat\eta\chi$, $\mathcal A_\mu\to\hat\eta
\mathcal A_\mu$ and $\gamma^5\to\hat\eta\gamma^5$.  It is the latter
transformation on $\gamma^5$ that highlights the orientation reversal
nature of this map.

The above supersymmetry variations simplify considerably in the round
$S^3\times S^3$ limit, given by (\ref{eq:4round}).  Here, we obtain
\begin{eqnarray}
\delta\psi_\mu&=&\left[\nabla_\mu-\fft{i}{16}e^{-3\alpha}F_{\nu\lambda}
\gamma^{\nu\lambda}\gamma_\mu\right]\epsilon,\nonumber\\
\delta\lambda_\alpha&=&\left[\gamma^\mu\partial_\mu\alpha
+\fft{i}8e^{-3\alpha}F_{\mu\nu}\gamma^{\mu\nu}-\eta e^{-\alpha}\right]
\epsilon,\nonumber\\
\delta\lambda_\beta&=&\left[\gamma^\mu\partial_\mu\beta
-\fft{i}8e^{-3\alpha}F_{\mu\nu}\gamma^{\mu\nu}\pm
i\hat\eta e^{-\beta}\gamma^5\right]\epsilon,
\label{eq:4rounds3}
\end{eqnarray}
where the $+$ sign corresponds to the $U(1)$ charged Killing spinor
case, and the $-$ sign to the uncharged case.  These expressions reproduce
the supersymmetry variations of the LLM construction,
\cite{Lin:2004nb,Liu:2004ru}, as they must.  Here we see that the
sign choice in the last term of the $\delta\lambda_\beta$ variation
comes from the two types of Killing spinors on the (un)squashed sphere,
and not from the orientation sign $\hat\eta$ (which can be absorbed by
a redefinition of $\gamma^5$).

%%%%%%%%%%%%%%%%%%%%%%%%%%%%%%%%%%%%%%%%
\section{Supersymmetry analysis}
\label{sec:susyanal}
%%%%%%%%%%%%%%%%%%%%%%%%%%%%%%%%%%%%%%%%

%%%%%%%%%%%%%%%%%%%%%%%%%%%%%%%%%%%%%%%%
\subsection{1/8 BPS configurations}
\label{sec:d=7susy}

We begin with the general 1/8 BPS bubbling case, which only an $S^3$
inside AdS$_5$ is preserved.  In this case, the relevant supersymmetry
variations are (\ref{eq:7gto}) and (\ref{eq:7dto}).  A double Wick
rotated version of this system ({\it i.e.}~one with AdS$_3$ instead of
$S^3$ isometry) was recently investigated in \cite{Kim:2005ez},
and the results are directly applicable to the present case.

The analysis of \cite{Kim:2005ez} demonstrated that the seven-dimensional
metric may be written as time fibered over a six (real) dimensional
K\"ahler base which satisfies an appropriate geometric condition.  Here
we briefly review this construction.

For a Dirac spinor $\epsilon$ in seven dimensions, we start by forming a
set of Dirac bilinears
\begin{equation}
f=i\overline\epsilon\epsilon,\qquad K^\mu=\overline\epsilon\gamma^\mu
\epsilon,\qquad V^{\mu\nu}=\overline\epsilon\gamma^{\mu\nu}\epsilon,
\qquad Z^{\mu\nu\lambda}=i\overline\epsilon\gamma^{\mu\nu\lambda}\epsilon.
\label{eq:7dbil}
\end{equation}
The factors of $i$ are chosen to make these quantities real.  In
addition to the above, we may also form a set of (complex) Majorana
bilinears
\begin{equation}
f^m=\epsilon^c\epsilon,\qquad Z^m_{\mu\nu\lambda}=\epsilon^c
\gamma_{\mu\nu\lambda}\epsilon.
\label{eq:7mbil}
\end{equation}
Counting the individual tensor components of the above, we find
64 real Dirac bilinear components and 36 complex Majorana bilinear
components, giving rise to $136=\fft12(16\cdot17)$ total real
components.  Since this matches the number of bilinears formed out of a
spinor $\epsilon$ with 16 real components, we see that this set of bilinears
is complete.

Of course, these tensor quantities are highly constrained by the
algebraic identities (corresponding to Fierz rearrangement).  Here
we do not aim to be comprehensive, but simply list some relevant identities.
First we have the normalization conditions
\begin{equation}
K^2=-f^2-|f^m|^2,\qquad V^2=6f^2+6|f^m|^2,
\qquad Z^2=-18f^2+24|f^m|^2,\qquad |Z^m|^2=48f^2+6|f^m|^2.
\label{eq:7sfier}
\end{equation}
Then there are the orthogonality conditions
\begin{equation}
K^\mu V_{\mu\nu}=0,\qquad
K^\mu Z^m_{\mu\nu\lambda}=f^mV_{\nu\lambda}.
\end{equation}
Finally, there are the identities which are directly useful for
determining the structure
\begin{eqnarray}
&&fZ+K\wedge V+\Re(f^{m\,*}Z^m)=0,\label{eq:7fz}\\
&&V\wedge Z^m=-2f^m*V,\label{eq:7yz}\\
&&V\wedge V=-2*(K\wedge V),\label{eq:7yy}\\
&&K\wedge Z^m=-i*(fZ^m-f^mZ),\\
&&Z^m\wedge Z^{m\,*}=8if*K.\label{eq:7zz}
\end{eqnarray}
Here $f^{m\,*}$ and $Z^{m\,*}$ denote the complex conjugates of $f^m$
and $Z^m$, respectively.

As shown in \cite{Kim:2005ez}, backgrounds preserving (at least) 1/8 of
the supersymmetries necessarily have SU(3) structure.  To see this,
we first note that (\ref{eq:7sfier}) constrains the norm of $K^\mu$ to
be non-positive.  Furthermore, from (\ref{eq:7killing}), we see that
$K^\mu$ satisfies the Killing equation.  We may thus choose $K^\mu$ as a
preferred time like Killing vector $K^\mu\partial_\mu=\partial/\partial t$.
(Although the null possibility may be of interest, we do not pursue it here,
as we are mainly interested in bubbling AdS configurations.)  In fact, we
may deduce a fair bit more
about the structure by noting from (\ref{eq:7fmzero}) that the Majorana
scalar invariant $f^m$ necessarily vanishes.  This gives us the norms of
the tensors
\begin{equation}
K^2=-f^2,\qquad V^2=6f^2,\qquad Z^2=-18f^2,\qquad |Z^m|^2=48f^2,
\end{equation}
as well as the conditions that $V$ and $Z^m$ are orthogonal to $K^\mu$
\begin{equation}
i_KV=i_KZ^m=0.
\end{equation}
Using (\ref{eq:7fz}), we may also solve for $Z$
\begin{equation}
Z=-f^{-1}K\wedge V,
\end{equation}
demonstrating that $Z$ is not an independent tensor quantity.
As a result, the structure is implicitly defined by the time-like Killing
vector $K^\mu$ along with a real 2-form $V$ and complex 3-form $Z^m$.
Using (\ref{eq:7yz}), (\ref{eq:7yy}) and (\ref{eq:7zz}), it as easy
to see that
\begin{equation}
V\wedge Z^m=0,\qquad
V\wedge V\wedge V=\ft{3i}4 fZ^m\wedge Z^{m\,*}=-6f^2*K.
\label{eq:7struct}
\end{equation}
But this is simply the requirement for SU(3) structure in $6+1$ dimensions.
Thus the seven-dimensional space splits naturally into time and a six (real)
dimensional base with SU(3) structure.

To proceed with an explicit construction, we may now solve (\ref{eq:7emaf})
to obtain $f=e^\alpha$.  We then make a choice of metric of the form
\begin{equation}
ds_7^2=-e^{2\alpha}(dt+\omega)^2+e^{-2\alpha}h_{ij}dx^idx^j.
\label{eq:7metc}
\end{equation}
The one-form associated with the Killing vector $K^\mu\partial_\mu=\partial_t$
is then $K_\mu dx^\mu=-e^{2\alpha}(dt+\omega)$.  Following \cite{Kim:2005ez},
we define the canonical
two-form $J$ and the holomorphic three-form
\begin{equation}
J= e^\alpha V,\qquad \Omega=e^{2\alpha}e^{-2i\eta t}Z^m.
\label{eq:7jomega}
\end{equation}
Note that $\Omega$ is independent of time.  The restriction (\ref{eq:7struct})
onto the six-dimensional base gives the usual SU(3) structure conditions
\begin{equation}
J\wedge\Omega=0,\qquad J\wedge J\wedge J=\ft{3i}4\Omega\wedge\Omega^*
=-6*_61,
\end{equation}
while the differential identities (\ref{eq:7eay}) and (\ref{eq:7e2azm})
give the integrability equations
\begin{equation}
dJ=0,\qquad d\Omega=2i\eta\omega\wedge\Omega.
\label{eq:7kah}
\end{equation}
This ensures that the six-dimensional base has U(3) holonomy.  In other
words, it is K\"ahler, with the K\"ahler form
\begin{equation}
J=i h_{i\bar j} dz^i\wedge d\bar z^{\bar j}=
\frac{1}{2} J_{ij}dx^i \wedge dx^j,
\end{equation}
and the Ricci form
\bea
\mathcal R=iR_{i\bar j} dz^i \wedge dz^{\bar j}\nn=2\eta d\omega.
\eea
In addition, the differential identities constrain the two-form $F$ and
scalar $\alpha$ to satisfy
\begin{equation}
F=d[e^{4\alpha}(dt+\omega)]-2\eta J,\qquad
e^{-4\alpha}=-\ft18R,
\label{eq:7fal}
\end{equation}
where $R$ is the scalar curvature of $h_{ij}$ \cite{Kim:2005ez}.

Finally, to guarantee that the above is a true solution to the equations
of motion, we may apply the Bianchi identity and equation of motion
for $F_{(2)}$.  From (\ref{eq:7fal}) along with $dJ=0$ the Bianchi
identity turns out to be trivial, while the $F_{(2)}$ equation of
motion gives
\begin{equation}
\square_6e^{-4\alpha}=\ft18(R_{ij}R^{ij}-\ft12R^2),
\end{equation}
where $\square_6$ as well as the tensor contraction is with respect
to the base metric $h_{ij}$.  Substituting in the expression for
$e^{-4\alpha}$ in (\ref{eq:7fal}) then gives a condition on the curvature
\begin{equation}
\square_6R=-R_{ij}R^{ij}+\ft12R^2.
\label{eq:7cond}
\end{equation}

In summary, 1/8 BPS configurations preserving an $S^3$ isometry may
be described by a seven-dimensional metric (\ref{eq:7metc}) with form
field and scalar given by (\ref{eq:7fal}).  The one-form $\omega$ is
defined according to $\mathcal R=2\eta d\omega$, where the sign $\eta$
is related to the orientation of the Killing spinor on $S^3$.  The
full solution is determined in terms of a six-real dimensional K\"ahler
metric $h_{ij}$ satisfying the curvature condition (\ref{eq:7cond}).

{}From a ten-dimensional point of view, the solution is essentially
given by time and $S^3$ fibered over the six-dimensional base.
In order to ensure regularity, we may focus on regions on the base
where the $S^3$ fiber shrinks to zero size.  This corresponds to
regions where $e^\alpha\to0$, which by (\ref{eq:7fal}) corresponds
to $R\to\infty$.  Thus the six-dimensional base generally will be
bounded by surfaces of infinite curvature where the $S^3$ degenerates.
At the same time, the $e^{-2\alpha}$ factor in front of the six-dimensional
metric ought to be such that the physical ten-dimensional metric
remains regular.  Furthermore, the collapsing $S^3$ along with the transverse
direction to the degeneration surface must locally yield $\mathbb R^4$
to ensure the absence of conical singularities.  Examination of these
boundary conditions will be taken up in Sections~\ref{sec:bubble} and
\ref{regular} below.

%%%%%%%%%%%%%%%%%%%%%%%%%%%%%%%%%%%%%%%%
\subsection{1/4 BPS configurations}
\label{sec:d=6susy}

Following the above analysis, we now turn to the 1/4 BPS case preserving
$S^3\times S^1$ isometry.  Here there are at least two possible
approaches that may be taken.  The first is to realize that, since
1/4 BPS configurations form a subset of all 1/8 BPS solutions, we
may simply take the above 1/8 BPS analysis and demand that
the resulting geometry admits a further $U(1)$ isometry.  The second
is to directly analyze the effective six-dimensional supersymmetry
variations (\ref{eq:6gto}) and (\ref{eq:6dto}).  The advantage of
this method, which was recently employed in \cite{Donos:2006iy,Donos:2006ms},
is that it leads to a natural choice of coordinates with which to
parameterize the solution.

Before turning to the full supersymmetry analysis of
\cite{Donos:2006iy,Donos:2006ms}, we first examine the possibility of
imposing an additional $U(1)$ isometry on the 1/8 BPS solutions
described above.  Noting that the generic solution is given in terms
of a complex three-dimensional K\"ahler base identified by (\ref{eq:7kah})
and with curvature satisfying (\ref{eq:7cond}), we may locally choose
an appropriate set of complex coordinates
\begin{equation}
z_1,\quad z_2,\quad z_3\equiv r e^{i\psi},
\label{eq:6reds1}
\end{equation}
and impose symmetry under $\psi$ translation ({\it i.e.}~by demanding
that $\partial/\partial\psi$ is a Killing vector).  This indicates that
the K\"ahler potential ought to be of the form
\begin{equation}
K(z_i,\overline z_i,r^2)\qquad i=1,2.
\end{equation}
This K\"ahler potential leads to a metric on the base of the form
\begin{eqnarray}
h_{ij} dx^i dx^j&=&h_{i\bar j} dz^i dz^{\bar j}+c.c.=2h_{i\bar j} dz^i
dz^{\bar j}=2\partial_i\partial_{\bar j}K dz^i dz^{\bar j}\nn\\
&=&
2\partial_{i}\partial_{\bar j}K
dz_id\overline z_j +2(r^2K')'(dr^2+r^2d\psi^2)
+4rdr\Re(\partial_{i}K'dz_i)
+4r^2d\psi\Im(\partial_{i}K'dz_i),\nn\\
\end{eqnarray}
where a prime denote partial differentiation with respect to $r^2$,
and $\Re$ and $\Im$ denote real and imaginary parts, respectively.
After completing the square, this may be rewritten as
\begin{eqnarray}
h_{ij}dx^idx^j&=&2\left(\partial_{i}\partial_{\bar j}K
-\fft{r^2}{(r^2K')'}\partial_{i}K'\partial_{\bar j}K'\right)
dz_id\overline z_j
+\fft1{2r^2(r^2K')'}d(r^2K')^2\nonumber\\
&&+2r^2(r^2K')'\left(d\psi+\fft1{(r^2K')'}\Im(\partial_{i}K'dz_i)
\right)^2.
\end{eqnarray}
A change of variables $y^2=2r^2K'$ brings this to the form
\begin{eqnarray}
&&h_{ij}dx^idx^j=2\left(\partial_{i}\partial_{\bar j}K
-\fft{2r^2}{(y^2)'}\partial_{i}K'\partial_{\bar j}K'\right) dz_id\overline
z_j+\frac{y^2}{r^2(y^2)'}dy^2+r^2(y^2)'(d\psi+\mathcal
A)^2,\nonumber\\ &&\mathcal
A=\fft2{(y^2)'}\Im(\partial_{i}K'dz_i)\, ,
\end{eqnarray}
where $(y^2)'=(2r^2K')'$, and $r$ is to be eliminated by inverting the
above transformation.

Although this form of the metric is suggestive that the complex
three-dimensional base splits into a two-dimensional piece along
with a `radial' coordinate $y$ and fiber direction $\psi$, the
physical understanding of this solution is somewhat obscure.  For
this reason, it is instructive to perform the supersymmetry
analysis directly with the actual variations (\ref{eq:6gto}) and
(\ref{eq:6dto}).  This analysis, which was initiated in
\cite{Donos:2006iy,Donos:2006ms}, starts with the definition of the
(Dirac and Majorana) spinor bilinears
\begin{eqnarray}
&&f_1=\overline\epsilon\gamma^7\epsilon,\qquad
f_2=i\overline\epsilon\epsilon,\qquad
K^\mu=\overline\epsilon\gamma^\mu\epsilon,\qquad
L^\mu=\overline\epsilon\gamma^\mu\gamma^7\epsilon,\nonumber\\
&&V^{\mu\nu}=\overline\epsilon\gamma^{\mu\nu}\epsilon,\qquad
Y^{\mu\nu}=i\overline\epsilon\gamma^{\mu\nu}\gamma^7\epsilon,\qquad
Z^{\mu\nu\lambda}=i\overline\epsilon\gamma^{\mu\nu\lambda}\epsilon,\nonumber\\
&&f^m=\epsilon^c\epsilon,\qquad
Y_{\mu\nu}^m=\epsilon^c\gamma_{\mu\nu}\gamma^7\epsilon,\qquad
Z_{\mu\nu\lambda}^m=\epsilon^c\gamma_{\mu\nu\lambda}\epsilon.
\label{eq:6bil}
\end{eqnarray}
We have highlighted the close relation between six and seven-dimensional
Dirac spinors by using an identical notation with the bilinears defined
above in (\ref{eq:7dbil}) and (\ref{eq:7mbil}), except for the cases
where $\gamma^7$ is involved (and with a rewriting $f\to f_2$ consistent
with the LLM notation).  The `new' bilinears with $\gamma^7$ are of
course the components of the seven-dimensional bilinears (\ref{eq:7dbil})
and (\ref{eq:7mbil}) along the circle direction.

Although the six-dimensional Fierz identities may in principle
be derived from the seven-dimensional ones, some of the expressions
we are interested in cannot be written in a seven-dimensional covariant
manner.  Thus we work directly with the above bilinears in six
dimensions.  In this case, we have the normalization conditions
\begin{eqnarray}
&&K^2=-L^2=-f_1^2-f_2^2-|f^m|^2,\qquad
V^2=-2f_1^2+4f_2^2+4|f^m|^2,\nonumber\\
&&Y^2=4f_1^2-2f_2^2+4|f^m|^2,\qquad
Z^2=-12f_1^2-12f_2^2+12|f^m|^2,\nonumber\\
&&|Y^m|^2=8f_1^2+8f_2^2+2|f^m|^2,\qquad
|Z^m|^2=-24f_1^2+24f_2^2.
\end{eqnarray}
We also have identities related to the projection of the various
tensors onto $K^\mu$ and $L^\mu$
\begin{eqnarray}
&&K\cdot L=0,\nonumber\\
&&K^\mu V_{\mu\nu}=f_1L_\nu,\qquad L^\mu V_{\mu\nu}=f_1K_\nu,\nonumber\\
&&K^\mu Y_{\mu\nu}=f_2L_\nu,\qquad L^\mu Y_{\mu\nu}=f_2K_\nu,\nonumber\\
&&K^\mu Y^m_{\mu\nu}=f^mL_\nu,\qquad L^\mu Y^m_{\mu\nu}=f^mK_\nu,\nonumber\\
&&K^\mu Z_{\mu\nu\lambda}=-f_1Y_{\nu\lambda}+f_2V_{\nu\lambda},\qquad
L^\mu Z_{\mu\nu\lambda}=\Im(f^mY^{m\,*}_{\nu\lambda}),\nonumber\\
&&K^\mu Z^m_{\mu\nu\lambda}=-f_1Y^m_{\nu\lambda}+f^mV_{\nu\lambda},\qquad
L^\mu Z^m_{\mu\nu\lambda}=-if_2Y^m_{\nu\lambda}+if^mY_{\nu\lambda}.
\end{eqnarray}
Finally, the following Fierz identities are
useful for determining the structure
\begin{eqnarray}
&&f_1V+f_2Y+\Re(f^mY^{m\,*})=-K\wedge L,
\nonumber\\
&&K\wedge Z=*\Im(f^mY^{m\,*}),\qquad
L\wedge Z=*(f_2V-f_1Y),\nonumber\\
&&K\wedge Z^m=-i*(f_2Y^m-f^mY),\qquad
L\wedge Z^m=-*(f_1Y^m-f^mY).
\end{eqnarray}

Since the six-dimensional bilinears parallel those of the seven-dimensional
case, it is not surprising to see from (\ref{eq:6fmzero}) that the Majorana
scalar invariant $f^m$ vanishes in this case as well.  Setting $f^m=0$, we
now obtain
\begin{equation}
K^2=-L^2=-f_1^2-f_2^2,\qquad K\cdot L=0, \label{eq:6ksqlsq}
\end{equation}
which we note is identical to the LLM case, even though we are
working in six dimensions instead of four.  This ensures that $K^\mu$
is time-like while $L_\mu$ is space-like and orthogonal to $K^\mu$.
This gives rise to a natural decomposition of the six-dimensional
space into a four-dimensional base along with a preferred time-like
and a preferred space-like direction.

Furthermore, the above identities allow us to decompose the bilinears
into components along $K^\mu$ and $L_\mu$ and those orthogonal to them.
The result is
\begin{eqnarray}
&&V=-\fft{f_1}{f_1^2+f_2^2}K\wedge L-f_2I^3,\qquad
Y=-\fft{f_2}{f_1^2+f_2^2}K\wedge L+f_1I^3,\qquad
Z=K\wedge I^3,\nonumber\\
&&Y^m=-{\textstyle\sqrt{f_1^2+f_2^2}}(I^1-iI^2),\qquad
Z^m=-\fft1{\sqrt{f_1^2+f_2^2}}(f_1K-if_2L)\wedge(I^1-iI^2),
\label{eq:6struct}
\end{eqnarray}
where the triplet of %real
two-forms $I^i$ are orthogonal to both $K^\mu$
and $L_\mu$ and satisfy the $\mathrm{SU}(2)$ structure equation
\begin{equation}
I^i_{ab}I^j_{bc}=-\delta_{ac}\delta^{ij}-\epsilon^{ijk}I^k_{ac},
\label{eq:6iistruct}
\end{equation}
as well as the self-duality condition
\begin{equation}
I^i_{ab}=\ft12\epsilon_{abcd}I^i_{cd},
\end{equation}
on the four-dimensional base.  It should be noted, however that
since the Majorana bilinears are charged under the $U(1)$ gauge
symmetry carried by $\mathcal A_\mu$, the two-forms $I^1\pm iI^2$
carry $U(1)$ charge, while only $I^3$ is neutral.  The implication
of this is that only $I^3$ is gauge invariant, and as a result we
conclude that the system has $U(2)$ structure in $5+1$ dimensions,
except for backgrounds with vanishing $\mathcal A_\mu$, which instead
carry $SU(2)$ structure.  In either case, the structure group is a
subgroup of $SU(3)$, which showed up as the structure group pertaining
to the 1/8 BPS solutions found above.

In contrast to the 1/8 BPS analysis given above, an explicit construction
of 1/4 BPS configurations is complicated by the fact that many more
field components now need to be specified.  In addition to the six-dimensional
metric $g_{\mu\nu}$, we have the three scalars $\alpha$, $\beta$ and
$\chi$ as well as the field strengths $F_{(2)}$ and $\mathcal F_{(2)}$.
We note, however, that the axionic scalar $\chi$ is related to the IIB
five-form flux threading both $S^3$ and $S^1$ in the reduction in the
sense that
\begin{equation}
{}^{10}F_{(5)}=d\chi\wedge (d\psi+\mathcal A)\wedge \omega_3+\cdots.
\end{equation}
While this is certainly allowed by the isometries, any excitation of
$\chi$ necessarily falls outside of the `bubbling AdS' interpretation,
as non-zero $\chi$ corresponds to mixed components of five-form flux
(where $S^3$ is inside AdS$_5$ and $S^1$ is inside $S^5$).  We thus
specialize the analysis by taking $\chi=0$.  At the same time, we recall
that such a truncation leads to the requirement $F_{\mu\nu}\mathcal F^{\mu\nu}
=0$, which will be expected to show up as additional constraints on
the solution.

Following \cite{Donos:2006iy,Donos:2006ms}, the supersymmetry
analysis begins by using the one-form identities given in
(\ref{eq:6emaf2}) through (\ref{eq:6em2abf1}) to obtain the scalar
bilinears $f_1$ and $f_2$ in terms of the fields $\alpha$ and
$\beta$ and then to solve for the components of the field
strengths $F_{(2)}$ and $\mathcal F_{(2)}$.  Noting from
(\ref{eq:6emaf2}) that $d(e^{-\alpha}f_2)=0$, we may immediately
write $f_2=ae^\alpha$ for some constant $a$.  However, obtaining
an expression for $f_1$ is somewhat more involved.  To proceed, we
make the simplifying assumption that $i_K\mathcal F=0$, which was
also imposed in \cite{Donos:2006ms}.  This assumption that the
electric component of $\mathcal F_{(2)}$ vanishes ensures that the
$U(1)$ bundle is only fibered over the spatial components of the
metric. This is consistent with taking the gauged $U(1)$ to be
contained inside the original $S^5$ as opposed to AdS$_5$, so we
do not believe this assumption to be overly restrictive, at least
as far as bubbling geometries are concerned.  In any case, we keep
in mind that the following supersymmetry analysis only pertains to
the specialization of the most general $S^3\times S^1$ system to
the case when
\begin{equation}
\chi=0,\qquad i_K\mathcal F=0.
\end{equation}

Having imposed $i_K\mathcal F=0$, (\ref{eq:6embf1}) may then be
solved to yield $f_1=be^\beta$ for constant $b$.  As a result, all
scalar bilinears are now fully determined
\begin{equation}
f^m=0,\qquad f_1=be^\beta,\qquad f_2=ae^\alpha.
\label{eq:6f1f2}
\end{equation}
At this point, it is useful to specialize the form of the six-dimensional
metric.  Noting from (\ref{eq:6killing}) that $K^\mu$ is a Killing vector,
we take $K^\mu\partial_\mu=\partial_t$.  Furthermore, (\ref{eq:6e2abf1})
then gives $L=-\eta b\,de^{\alpha+\beta}$, so that $L$ is a closed one-form.
In particular, using (\ref{eq:6f1f2}), we may express $y=-\eta a^{-1}f_1f_2$
if desired.  From (\ref{eq:6ksqlsq}), we may now specialize the choice of
coordinates to take $L=dy$.  As a result, we now make a choice of metric of
the form
\begin{equation}
ds_6^2=-h^{-2}(dt+\omega)^2+f_2^{-2}h_{ij}dx^idx^j+h^2dy^2,
\label{eq:6metans}
\end{equation}
where
\begin{equation}
h^{-2}=f_1^2+f_2^2,\qquad
K=-h^{-2}(dt+\omega),\qquad L=dy,
\end{equation}
and we have included a factor of $f_2^{-2}$ in front of the
four-dimensional metric $h_{ij}$ for latter convenience.

Given the above, the remaining one-form differential identities
(\ref{eq:6e3af2}) through (\ref{eq:6em2abf1}) allow us to determine
most components of $F_{(2)}$ and $\mathcal F_{(2)}$.  We find
\begin{eqnarray}
a^3F_{(2)}&=&d(f_2^4)\wedge(dt+\omega)+4 h^2f_2^5I^3_i{}^j\partial_jf_1\,dx^i
\wedge dy+\ft12a^3F_{ij}dx^i\wedge dx^j,\nonumber\\
\mathcal F_{(2)}&=&\ft12\mathcal F_{ij}dx^i\wedge dx^j,
\end{eqnarray}
where
\begin{equation}
a^3I^3_{ij}F^{ij}=-8f_2^{-1}\partial_yf_1,\qquad
I^3_{ij}\mathcal F^{ij}=-4b^2f_1^{-2}f_2^{-4}\left(\fft{a}b\eta-n\right).
\label{eq:6ifs}
\end{equation}
Note that indices on the four-dimensional base are raised and lowered
with the metric $h_{ij}$.

Before completing the determination of the two-form field strengths,
we examine the content of the three-form identity (\ref{eq:6eav}),
which states $d(f_2V)=b^{-1}f_1f_2\mathcal F\wedge dy$.  Using the
structure identities (\ref{eq:6struct}), we may write
$V=f_1(dt+\omega)\wedge dy-f_2I^3$.  As a result, (\ref{eq:6eav}) leads
to the identities
\begin{equation}
\tilde d(f_2^2I^3)=0,\qquad \tilde d\omega=b^{-1}\mathcal F+(f_1f_2)^{-1}
\partial_y(f_2^2I^3),
\end{equation}
where $\tilde d=dx^i\partial_i$ acts only on the four-dimensional base.
At this stage, it ought to be clear why we have chosen a prefactor $f_2^{-2}$
in front of the base metric $h_{ij}$ in (\ref{eq:6metans}).  This is because,
by defining
\begin{equation}
I^3=f_2^{-2}J,
\label{eq:6i3j}
\end{equation}
we obtain the canonical two-form $J$ which is closed ($\tilde dJ=0$), and which
satisfies $J\wedge J=2*_41$, where the volume form is given in terms of
$h_{ij}$.  This in particular indicates that the four-dimensional base is
K\"ahler.

Additional information on the form of the solution remains to be extracted
from the $\nabla_\mu V_{\nu\lambda}$ identity, (\ref{eq:6dmuv}).
Examining $\nabla_y V_{ij}$ and $\nabla_i V_{jk}$ yield the identities
\begin{equation}
\nabla^4_iJ_{jk}=0,\qquad\partial_yJ_i{}^j=0,
\end{equation}
confirming that $J$ is covariantly constant with respect to the metric
$h_{ij}$.  Note, however, that while $J_i{}^j$ is independent of $y$,
in general both $J_{ij}$ and $h_{ij}$ are highly non-trivial functions
of $y$.  The remaining components of (\ref{eq:6dmuv}) serve to complete
the determination of the two-forms
\begin{eqnarray}
F&=&\fft1{a^3}d[f_2^4(dt+\omega)]+\fft{y^2}a(d\omega-b^{-1}\mathcal F)
+\fft{2\eta}{a^2}J,\nonumber\\
\mathcal F&=&\ft12\mathcal F_{ij}dx^i\wedge dx^j,\qquad
\mathcal F_{ij}^{(+)}=-\fft{b^2}{a^2y^2}\left(\fft{a}b\eta-n\right)J_{ij},
\nonumber\\
d\omega&=&\fft1b\mathcal F-\fft\eta{ay}\left(\partial_yJ-
J_i{}^j\partial_jZdx^i\wedge dy\right).
\label{eq:6fss}
\end{eqnarray}
Here, as in \cite{Donos:2006iy,Donos:2006ms}, we have defined the
LLM function
\begin{equation}
Z=\fft12\fft{f_2^2-f_1^2}{f_2^2+f_1^2}.
\end{equation}
Note that the anti-self-dual part of $\mathcal F$ is unconstrained
by the differential identities.

Given these field strengths, the second expression in (\ref{eq:6ifs})
is identically satisfied.  On the other hand, compatibility of
$J_{ij}F^{ij}$ between the first expression in (\ref{eq:6ifs}) and
the form of $F$ given in (\ref{eq:6fss}) gives rise to an important
condition on the volume of the K\"ahler base
\begin{equation}
J^{ij}\partial_y J_{ij}\equiv\partial_y\log\det h_{ij}
=4h^2\left[\fft{2f_1^2}{f_2}\partial_yf_2+\fft{bf_2}{f_1}
\left(\fft{a}b\eta-n\right)\right].
\label{eq:6volcond}
\end{equation}
By substituting in
\begin{equation}
f_1f_2=-a\eta y,\qquad \fft{f_1}{f_2}=e^{-G},
\label{eq:6yg}
\end{equation}
the above expression may be brought into the form
\begin{equation}
\ft12\partial_y\log\det h_{ij}=\fft{2e^{-G}}{e^G+e^{-G}}\partial_yG
+\fft2{y(1+e^{2G})}\left(2-\fft{b}an\eta\right)
-\fft2y\left(1-\fft{b}an\eta\right),
\label{eq:6monge}
\end{equation}
originally given in \cite{Donos:2006ms}.  The factor of $1/2$ on the
left hand side arises because here we still take $h_{ij}$ as a real
metric.

To ensure a complete solution to the equations of motion, we now
apply the Bianchi identities and equations of motions (\ref{eq:6bieom}),
which for $\chi=0$ take on the simple form
\begin{equation}
0=dF=d\mathcal F=d(f_1f_2^{-3}*_6F)=d(f_1^3f_2^3*_6\mathcal F).
\label{eq:6fbeom}
\end{equation}
We begin with the Bianchi identities.  Since $\mathcal F$ is incompletely
specified, we are left with the requirement $d\mathcal F=0$, which admits
no particular simplification.  For $dF=0$, however, we see from
(\ref{eq:6fss}) that it is automatically satisfied, provided $\mathcal F$
and $d\omega$ are both closed.  Actually $d^2\omega=0$ is not
guaranteed in the above expression.  Instead, just as in the LLM case
\cite{Lin:2004nb}, it gives rise to the second-order condition
\begin{equation}
i y\partial_y\fft1y\partial_yJ_{ij}+2J_{[j}{}^k\nabla_{i]}\nabla_kZ=0.
\label{eq:6zcond}
\end{equation}
Introducing a K\"ahler potential $K$ with
\begin{equation}
h_{ij}=\ft12(\nabla_i\nabla_j+J_i{}^k J_j{}^l\nabla_k\nabla_l)K,
\label{eq:6realk}
\end{equation}
we see that the condition (\ref{eq:6zcond}) may be solved by taking
\begin{equation}
Z(x^i,y)=-\fft12y\partial_y\fft1y\partial_yK(x^i,y).
\label{eq:6zsoln}
\end{equation}
Note that, while an arbitrary harmonic function may be added to $Z$,
this may be absorbed by making an appropriate K\"ahler transformation
on $K$.

Turning to the equations of motion, we see that the $\mathcal F$
equation of motion given in (\ref{eq:6fbeom}) is equivalent to
$d(y^3*_6\mathcal F)=0$.  Through appropriate manipulations, and
using the fact that $*_4\mathcal F=\mathcal F^{(+)}-\mathcal F^{(-)}
=2\mathcal F^{(+)}-\mathcal F$, we may show that this is
equivalent to
\begin{equation}
\mathcal F_{ij}\mathcal F^{ij}=\fft{\eta b}{ay}\mathcal F^{ij}\partial_yJ_{ij}.
\end{equation}
Using the Bianchi identity $d\mathcal F=0$, and in particular $\partial_y
\mathcal F_{ij}=0$, we obtain
\begin{equation}
\mathcal F^{ij}\partial_yJ_{ij}=-\partial_y(\mathcal F^{ij}J_{ij})
=-\fft{8b^2}{a^2y^3}\left(\fft{a}b\eta-n\right).
\end{equation}
As a result, the $\mathcal F$ equation of motion reduces to
\begin{equation}
\mathcal F_{ij}\mathcal F^{ij}=-\fft{4b^4}{a^4y^4}\left(2\fft{a}b\eta\right)
\left(\fft{a}b\eta-n\right).
\label{eq:6calfsq}
\end{equation}
Since the self-dual component of $\mathcal F$ is known from (\ref{eq:6fss}),
the above may be rewritten in the equivalent form
\begin{equation}
\mathcal F_{ij}*_4\mathcal F^{ij}=\fft{8b^4}{a^4y^4}\left(\fft{a}b\eta-n\right)
\left(2\fft{a}b\eta-n\right),
\label{eq:6fwedgef}
\end{equation}
which is identical to the $\mathcal F\wedge\mathcal F$ constraint given
in \cite{Donos:2006ms}.  Incidentally, we note that the self-dual and
anti-self-dual components of $\partial_y J$ may be expressed as
\begin{eqnarray}
(\partial_yJ)^{(+)}&=&\ft14J\partial_y\log\det h_{ij},\nonumber\\
(\partial_yJ)^{(-)}&=&\partial_yJ-\ft14J\partial_y\log\det h_{ij}.
\end{eqnarray}
In addition, as a consequence of (\ref{eq:6calfsq}), we may verify that
both the $F$ equation of motion and
the $F_{\mu\nu}\mathcal F^{\mu\nu}=0$ constraint are automatically satisfied.

Finally, to complete the solution, we note that the $U(2)$ structure of
the base is highlighted by both the canonical two-form $J$ identified in
(\ref{eq:6i3j}) and a holomorphic two-form $\Omega$, which may be defined
by
\begin{equation}
\Omega=-i f_2^2(I^1-iI^2).
\end{equation}
The structure equation (\ref{eq:6iistruct}) along with self-duality is
then equivalent to the statement
\begin{equation}
J\wedge\Omega=0,\qquad J\wedge J=\ft12\Omega\wedge\Omega^*=2*_41.
\end{equation}
Along with $\tilde dJ=0$ shown above, we are also interested in the
integrability of $\Omega$.  This may be investigated by considering
(\ref{eq:6e2aby}), where $Y^m=i f_2^{-2}h\Omega$ according to
(\ref{eq:6struct}).  We find
\begin{equation}
D\Omega=\left[-ib\left(2\fft{a}b\eta-n\right)(dt+\omega)+\ft12\tilde d
\log(Z+\ft12)+\ft14\partial_y\log h\,dy\right]\wedge\Omega.
\label{eq:6Domega}
\end{equation}
To interpret this result, we examine each component separately.  Along
the time direction, we have
\begin{equation}
\partial_t\Omega=-ib\left(2\fft{a}b\eta-n\right)\Omega,
\end{equation}
indicating that we may take
\begin{equation}
\Omega=e^{-ib\left(2\fft{a}b\eta-n\right)t}\Omega_0,
\end{equation}
where $\Omega_0$ is independent of time.  Note that this time dependence
is analogous to that found in (\ref{eq:7jomega}) for the 1/8 BPS solutions
given above.  Along the $y$ direction, (\ref{eq:6Domega}) gives
\begin{equation}
\partial_y\Omega=\ft14\partial_y\log\det h_{ij}\,\Omega,
\end{equation}
which is compatible with $\Omega\wedge\Omega^*$ being proportional
to the volume form on the base.

What we are mainly interested in, of course, is $\tilde d\Omega$ on the base.
Taking into account that $D=d+in\mathcal A$, we see that
\begin{equation}
\tilde d\Omega=\left[-in\mathcal A-ib\left(2\fft{a}b\eta-n\right)\omega
+\ft12\tilde d\log(Z+\ft12)\right]\wedge\Omega.
\end{equation}
{}From this, we may extract the Ricci form on the base
\begin{eqnarray}
\mathcal R&=&\bigg(
-n\mathcal F-b\left(2\fft{a}b\eta-n\right)\tilde d\omega
-\ft12\tilde d\left(J_i{}^j\partial_j\log(Z+\ft12)dx^i\right)
\bigg)
\nonumber\\
&=&\bigg(
-2\fft{a}b\eta\mathcal F
-i
\fft{b}a\eta\left(2\fft{a}b\eta-n\right)
\fft1y\partial_yJ-\ft12\tilde d\left(J_i{}^j\partial_j\log(Z+\ft12)dx^i\right)
\bigg),
\label{eq:6riform}
\end{eqnarray}
where in the second line we have used the expression (\ref{eq:6fss}) for
$\tilde d\omega$.  For a K\"ahler metric $h_{ij}$, the Ricci form may
be given as
\begin{equation}
\mathcal R_{ij}=-\ft12J_{[j}{}^k\nabla_{i]}\nabla_k\log\det h_{lm}.
\end{equation}
In this case, we may take a $y$ derivative of (\ref{eq:6riform})
and substitute in the expression (\ref{eq:6volcond}) to obtain
\begin{eqnarray}
-2J_{[j}{}^k\nabla_{i]}\nabla_k\left[h^2\left(\fft{2f_1^2}{f_2}\partial_yf_2
+\fft{bf_2}{f_1}\left(\fft{a}b\eta-n\right)\right)\right]
&=&-2\fft{a}b\eta\partial_y\mathcal F_{ij}-i\fft{b}a\eta
\left(2\fft{a}b\eta-n\right)\partial_y\fft1y\partial_yJ_{ij}\nonumber\\
&&-J_{[j}{}^k\nabla_{i]}\nabla_k\partial_y\log(Z+\ft12).
\end{eqnarray}
Noting that $\partial_y\mathcal F_{ij}=0$, and using (\ref{eq:6zcond})
to rewrite $\partial_yy^{-1}\partial_yJ_{ij}$ in terms of derivatives of
$Z$, we may see that the above expression is automatically satisfied.
Thus compatibility of (\ref{eq:6riform}) with (\ref{eq:6volcond}) is
ensured.

As may be evidenced by the above discussion, the supersymmetry analysis
leading to the complete 1/4 BPS system is rather involved.  In order to
summarize the results, and to make a comparison with
\cite{Donos:2006iy,Donos:2006ms}, we may reexpress the scalars $\alpha$
and $\beta$ in terms of the coordinate $y$ and the function $G$ through
(\ref{eq:6yg}).  In this case, the full ten-dimensional metric takes the
form
\begin{equation}
ds_{10}^2=-h^{-2}(dt+\omega)^2+h^2[2(Z+\ft12)^{-1}\partial_i\partial_{\bar j}K
dz^id\bar z^{\bar j}+dy^2]
+y[e^Gd\Omega_3^2+e^{-G}(d\psi+\mathcal A)^2],
\end{equation}
where
\begin{equation}
h^{-2}=2y\cosh G,\qquad Z=\ft12\tanh G.
\end{equation}
In the equation above we have switched to a complex notation
for the K\"ahler base, so that in particular the metric is given by
\begin{equation}
ds_4^2=h_{ij} dx^i dx^j= 2h_{i\bar j} dz^i dz^{\bar j}=
2\partial_i\partial_{\bar j}K(z_i,\bar z_{\bar i};y) dz^i d\bar z^{\bar j},
\end{equation}
This is the complex form of the expression given previously in real
notation in (\ref{eq:6realk}).

The LLM function $Z$ is constrained according to (\ref{eq:6zsoln})
\beq
Z=-\frac{1}{2}y\partial_y\frac{1}{y}\partial_y K(z_i,\bar z_{\bar i};y),
\label{eq:ZKrel}
\eeq
and furthermore the K\"ahler metric must satisfy a Monge-Amp\`{e}re
type equation (\ref{eq:6monge})
\beq
\partial_y \log\det h_{i\bar j}=\frac{2
e^{-G}}{e^G+e^{-G}}\partial_y G+ \frac{2}{y(e^{2G}+1)}(2-n\eta)
-\frac 2y (1-n\eta).
\label{eq:MongeAmpere}
\eeq
Note that, for simplicity, we have set the constants $a=b=-\eta$.
This equation can be integrated to yield
\beq
\log\det h_{i\bar j}=\log(Z+\frac 12)+n\eta \log y+\frac 1{y}(2-n\eta)
\partial_y K + D(z_i,\bar z_{\bar j}),
\label{eq:sol_MA}
\eeq
where $D(z_i,\bar z_{\bar j})$ arises as an integration constant
as we peel off a $\partial_y$ derivative from (\ref{eq:MongeAmpere}).
Furthermore, the Ricci form on the base must satisfy the constraint
(\ref{eq:6riform}).  When expressed in complex coordinates, this
reduces to
\beq
{\mathcal R}=i \partial \bar\partial \log\det h_{i\bar j}
=i\left(2i\eta{\mathcal F}+(2-n\eta)\frac 1{y}
\partial\bar\partial\partial_y K+\partial\bar\partial\log(Z+\frac12)\right),
\label{eq:R14BPS}
\eeq
where the holomorphic and anti-holomorphic differential operators
$\partial$ and $\bar\partial$ are defined by
\beq
\partial=dz^i \partial_i, \qquad \bar\partial = d\bar z^{\bar j}
\partial_{\bar j},
\eeq
and where we recall that the K\"ahler form is $J=i h_{i\bar j}dz^i\wedge
d\bar z^{\bar j}=i\partial\bar\partial K$.
Substituting the solution to the Monge-Amp\`{e}re equation
(\ref{eq:sol_MA}) into (\ref{eq:R14BPS}), we find that
\beq
\partial\bar\partial D = 2i\eta{\cal F}\,,
\label{eq:Deqn}
\eeq
where  $\mathcal F=d\mathcal A$ is the field strength
corresponding to the gauging of the $S^1$ isometry.

Of course, the complete solution also involves the two-forms given in
(\ref{eq:6fss}).  In particular, with $a=b=-\eta$, we have
\begin{eqnarray}
\eta F&=&-d[y^2e^{2G}(dt+\omega)]-y^2(d\omega+\eta\mathcal F)+2i\partial
\bar\partial K,\nonumber\\
\mathcal F^{(+)}&=&-\fft{i}{y^2}(\eta-n)\partial\bar\partial K,\nonumber\\
d\omega&=&-\eta\mathcal F+\fft{i}y(\partial\bar\partial\partial_y K
-(\partial-\bar\partial)Z\wedge dy).
\end{eqnarray}
Note that only the self-dual part of $\mathcal F$ is determined.
Comparing $\mathcal F^{(+)}$ with (\ref{eq:Deqn}) then implies
\beq
(1+*_4)\partial\bar\partial D= \frac 4{y^2}(1-n\eta)
\partial\bar\partial K.
\label{eq:Deqn2}
\eeq
Finally, one last condition on the solution arises from the $\mathcal F$
equation of motion, namely the $\mathcal F\wedge\mathcal F$ constraint
(\ref{eq:6fwedgef})
\begin{equation}
\mathcal F\wedge\mathcal F=\fft4{y^4}(1-n\eta)(2-n\eta)*_41.
\label{eq:6fwf2}
\end{equation}

As demonstrated in \cite{Donos:2006ms}, the BPS solutions with $S^3\times S^1$
isometry fall into several families, depending on the $U(1)$ charge $n$ of
the Killing spinor.  A particularly simple case, first considered in
\cite{Donos:2006iy}, is the ungauged ansatz, where $\mathcal A=0$,
corresponding to $\psi$ being trivially fibered over the base.
In this case, $\mathcal F$ vanishes, and (\ref{eq:Deqn}) reduces to
\beq
\partial\bar\partial D = 0 \, .
\label{eq:Deqn_ungauged}
\eeq
This indicates that $D$ can be an arbitrary harmonic
function of $z_1, z_2$. Furthermore, from (\ref{eq:Deqn2}) we see that
this condition corresponds to having
\beq
n\eta=1,
\eeq
which is also consistent with the vanishing of the $\mathcal F\wedge
\mathcal F$ constraint in (\ref{eq:6fwf2}).  Curiously, this constraint
also takes on a simple form when $n\eta=2$.  As shown in \cite{Donos:2006ms},
this allows the embedding of the 1/2 BPS LLM ansatz into the gauged
1/4 BPS ansatz.  The case $n\eta=3$ is also interesting, as it allows for
solutions of the form AdS$_5$ times a Sasaki-Einstein space.

%%%%%%%%%%%%%%%%%%%%%%%%%%%%%%%%%%%%%%%%
\subsection{1/2 BPS configurations}
\label{sec:d=4susy}

Continuing along the chain of reductions, the final case to consider
corresponds to taking $S^3$ times squashed $S^3$ isometry, as
described in Section~\ref{sec:cp1red}, where the squashed $S^3$ is
written as $U(1)$ bundled over $CP^1$.  In general, squashing the
$S^3$ inside $S^5$ (while keeping the round $S^3$ inside AdS$_5$)
further reduces the supersymmetries of the original LLM system from
1/2 down to 1/8 BPS.  The complete analysis of the supersymmetry
variations (\ref{eq:4var1}) and (\ref{eq:4var2}) is quite involved,
and will not be pursued below.  The first system, (\ref{eq:4var1}),
corresponding to Killing spinors charged along the $U(1)$ fiber
was thoroughly analyzed in \cite{Gava:2006pu}.

We are of course more directly interested in the sequence of
1/2, 1/4 and 1/8 BPS states corresponding to the successive
turning on of $R$-charges $J_1$, $J_2$ and $J_3$.  In this case,
we limit our consideration to the round $S^3\times S^3$ reduction,
which is nothing but the original LLM system of \cite{Lin:2004nb}.
Although the supersymmetry analysis of this system has been thoroughly
investigated in \cite{Lin:2004nb} and subsequent work, for
completeness, and to highlight the complete 1/2, 1/4 and 1/8 BPS
family of solutions, we review the analysis here.

For the round $S^3\times S^3$ reduction, the relevant supersymmetry
variations are given by (\ref{eq:4rounds3}).  Replacing $\pm\hat\eta$
in (\ref{eq:4rounds3}) by $-\tilde\eta$ to  simplify notation,
the supersymmetry variations read
\begin{eqnarray}
\delta\psi_\mu&=&\left[\nabla_\mu-\fft{i}{16}e^{-3\alpha}F_{\nu\lambda}
\gamma^{\nu\lambda}\gamma_\mu\right]\epsilon,\nonumber\\
\delta\lambda_\alpha&=&\left[\gamma^\mu\partial_\mu\alpha
+\fft{i}8e^{-3\alpha}F_{\mu\nu}\gamma^{\mu\nu}-\eta e^{-\alpha}\right]
\epsilon,\nonumber\\
\delta\lambda_\beta&=&\left[\gamma^\mu\partial_\mu\beta
-\fft{i}8e^{-3\alpha}F_{\mu\nu}\gamma^{\mu\nu}-
i\tilde\eta e^{-\beta}\gamma^5\right]\epsilon.
\label{eq:4s3s3susy}
\end{eqnarray}
Since $\epsilon$ may be viewed as a Dirac spinor in four dimensions,
we may form the following bilinears \cite{Lin:2004nb}
\begin{eqnarray}
&&f_1=\overline\epsilon\gamma^5\epsilon,\qquad
f_2=i\overline\epsilon\epsilon,\qquad
K^\mu=\overline\epsilon\gamma^\mu\epsilon,\qquad
L^\mu=\overline\epsilon\gamma^\mu\gamma^5\epsilon,\qquad
Y^{\mu\nu}=i\overline\epsilon\gamma^{\mu\nu}\gamma^5\epsilon,\nonumber\\
&&K^m_\mu=\epsilon^c\gamma_\mu\epsilon,\qquad
Y^m_{\mu\nu}=\epsilon^c\gamma_{\mu\nu}\gamma^5\epsilon.
\end{eqnarray}
Note that $K^m$, viewed as a complex one-form, was denoted $\omega$ in
\cite{Lin:2004nb}.

The above bilinears are normalized according to the Fierz relations
\begin{equation}
K^2=-L^2=-f_1^2-f_2^2,\qquad
Y^2=2f_1^2-2f_2^2,\qquad
|K^m|^2=2f_1^2+2f_2^2,\qquad
|Y^m|^2=-4f_1^2+4f_2^2.
\end{equation}
In addition, they satisfy the identities
\begin{equation}
K\cdot L=K\cdot K^m=L\cdot K^m=0,\qquad
K^\mu Y_{\mu\nu}=f_2L_\nu,\qquad L^\mu Y_{\mu\nu}=f_2 K_\nu.
\end{equation}
Following \cite{Lin:2004nb}, we note that $K^\mu$ defines a time-like
(Killing) direction, while $L_\mu$ is space-like and orthogonal to $K^\mu$.
The four-dimensional space then splits into a two-dimensional base
(the LLM $x_1$--$x_2$ plane) along with a preferred time-like and a
preferred space-like (the LLM $y$ coordinate) direction.

The structure defined by the above bilinears is highlighted by noting that
they may be decomposed according to
\begin{equation}
Y=-\fft{f_2}{f_1^2+f_2^2}K\wedge L+f_1 I,\qquad
K^m=\sqrt{f_1^2+f_2^2}\,\widetilde\Omega,\qquad
Y^m=\fft1{\sqrt{f_1^2+f_2^2}}(f_1K-if_2L)\wedge\widetilde\Omega,
\label{eq:4struct}
\end{equation}
where
\begin{equation}
I_{ab}I_{bc}=-\delta_{ac},\qquad
|\widetilde\Omega|^2=2.
\end{equation}
These expressions are the analog of (\ref{eq:6struct}) for the present
case.  In particular, here the real
two-form $I$ along with the complex
one-form $\widetilde\Omega$ together define a preferred $U(1)$ structure.

The familiar analysis of \cite{Lin:2004nb} proceeds by solving the
one-form identities (\ref{eq:4debf1}) through (\ref{eq:4de3af2})
for the bilinears $f_1$ and $f_2$ as well as for the field strength
$F_{(2)}$.  For simplicity with signs, we choose
\begin{equation}
f_1=-\eta e^\beta,\qquad f_2=-\tilde\eta e^\alpha,
\end{equation}
so that
\begin{equation}
e^{\alpha+\beta}=y,
\end{equation}
where we have chosen to write $L=dy$, which is compatible with $L$ being
a closed one-form, as indicated by (\ref{eq:4dl}).  In this case, $F_{(2)}$
is given by
\begin{equation}
F_{(2)}=\tilde\eta(dt+\omega)\wedge de^{4\alpha}
-\eta h^2e^{3\alpha-3\beta}*_3de^{4\beta},
\label{eq:4f2exp}
\end{equation}
where we have chosen to write the four-dimensional metric as
\begin{equation}
ds_4^2=-h^{-2}(dt+\omega)^2+h^2[h_{ij}dx^idx^j+dy^2],
\label{eq:4met}
\end{equation}
with
\begin{equation}
h^{-2}=f_1^2+f_2^2=e^{2\alpha}+e^{2\beta},\qquad
K=-h^{-2}(dt+\omega),\qquad L=dy.
\end{equation}
Note that $*_3$ is the Hodge dual with respect to the three-dimensional
metric given inside the square brackets above.

We now note that (\ref{eq:4dk}) gives rise to the condition \cite{Lin:2004nb}
\begin{equation}
d\omega=-\eta\tilde\eta\fft1{y}*_3dZ,
\label{eq:4domeg}
\end{equation}
where
\begin{equation}
Z=\fft12\fft{f_2^2-f_1^2}{f_2^2+f_1^2}=\fft12\fft{e^{2\alpha}-e^{2\beta}}
{e^{2\alpha}+e^{2\beta}}.
\end{equation}
In terms of $\omega$ and $Z$, the expression (\ref{eq:4f2exp}) can be
rewritten as
\begin{equation}
F_{(2)}=-\tilde\eta d[e^{4\alpha}(dt+\omega)]-\tilde\eta y^2d\omega
-2\eta(Z+\ft12)*_3dy.
\end{equation}
It is now easy to see that $F_{(2)}$ is automatically closed, so long
as $d\omega$ is \cite{Lin:2004nb,Liu:2004ru}.  Of course, the requirement
$d\omega=0$ for $d\omega$ given in (\ref{eq:4domeg}) yields the LLM
condition that $Z$ be a harmonic function
\begin{equation}
d\left(\fft1y*_3dZ\right)=0,
\label{eq:4llmlap}
\end{equation}
which is the basis for the bubbling AdS picture. The Hodge dual is evaluated
with respect to the three-dimensional metric $h_{ij} dx^i dx^j + dy^2$.

To complete the 1/2 BPS picture, it is worth noting that the metric
$h_{ij}$ on the two-dimensional base can be specified by defining the
canonical-two form $J$ and holomorphic one-form $\Omega$
\begin{equation}
I=h^2J,\qquad \widetilde\Omega=h\Omega.
\label{eq:4iomega}
\end{equation}
where $I$ and $\widetilde\Omega$ are given in (\ref{eq:4struct}).
Using the decomposition of $Y$ in (\ref{eq:4struct}) and
the differential identities (\ref{eq:4eby}) and (\ref{eq:4ea*y})
\begin{equation}
d(f_1Y)=0,\qquad d(f_2*Y)=0,
\end{equation}
we see that
\begin{equation}
d(f_1^2I)=*_3dZ\wedge dy,\qquad d(f_2^2I)=-*_3dZ\wedge dy,
\end{equation}
so that
\begin{equation}
dJ=d(h^{-2}I)=0.
\end{equation}
Furthermore, comparing (\ref{eq:4iomega}) with (\ref{eq:4struct}) demonstrates
that $\omega=K^m$, in which case (\ref{eq:4dkm}) immediately shows that
\begin{equation}
d\Omega=0.
\end{equation}
This combination of $dJ=0$ and $d\Omega=0$ now demonstrates that the
two-dimensional base is flat, in which case we can rewrite (\ref{eq:4met})
using the trivial base metric
\begin{eqnarray}
ds_4^2&=&-h^{-2}(dt+\omega)^2+h^2[dx_1^2+dx_2^2+dy^2]\nonumber\\
&=&-h^{-2}(dt+\omega)^2+h^2[dzd\overline z+dy^2].
\end{eqnarray}
This essentially completes the summary of the LLM analysis \cite{Lin:2004nb}.
In the remaining sections of this paper, we will make use of the results of
the above supersymmetry analyses to develop a universal picture of bubbling
AdS geometries.

%%%%%%%%%%%%%%%%%%%%%%%%%%%%%%%%%%%%%%%%
\section{Bubbling AdS}
\label{sec:bubble}
%%%%%%%%%%%%%%%%%%%%%%%%%%%%%%%%%%%%%%%%

The above reductions on $S^3$, $S^3\times S^1$ and $S^3\times S^3$ and
the supersymmetry analyses provide a uniform framework for describing
the corresponding 1/8, 1/4 and 1/2 BPS configurations in IIB supergravity.
However, we are interested in much more than simply a useful means of
characterizing the supergravity solutions.  What we desire is a
complete understanding of the geometries and how they are mapped into
states in the dual $\mathcal N=4$ Yang Mills theory.

The best developed picture for these bubbling AdS states is of course
in the 1/2 BPS sector, where the $x_1$--$x_2$ plane of \cite{Lin:2004nb}
has a direct counterpart in the phase plane of the dual free fermion
picture of the 1/2 BPS sector of the $\mathcal N=4$ Yang Mills theory
\cite{Corley:2001zk,Berenstein:2004kk}.  Furthermore, `droplets' in the
LLM plane are related to non-trivial topology of the gravity solution,
and are directly equivalent to giant gravitons expanding either in AdS$_5$
or $S^5$.

What we would like to obtain is a similar understanding of the 1/4
and 1/8 BPS sectors of the theory.  However, this task is made rather
more complicated for several reasons. For one thing, on general
grounds, we expect that the 1/2 BPS states (which preserve 16 real
supersymmetries) are described by wave-functions of a
non-interacting free fermion system.  (Note, however, that the
system appears to be interacting when the fermionic degrees of
freedom are changed to bosons.)  The reduced supersymmetry cases do
of course admit descriptions as {\it e.g.}~multi-matrix models on the
gauge theory side.  However, we expect the resulting system to be a
system of interacting bosons without a dual free fermion description, 
and hence more complicated to describe
on the gravity side of the duality. This is in fact borne out by the
explicit 1/8 and 1/4 BPS analysis of
\cite{Kim:2005ez,Donos:2006iy,Donos:2006ms}, as reviewed above in
Section~\ref{sec:susyanal}.  In particular, both the 1/8 and 1/4
cases involve non-linear equations, in contrast with the linear
LLM equation (\ref{eq:4llmlap}), which is the basis for harmonic
superposition of 1/2 BPS states.

Nevertheless, there is an elegant structure underlying the sequence of
1/2, 1/4 and 1/8 BPS states.  As discussed in Section~\ref{sec:susyanal},
these configurations are characterized by $U(1)$, $U(2)$ and $SU(3)$
structure, respectively, and are described by specifying appropriate
field configurations on the corresponding one-, two- and three-complex
dimensional base manifolds.  Since these manifolds are K\"ahler, they
can also be considered symplectic, which is perhaps more natural for a
phase-space description.  In the 1/2 and 1/4 BPS cases, there is an
additional $y$ direction where $y$ is directly related to the volume of
$S^3\times S^3$ for the 1/2 BPS case, or somewhat indirectly related to
the volume of $S^3\times S^1$ in the 1/4 BPS case.  Although the 1/8 BPS
metric, (\ref{eq:7metc}), has no room for an extra $y$ coordinate, we
may nevertheless define $y\equiv e^\alpha$, and thereby obtain an
effective $y$ variable related to the volume of $S^3$.

At this point, it is perhaps worthwhile to summarize the main features of
the 1/2, 1/4 and 1/8 BPS geometries.  From (\ref{eq:4met}), (\ref{eq:6metans})
and (\ref{eq:7metc}), along with the liftings of Section~\ref{sec:breathing},
we have
\begin{eqnarray}
\hbox{1/2 BPS:}&&\qquad ds^2=-h^{-2}(dt+\omega)^2+h^2[h_{ij}dx^idx^j+dy^2]
+e^{2\alpha}d\Omega_3^2+e^{2\beta}d\widetilde\Omega_3^2,\nonumber\\
\hbox{1/4 BPS:}&&\qquad ds^2=-h^{-2}(dt+\omega)^2+e^{-2\alpha}h_{ij}dx^idx^j
+h^2dy^2+e^{2\alpha}d\Omega_3^2+e^{2\beta}(d\psi+\mathcal A)^2,\nonumber\\
\hbox{1/8 BPS:}&&\qquad ds^2=-e^{2\alpha}(dt+\omega)^2+e^{-2\alpha}
h_{ij}dx^idx^j+e^{2\alpha}d\Omega_3^2,
\label{eq:bubblemets}
\end{eqnarray}
where in all cases $h^{-2}=e^{2\alpha}+e^{2\beta}$.  In addition
\begin{equation}
y=e^{\alpha+\beta}\quad\hbox{(for 1/2 and 1/4 BPS)}\quad\hbox{or}
\quad y=e^\alpha\quad\hbox{(for 1/8 BPS)}.
\label{eq:ydef}
\end{equation}
Although the metric and form fields must satisfy various local
conditions (some of which may be rather complicated, especially
in the 1/4 BPS case) in order to ensure a valid solution, the global
features that we are mainly interested in are encoded by the boundary
conditions imposed to ensure regularity of the above metrics. As in
the LLM analysis \cite{Lin:2004nb}, we are concerned with regularity
as any one of the spheres (or circle) in (\ref{eq:bubblemets}) shrinks
to zero size.  Since this occurs at $y=0$, we obtain a natural generalization
of the LLM condition (\ref{eq:llmbc})
\begin{equation}
Z(x_i,y=0)=\pm\ft12\quad\hbox{(for 1/2 and 1/4 BPS)},
\end{equation}
where in both cases
$Z=\fft12(e^{2\alpha}-e^{2\beta})/(e^{2\alpha}+e^{2\beta})$.  The analogous
$y=0$ boundary condition for the 1/8 BPS system is more difficult to
characterize, but is similar in spirit to the above.

In the 1/2 BPS (LLM) case, for geometries asymptotic to AdS$_5\times S^5$,
the $y=0$ plane consists of regions of $Z=-1/2$ (shrinking $S^3$ inside
AdS$_5$) in a background of $Z=1/2$ (shrinking $S^3$ inside $S^5$).  The
AdS$_5\times S^5$ `ground state' corresponds to a circular disk of $Z=-1/2$;
at $y=0$, the interior of this disk is mapped to the `center' of AdS, while
the exterior is mapped to the point where $S^3$ shrinks inside $S^5$.  In
general, the boundary between $Z=1/2$ and $Z=-1/2$ is the locus where both
of the three-spheres simultaneously shrink to zero size.  As a result,
the LLM solution essentially maps the non-trivial topology of the 1/2 BPS
background entirely onto a plane (the $y=0$ plane).  The configuration
is then fully determined by specifying one-dimensional curves in the plane,
corresponding to the boundary between the $Z=1/2$ and $Z=-1/2$ regions.
This is of course the dual picture of the `droplet' description where
regions, or droplets, are specified.

The extension of this picture to the 1/4 BPS case is then straightforward.
In this case, the topology of the background is again determined by the
structure of the solution on the $y=0$ hyperplane.  This time, the hyperplane
is four-dimensional, and may be divided into regions of $Z=1/2$ and $Z=-1/2$
by three-dimensional surfaces.  This time, however, $Z=1/2$ corresponds
to a shrinking one-cycle in $S^5$, while $Z=-1/2$ corresponds as usual to
shrinking $S^3$ inside AdS$_5$.  As we show below, the AdS$_5\times S^5$
ground state in this case consists of a ball of $Z=-1/2$ in a background
of $Z=1/2$.  We do note, however, that in contrast with the LLM picture,
this $y=0$ hyperplane has a non-trivial (K\"ahler) metric, and hence is
not flat.  Nevertheless, so long as the bubbling picture relies only on
the topology of the droplets, it will remain valid.  This distortion of
the geometry is of course to be expected for reduced supersymmetry
configurations, which can no longer be treated as non-interacting
collective modes.

The 1/8 BPS case is particularly interesting, both because it no longer
incorporates a $y=0$ hyperplane, and because it is the most general
case encompassing the other two in appropriate limits.  Defining the
variable $y=e^\alpha$, as in (\ref{eq:ydef}), the locus of shrinking $S^3$
inside AdS$_5$ then corresponds to five-dimensional surfaces of $y=0$
within the six-dimensional base.  In order to obtain a regular geometry,
the 1/8 BPS metric in (\ref{eq:bubblemets}) must then approach a
solution of the form
\begin{equation}
ds^2=\cdots + (dy^2+y^2d\Omega_3^2),\qquad\hbox{as}\quad y\to0.
\label{eq:locsurf}
\end{equation}
In other words, the shrinking $S^3$ combines with the $y$ direction to
locally form $\mathbb R^4$.  In this case, $y$ is non-negative,
and may be considered as a local coordinate normal to the five-dimensional
boundary surfaces.  Viewed in this manner, since $y$ terminates at zero
and does not become negative, the six-dimensional base space ends at
these five-dimensional surfaces.  In particular, the interiors are
unphysical; they simply do not exist.  Another way to understand this is
to note from (\ref{eq:7fal}) that $y$ is related to the scalar curvature
of the base according to $R=-8/y^4$.  Thus these five-dimensional
surfaces of vanishing $y$ are singular (from the six-dimensional point
of view), and space simply ends there, as there is no natural extension
for going past such singularities.  Of course, the full ten-dimensional
solution remains regular, so long as the $y=0$ surfaces are locally of
the form (\ref{eq:locsurf}).

The general picture of 1/8 BPS states is thus one of $S^3$ and time
fibered over a six-dimensional base, where various regions ({\it i.e.}~droplets)
have been excised.  Since the $S^3$ inside AdS$_5$ shrinks on the (in general
disconnected) five-dimensional boundary surface, this surface may be
related to the locus of D3-branes wrapped on the $S^3$, which are simply
dual giant gravitons expanding in AdS$_5$ \cite{Berenstein:2007wz}.
In cases with additional supersymmetries (1/4 or 1/2 BPS), this
six-dimensional base admits an additional $S^1$ or $S^3$ isometry.  In
such cases, the $S^1$ or $S^3$ can be pulled out explicitly, along with
the $y$ variable, which can be promoted to an actual coordinate normal
to the shrinking $S^3$ inside AdS$_5$.  This transformation, which maps
the five-dimensional boundary surfaces to the $y=0$ hyperplane, is
highly non-trivial, but has the feature of placing much of the interesting
topological data onto a single hyperplane within the full ten-dimensional
space-time.

Abstracting the details for a moment, we see a uniform picture emerging,
where 1/2, 1/4 and 1/8 BPS configurations are described by one, three
and five-dimensional surfaces embedded within two, four and six-dimensional
hyperplanes.  Equivalently, we may use a dual description of two, four
and six-dimensional droplets.  Only in the 1/2 BPS case is the $y=0$
hyperplane actually flat.  In the other cases, we expect them to be
diffeomorphic to $\mathbb R^4$ and $\mathbb R^6$ \cite{Berenstein:2007wz},
although such global properties cannot be seen directly from the local
supersymmetry analysis of Section~\ref{sec:susyanal}.  In particular,
bubbling orientifold models \cite{Mukhi:2005cv} can be constructed by
making appropriate discrete identifications on the base spaces.

{}From the $\mathcal N=4$ Yang-Mills side of the duality, the 1/2,
1/4 and 1/8 BPS configurations may be described by one, two and three
(complex) matrix models corresponding to the three complexified
adjoint scalars $X=\phi_1+i\phi_2$, $Y=\phi_3+i\phi_4$ and
$Z=\phi_5+i\phi_6$ of the $\mathcal N=4$ theory
\cite{Corley:2001zk}.  As a result, there is a natural map between
the space of matrix eigenvalues ({\it i.e.}~the free fermion phase
space in the 1/2 BPS case) and the corresponding one, two and
three complex dimensional base spaces $ds^2=h_{ij}dx^idx^j$ in
(\ref{eq:bubblemets}).  In all such cases, the AdS$_5\times S^5$
ground state corresponds to taking a round ball in the base space
(at $y=0$ when appropriate).  Turning on giant graviton excitations
on top of the ground state then corresponds to introducing disconnected
droplets, either inside the ball (giant gravitons expanding in $S^5$)
or outside (dual giant gravitons expanding in AdS$_5$).  Of course, for
1/8 BPS configurations, only the giant gravitons expanding in AdS$_5$
are manifest, as the interior of the ball is completely removed.

Until now, we have said very little about the non-linear equations
characterizing the 1/8 and 1/4 BPS solutions.  For the former, the
main condition on the solution is given by (\ref{eq:7cond}), while
for the latter, one has (\ref{eq:sol_MA}), along with the subsidiary
conditions (\ref{eq:Deqn}), (\ref{eq:Deqn2}) and (\ref{eq:6fwf2}).
In general, these conditions are difficult to work with, and hence
we are unable to present an explicit construction of these reduced
supersymmetry bubbling AdS geometries.  We do note, however, that
in the case of LLM, the 1/2 BPS geometries are fully characterized
by the LLM boundary condition (\ref{eq:llmbc}) $Z=\pm\fft12$ at
$y=0$.  In particular, the LLM Laplacian (\ref{eq:4llmlap}) is only
of secondary importance in developing the bubbling AdS interpretation
of the solutions.  This linear equation does of course facilitate the
writing of explicit solutions, and furthermore is presumably intimately
tied to the non-interacting nature of 1/2 BPS states.  Nevertheless,
the topology of the system, and hence much of the information on
giant gravitons, is contained in the LLM boundary condition itself,
and not necessarily the harmonic superposition rule derived from
(\ref{eq:4llmlap}).  Of course, this was already noted in \cite{Lin:2004nb}
in the case of 1/2 BPS configurations of M-theory, where a droplet
picture emerged from consideration of the boundary conditions,
despite the fact that the full solution involves the Toda equation.

Likewise for the 1/8 and 1/4 BPS systems, we expect that each choice
of boundary conditions (specified either as $y=0$ surfaces in a
six-dimensional base, or as droplets in the $y=0$ hyperplane) gives
rise to a unique bubbling AdS geometry.  Because of the non-linear
nature of the expressions involved, however, we do not envision a
simple proof of either the existence or uniqueness of the solutions.
We certainly expect large classes of solutions to exist, although it
would also be interesting to see if the conditions on the solutions
preclude any particular classes of droplets from existing as regular
bubbling AdS geometries.

%%%%%%%%%%%%%%%%%%%%%%%%%%%%%%%%%%%%%%%%
\section{Examples fitting into the 1/8 BPS case}
\label{EightBPSsection}
%%%%%%%%%%%%%%%%%%%%%%%%%%%%%%%%%%%%%%%%

Although we have not been able to solve the 1/8 and 1/4 BPS conditions
(\ref{eq:7cond}) and (\ref{eq:6volcond}) completely, we may nevertheless
use the existing (known) solutions, as well as a specific class of new
1/4 BPS solutions, to present evidence for the general droplet picture.
We start with several 1/8 BPS (actually $S^3$ isometry) examples
before turning, in Section~\ref{QuarterBPSsection}, to 1/4 BPS geometries.
We should also note that in Section~\ref{regular} we will analyze the
regularity conditions for a rather generic class of 1/8 BPS solutions,
and see that a picture of six-dimensional droplets will emerge by requiring
their ten-dimensional metric to be regular.

The general 1/8 BPS system falls into the $S^3$ isometry analysis
of Section~\ref{sec:d=7susy}.  This solution is presented in terms
of a seven-dimensional metric $g_{\mu\nu}$, two-form $F_{(2)}$ and
scalar $\alpha$, which are given by (\ref{eq:7metc}) and (\ref{eq:7fal}).
Our main concern here is with the metric, which when lifted to ten
dimensions takes the form (\ref{eq:bubblemets})
\begin{equation}
ds^2=-y^2(dt+\omega)^2+\fft1{y^2}h_{ij}dx^idx^j+y^2d\Omega_3^2,
\label{eq:7met2}
\end{equation}
where we have made explicit the identification of $y(x_i)$ with
$e^{\alpha(x_i)}$, as in (\ref{eq:ydef}).  The complete solution is determined
(at least up to diffeomorphisms) in terms of a K\"ahler metric $h_{ij}$
with curvature satisfying (\ref{eq:7cond})
\begin{equation}
\square_6R=-R_{ij}R^{ij}+\ft12R^2,
\label{eq:7cond2}
\end{equation}
and with $y=(-8/R)^{1/4}$.  Note that this identification of $y$ demands
that the K\"ahler base has non-vanishing negative scalar curvature,
with $R\to-\infty$ on the five-dimensional degeneration surfaces where
$y\to0$.  Given these preliminaries, we now turn to some examples.

%%%%%%%%%%%%%%%%%%%%%%%%%%%%%%%%%%%%%%%%
\subsection{AdS$_3\times S^3\times T^4$}

While we are mainly interested in geometries which are asymptotically
connected to AdS$_5\times S^5$, we note that (\ref{eq:7cond2}) admits
a simple solution where the base is taken to be the direct product of
a hyperbolic space with a torus, $\mathbb H^2\times T^4$, with curvature
given by
\begin{equation}
R_{ij}=\begin{cases}-4h_{ij}&i,j=1,2,\\
0&i,j=3,\ldots,6\end{cases}
\end{equation}
(using real coordinates).  This base can be obtained from a K\"ahler potential
\begin{equation}
K(z_1,z_2,z_3)=-\ft12\log(1-|z_1|^2)+\ft12(|z_2|^2+|z_3|^2).
\end{equation}
Because $y$ is a constant (which in our normalization is simply $y=1$), this
solution has constant scalar curvature, and hence no shrinking three-cycles.
Of course, we recall that, since here $y$ is a function and not a coordinate,
there is no problem with setting it to a constant.

When this $\mathbb H^2\times T^4$ base is incorporated into the full metric
(\ref{eq:7met2}), it is easy to see that the resulting geometry is that of
AdS$_3\times S^3\times T^4$.  In particular, by writing the metric on
$\mathbb H^2$ as
\begin{equation}
ds_2^2=d\rho^2+\ft14\sinh^2(2\rho)d\psi^2,
\end{equation}
and by taking
\begin{equation}
\omega=\sinh^2\rho\,d\psi,
\end{equation}
(which is compatible with the condition $\mathcal R=2d\omega$),
we end up with AdS$_3\times S^3\times T^4$ written as
\begin{equation}
ds_{10}^2=-\cosh^2\rho\,dt^2+d\rho^2+\sinh^2\rho\,(d\psi-dt)^2+d\vec x_4^2
+d\Omega_3^2.
\end{equation}
Note that the natural coordinates implicit in the fibration of time over
the K\"ahler base involve motion at the speed of light along the angular
direction in AdS$_3$.

This example is of course the double analytic continuation of the
similar example given in \cite{Kim:2005ez}, which realized
AdS$_3\times S^3\times T^4$ using an $S^2\times T^4$ base.

%%%%%%%%%%%%%%%%%%%%%%%%%%%%%%%%%%%%%%%%
\subsection{AdS$_5\times S^5$}
\label{ads5xs5}

Our primary interest is of course with developing a droplet picture
for excitations on top of AdS$_5\times S^5$.  To proceed in this direction,
we first consider the realization of the AdS$_5\times S^5$ ground state
itself.  In this case, we take the ten-dimensional metric%
\footnote{Note that here we have taken the AdS$_5$ radius $L$ to be unity.}
\begin{equation}
ds_{10}^2=-\cosh^2\rho\,dt^2+d\rho^2+\sinh^2\rho\,d\Omega_3^2
+d\Omega_5^2,
\end{equation}
and identify the $S^3$ in AdS$_5$ with the $S^3$ of (\ref{eq:7met2}).
This determines
\begin{equation}
y=\sinh\rho
\end{equation}
along with the remaining seven-dimensional metric
\begin{equation}
ds_7^2=-\cosh^2\rho\,dt^2+d\rho^2+d\Omega_5^2.
\label{eq:7as5}
\end{equation}
Here there are multiple ways of proceeding.  What we would like, of
course, is to rewrite this metric using giant graviton speed of light
angular coordinates of the form
\begin{equation}
\phi=\psi-t,
\label{eq:phipsit}
\end{equation}
where $\phi$ is a rotation angle in $S^5$, and $\psi$ its natural
giant graviton counterpart.  Because of the
symmetry of the five-sphere, it is natural to parameterize it in terms
of three rotation planes (with three angular coordinates $\phi_i$ and
corresponding angular momenta $J_i$).  However, it is also possible,
and perhaps more convenient, to write $S^5$ as $U(1)$ bundled over
$CP^2$.  While $CP^2$ does not admit a spin-structure,
it nevertheless admits a spin$^c$-structure, and that is the main
reason why we must allow for charged Killing spinors along the fiber
when reducing to six dimensions.

Writing the $S^5$ metric as
\begin{equation}
d\Omega_5^2=ds^2(CP^2)+(d\phi+\mathcal A)^2,\qquad d\mathcal A=2J,
\label{eq:7u1bundle}
\end{equation}
and performing the angular shift (\ref{eq:phipsit}) yields the
seven-dimensional metric
\begin{equation}
ds_7^2=-\sinh^2\rho\left(dt+\sinh^{-2}\!\rho(d\psi+\mathcal A)\right)^2
+\sinh^{-2}\!\rho\left(\sinh^2\rho(d\rho^2+ds^2(CP^2))+\cosh^2\rho
(d\psi+\mathcal A)^2\right).
\end{equation}
As a result, the six-dimensional metric on the base is
\begin{equation}
ds_6^2=(r^2-1)ds^2(CP^2)+dr^2+r^2(d\psi+\mathcal A)^2,
\end{equation}
where we have defined $r=\cosh\rho$.  The Ricci tensor is
\begin{equation}
R_{ij}=\begin{cases}-4(r^2-1)^{-2}h_{ij}&i,j=3,\ldots,6\mbox{ ($CP^2$)},\\
4(r^2-1)^{-2}h_{ij}&i,j=1,2\mbox{ ($r$ and $\psi$)}.\end{cases}
\end{equation}

The alternate more symmetrical decomposition of $S^5$ follows by
introducing the complex coordinates
\begin{eqnarray}
z_1&=&r\cos\theta_1 e^{i\phi_1},\nonumber\\
z_2&=&r\sin\theta_1\cos\theta_2e^{i\phi_2},\nonumber\\
z_3&=&r\sin\theta_1\sin\theta_2e^{i\phi_3}.
\end{eqnarray}
In this case, we have
\begin{eqnarray}
|dz_i|^2&=&dr^2+r^2d\Omega_5^2,\nonumber\\
|\overline z_idz_i|^2&=&r^2dr^2+r^4(\cos^2\theta_1 d\phi_1+\sin^2\theta_1
\cos^2\theta_2d\phi_2+\sin^2\theta_1\sin^2\theta_2d\phi_3)^2.
\end{eqnarray}
Taking the seven-dimensional metric (\ref{eq:7as5}) and shifting
\begin{equation}
\phi_i=\psi_i-t
\label{eq:7shiftpsi}
\end{equation}
we obtain
\begin{equation}
ds_7^2=-\sinh^2\rho(dt+\omega)^2+\sinh^{-2}\!\rho\left(
\sinh^2\rho\,d\rho^2-\cosh^2\rho\fft{dr^2}{r^2}+\sinh^2\rho
\fft{|dz_i|^2}{r^2}+\fft{|\overline z_idz_i|^2}{r^4}\right),
\end{equation}
where
\begin{equation}
\omega=\sinh^{-2}\!\rho\left(\cos^2\theta_1d\psi_1+\sin^2\theta_1\cos^2\theta_2
d\psi_2+\sin^2\theta_1\sin^2\theta_2d\psi_3\right),
\end{equation}
and now the $z_i$'s are defined with the angles $\psi_i$.

In order to eliminate the original $\rho$ coordinate, we may define
\begin{equation}
r=\cosh\rho.\label{rgeq1}
\end{equation}
The resulting six-dimensional base metric has the simple form
\begin{equation}
ds_6^2=(|z_i|^2-1)\fft{|dz_i|^2}{|z_i|^2}+\fft{|\overline
z_idz_i|^2}{(|z_i|^2)^2}, \label{eq:6gsbase}
\end{equation}
and may be obtained from a K\"ahler potential
\begin{equation}
K=\ft12|z_i|^2-\ft12\log(|z_i|^2).
\label{eq:7ads5kpot}
\end{equation}
For completeness, we note that
\begin{equation}
\omega=\fft1{|z_i|^2-1}\fft{\Im(\overline z_idz_i)}{|z_i|^2}.
\end{equation}

\subsubsection{Boundary conditions}
\label{sec:adsbc}

It is now instructive to examine the form of the six-dimensional base
given by (\ref{eq:6gsbase}).  The complex coordinates $z_i$ cover the
space completely, and are furthermore restricted to the region
$|z_i|^2\ge1$, as it is evident from (\ref{rgeq1}).
Moreover, since
\begin{equation}
y^2=|z_i|^2-1,
\end{equation}
we see that $y$ naturally parameterizes the radial direction in $\mathbb C^3$
starting from the unit five-sphere on outward.  This confirms the picture
developed above in Section~\ref{sec:bubble} that the AdS$_5\times S^5$
vacuum corresponds to removing a round ball from the K\"ahler base
which, while not flat, is nevertheless diffeomorphic to $\mathbb C^3$.
Note also that this description matches perfectly with the matrix
wave-function picture explored recently in \cite{Berenstein:2007wz}.

%%%%%%%%%%%%%%%%%%%%%%%%%%%%%%%%%%%%%%%%
\subsection{Three-charge smooth solutions}
\label{sec:3Q}

Given the picture of the AdS$_5\times S^5$ ground state as a round ball
removed from $\mathbb C^3$, we may in general consider two types of
excitations.  As in \cite{Lin:2004nb}, the first consist of deformations
of the surface of the ball, corresponding to Kaluza-Klein excitations
(gravitational ripples), and the second consists of introducing topology
changing droplets, corresponding to giant gravitons.

In principle, excitations corresponding to ripples on the Fermi surface
can be fully explored in the linearized regime.  By consistency, the result
must reproduce the subsector of Kaluza-Klein modes of IIB theory on
AdS$_5\times S^5$ \cite{Kim:1985ez} that is consistent with the 1/8 BPS
condition.  In the 1/2 BPS case, this connection was explicitly demonstrated
in \cite{Grant:2005qc}.

Here we choose not to carry out the complete linearized analysis at this
time.  Instead, we consider a class of smooth, three-charge `AdS bubble'
solutions which were studied in \cite{Chong:2004ce}.  These solutions
are smoothed out (no horizon) versions of the $R$-charged black holes
({\it i.e.}~superstars), and are described by a five-dimensional field
configuration
\bea
\label{5D3Q}
&&ds_5^2 = - (H_1 H_2 H_3)^{-2/3} \,f\, dt^2
+ (H_1 H_2 H_3)^{1/3}(f^{-1} dr^2 + r^2 d\Omega_3^2), \nn \\
&&A_{(1)}^i = -H_i^{-1} dt,\qquad X_i = (H_1 H_2 H_3)^{1/3}\, H_i^{-1},
\qquad \cosh\varphi_i = (RH_i)' , \nn \\
&&f=1+r^2 H_1 H_2 H_3,
\eea
where $R\equiv r^2$, and where a prime denotes a derivative with respect
to $R$.  Furthermore, the functions $H_i$ obey the equation
\beq
\label{Heq}
f (RH_i)''=[1-(RH_i)'^2 ] \,(H_1 H_2 H_3) H_i^{-1}.
\eeq
This non-linear coupled set of equations admits the trivial solution
$(RH_i)'=1$ (for all $i=1,2,3$), in which case $\varphi_i=0$ and
$H_i=1+Q_i/R$.  This simply reproduces the three-charge superstar solutions
of \cite{Behrndt:1998ns,Behrndt:1998jd}.  On the other hand, while
the general exact solution to this system of equations is not known
(except in the one-charge, {\it i.e.}~LLM, case), numerical investigations
indicate that it admits a six-parameter family of solutions, corresponding
to three charges $Q_i$ and three corresponding scalar deformations
related to turning on $\varphi_i\ne0$.  For fixed charges, the three scalar
parameters may then be adjusted to ensure regularity of the solution
as $R\to0$.  In particular, regularity here means that both $H_i$ and
its derivatives $H_i'$ remain bounded as $R\to0$.

These three-charge solutions preserve 1/8 of the supersymmetries, and
are furthermore regular without horizons.  As such, they must fall under
the classification of Section~\ref{sec:d=7susy}.  To see how they
may be expressed in the bubbling metric form of (\ref{eq:7met2}), we
first lift (\ref{5D3Q}) to ten dimensions following the procedure
outlined in \cite{Cvetic:2000nc}:
\beq
ds^2_{10}= \sqrt{\Delta} \, ds_5^2 + \frac{1}{\sqrt{\Delta}}\,
T_{IJ}^{-1} \, D\mu^I D\mu^J,
\eeq
where
\beq
\Delta \equiv T_{IJ} \, \mu^I \mu^J,\qquad \sum_{I=1}^6 \, \mu^I \mu^I=1,
\qquad D\mu^I\equiv d\mu^I+A^{IJ}_{(1)}\mu^J.
\eeq
The constrained scalars $X_i$, along with the fields $\varphi_i$ are given by
the decomposition
\beq
T_{IL} = \text{diag}(X_1 \, e^{-\varphi_1},X_1 \, e^{\varphi_1},
X_2 \,e^{-\varphi_2},X_2 \,e^{\varphi_2},
X_3 \,e^{-\varphi_3}, X_3 \,e^{\varphi_3}),
\eeq
and the $U(1)^3$ gauge fields are
\beq
\label{As}
A^{12}_{(1)}=A^1_{(1)}, \qquad A^{34}_{(1)}=A^2_{(1)}, \qquad
A^{56}_{(1)}=A^3_{(1)}.
\eeq
More explicitly, using (\ref{As}) and $A^i_{(1)}=-H_i^{-1}\, dt$ we have
the three pairs of expressions
\bea
D\mu_1 &=& d\mu_1 - \mu_2\, H_1^{-1}\, dt, \qquad
D\mu_2 = d\mu_2 + \mu_1\, H_1^{-1}\, dt, \nn \\
D\mu_3 &=& d\mu_3 -  \mu_4\, H_2^{-1}\, dt, \qquad
D\mu_4 =   d\mu_4 +  \mu_3\, H_2^{-1}\, dt, \nn \\
D\mu_5 &=& d\mu_5 -  \mu_6\, H_3^{-1}\, dt, \qquad
D\mu_6 =   d\mu_6 +  \mu_5\, H_3^{-1}\, dt.
\eea
Also, we find that
\beq
\Delta = X_1 (e^{-\varphi_1}\mu_1^2 + e^{\varphi_1}\mu_2^2) +  X_2 (e^{-\varphi_2}\mu_3^2 + e^{\varphi_2}\mu_4^2)
+ X_3 (e^{-\varphi_3}\mu_5^2 + e^{\varphi_3}\mu_6^2).
\eeq
The uplifted metric can then be written as
\bea
\label{3Qmetric}
ds^2_{10}&=& \sqrt{\Delta}\, \Bigl[-\frac{f}{(H_1 H_2 H_3)^{2/3}} \, dt^2
+ (H_1 H_2 H_3)^{1/3}(f^{-1}dr^2 +r^2d\Omega_3^2)\Bigr] \nn \\
&&+ \frac{1}{\sqrt{\Delta}} \,
\Biggl[H_1\frac{e^{\varphi_1}(D\mu_1)^2
+e^{-\varphi_1}(D\mu_2)^2}{(H_1 H_2 H_3)^{1/3}}
+ H_2 \frac{e^{\varphi_2}(D\mu_3)^2
+e^{-\varphi_2}(D\mu_4)^2}{(H_1 H_2 H_3)^{1/3}}\nn \\
&&\kern4em +H_3 \frac{e^{\varphi_3}(D\mu_5)^2
+e^{-\varphi_3}(D\mu_6)^2}{(H_1 H_2 H_3)^{1/3}}\Biggr].
\eea
In addition, we make the following explicit choice of coordinates on
the five-sphere
\beq
\label{mus}
{\mu}=(\tilde{\mu}_1\sin{\phi_1},\tilde{\mu}_1\cos{\phi_1},\tilde{\mu}_2\sin{\phi_2},\tilde{\mu}_2\cos{\phi_2},
\tilde{\mu}_3\sin{\phi_3},\tilde{\mu}_3\cos{\phi_3}),
\eeq
where
\beq
\tilde{\mu}_1 = \sin{\theta},\qquad\tilde{\mu}_2 = \cos{\theta} \sin{\alpha},
\qquad\tilde{\mu}_3 = \cos{\theta} \cos{\alpha}.
\eeq
These `direction cosines' obey
\beq
\sum_{I=1}^6 \, \mu_I^2 = \sum_{i=1}^3 \tilde{\mu}_i^2 =1.
\eeq

The first step in transforming this solution into the 1/8 BPS form
(\ref{eq:7met2}) is to identify the three-sphere inside AdS$_5$.
In this case, examination of (\ref{3Qmetric}) directly yields
\beq
y^2= \sqrt{\Delta} \, r^2 \, (H_1 H_2 H_3)^{1/3}.
\label{eq:3cysqfunc}
\eeq
Next, by properly collecting the time components, we may write the
remaining seven-dimensional part of the metric in the standard form
\beq
ds_7^2 = -y^2(dt+\omega)^2+y^{-2}h_{mn}\, dx^m \, dx^n ,
\label{Gen1/8Met}
\eeq
where
\bea
\omega_{\phi_i} &=& -\frac{\mut_i^2}{r^2 \, \Delta \,(H_1 H_2 H_3)^{2/3}}
\bigl[ (\cos\phi_i)^2 \, e^{-\varphi_i} + (\sin\phi_i)^2 \, e^{\varphi_i}
\bigr] ,\nn\\
\omega_{\mut_i} &=& \frac{2 \mut_i \sin\phi_i \cos\phi_i \sinh\varphi_i}
{(H_1 H_2 H_3)^{2/3}},
\label{omegaphi}
\eea
and the metric on the six-dimensional base is given by
\bea
\label{1/8base}
h_{rr} &=& \frac{r^2 \, (H_1 H_2 H_3)^{2/3} \, \Delta}{f} , \nn \\
h_{\phi_i \, \phi_j} &=& \delta_{ij} \, r^2 \, H_i \, \mut_i^2
\,[\cos^2\phi_i e^{-\varphi_i} + \sin^2\phi_i e^{\varphi_i} ] \nn\\
&&+  \frac{\mut_i^2 \, \mut_j^2}{\Delta \, (H_1 H_2 H_3)^{2/3} }
\,[\cos^2\phi_i e^{-\varphi_i} + \sin^2\phi_i e^{\varphi_i} ]
\,[\cos^2\phi_j e^{-\varphi_j} + \sin^2\phi_j e^{\varphi_j} ], \nn \\
h_{\mut_i \, \mut_j} &=& \delta_{ij} r^2 \, H_i \,
[\cos^2\phi_i e^{\varphi_i} + \sin^2\phi_i e^{-\varphi_i} ] \nn \\
&&+ \frac{4\mut_i \, \mut_j \, \cos\phi_i \, \sin\phi_i \, \cos\phi_j \,
\sin\phi_j }{\Delta \, (H_1 H_2 H_3)^{2/3}}\sinh{\varphi_i} \,
\sinh{\varphi_j}, \nn \\
h_{\mut_i \, \phi_j} &=&
-\delta_{ij} \, 2r^2 \mut_i \cos\phi_i \sin\phi_i H_i
\sinh{\varphi_i}\nn \\
&&- \frac{2\cos\phi_i \, \sin\phi_i}{\Delta \, (H_1 H_2
H_3)^{2/3}}\,[\cos^2\phi_j e^{-\varphi_j} + \sin^2\phi_j e^{\varphi_j} ]
\sinh{\varphi_i}.
\eea

To show that $h_{mn}$ is K\"ahler, we introduce complex coordinates
\beq
z_i = \rho_i(r^2) \, \mut_i \,
\bigl[\cos\phi_i e^{\varphi_i/2}+i \, \sin\phi_i e^{-\varphi_i/2} \bigr],
\qquad i=1,2,3 .
\label{eq:3ccc}
\eeq
The functions $\rho_i$ are implicitly defined through the equation
\beq
\partial_R \, \log\rho_i^2 = \frac{H_1 H_2 H_3}{H_i f}\, \cosh\varphi_i,
\qquad R\equiv r^2 .
\label{eq:rhoieqn}
\eeq
For the K\"ahler potential, we postulate the following dependence on the
complex coordinates
\beq
K=K(\ft12 (z_i^2+\zb_i^2),|z_i|^2) ,
\eeq
and for convenience we define the quantities
\bea
x_i &=& \ft12 (z_i^2+\zb_i^2)\, , \nn \\
y_i &=& |z_i|^2.
\eea
One can then read off from the $\mut_i$ and $\phi_i$ metric components
in (\ref{1/8base}) the following differential conditions for the K\"ahler
potential:
\bea
\partial_{y_i} K (x_i,y_i) &=& \frac{RH_i}{2\rho_i^2} ,\nn \\
\partial_{x_i}\partial_{x_j} K (x_i,y_i) &=& \frac{1}{2\Lambda \, H_1 H_2 H_3}
\frac{\sinh\varphi_i \, \sinh\varphi_j}{\rho_i^2 \, \rho_j^2} , \nn\\
\partial_{x_i}\partial_{y_j} K (x_i,y_i) &=& -\frac{1}{2\Lambda \, H_1 H_2 H_3}
\frac{\sinh\varphi_i \, \cosh\varphi_j}{\rho_i^2 \, \rho_j^2} ,\nn \\
\partial_{y_i}\partial_{y_j} K (x_i,y_i) &=& \frac{1}{2\Lambda \, H_1 H_2 H_3}
\frac{\cosh\varphi_i \, \cosh\varphi_j}{\rho_i^2 \, \rho_j^2} ,
\label{KDEQ14}
\eea
where
\beq
\Lambda = \frac{\Delta}{(H_1 H_2 H_3)^{1/3}}.
\eeq
Furthermore, consistency of the above equations implies the following differential conditions for the function $R(x_i,y_i)$:
\beq
\partial_{x_i} R = - \frac{f\, \sinh\varphi_i}{\rho_i^2\,\Lambda\,H_1 H_2 H_3},
\qquad
\partial_{y_i} R = \frac{f\, \cosh\varphi_i}{\rho_i^2\,\Lambda\,H_1 H_2 H_3}.
\eeq

In the end, we have only been able to obtain the K\"ahler potential implicitly
in terms of its derivatives (\ref{KDEQ14}).  To check
that we have obtained the correct metric, we compute
\bea
d{s}_6^2 &=& 2\partial_{z_i} \partial_{\zb_j} K \, dz_i \, d\zb_j \nn \\
&=& \sum_i \frac{R \, H_i}{\rho_i^2} \,dz_i d\zb_i + \sum_{i,j}
\frac{1}{\Lambda \,H_1 H_2 H_3 \, \rho_i^2 \, \rho_j^2} \nn \\
&&\times
\bigl[(\zb_i \, \cosh\varphi_i -z_i \, \sinh\varphi_i )\, dz_i \bigr]
\bigl[(z_j \, \cosh\varphi_j -\zb_j \, \sinh\varphi_j )\, d\zb_j \bigr].
\eea
After some algebra, and using
\beq
dz_i = z_i
\frac{d\mut_i}{\mut_i}+i(z_i\cosh\varphi_i-\zb_i\sinh\varphi_i)\,d\phi_i
+ \frac{r\,H_1 H_2 H_3}{H_i \,
f}(z_i\cosh\varphi_i-\zb_i\sinh\varphi_i)\,dr,
\eeq
one can recover the metric components listed in (\ref{1/8base}).

As a special limit of the regular three-charge solution discussed above,
we may consider the three-charge extremal black hole (superstar)
obtained by setting all the scalar fields $\varphi_i$ to zero.
The resulting singular solution has
\beq
H_i = 1+ \frac{Q_i}{r^2},
\label{eq:sshi}
\eeq
with $Q_i$ representing the black hole charges.  Thus, the
three-charge black hole can be embedded into the $1/8$ BPS ansatz
simply by taking the $\varphi_i=0$ limit of the K\"{a}hler metric
found above. Complex coordinates will now take the form
\beq
z_i = \rho_i(r^2) \, \mut_i \, e^{i\phi_i} \, , \qquad i=1,2,3,
\eeq
with the functions $\rho_i$ defined through
\beq
\partial_R \log\rho_i^2 = \frac{H_1 H_2 H_3}{f\, H_i}.
\label{eq:ssrhoi}
\eeq
Defining again $y_i=|z_i|^2$, we find that the K\"ahler potential is
now only a function of the magnitudes
\beq
K=K(|z_i|^2)=K(y_i).
\eeq
The differential equations for the K\"ahler potential reduce to
\bea
\partial_{y_i} K &=& \frac{RH_i}{2\rho_i^2} , \nn\\
\partial_{y_i}\partial_{y_j}K&=&\frac{1}{2\Lambda\,H_1 H_2 H_3\,\rho_i^2\,
\rho_j^2},
\label{k_2nd_deriv}
\label{KDEQbh}
\eea
where
\beq
\Lambda=\sum_{i=1}^3 \, \frac{y_i}{\rho_i^2 \, H_i}
=\frac{\Delta}{(H_1 H_2 H_3)^{1/3}}. \label{eqn_lambda} \eeq
Consistency of the equations above yields the equation for the
function $r^2(y_i)$:
\beq
\partial_{y_i}r^2 = \frac{f}{\Lambda \, H_1 H_2 H_3 \, \rho_i^2}.
\label{r2_deriv} \eeq

\subsubsection{Boundary conditions}

Our main interest in examining the three-charge smooth solutions is
of course to explore the boundary surface where the $S^3$ inside AdS$_5$
collapses.  As indicated by (\ref{eq:3cysqfunc}), the $y$ function
is given by
\beq
y^2=\sqrt{\Delta}\, r^2 (H_1 H_2 H_3)^{1/3},
\eeq
and we are interested in the locus where this vanishes.  Although this
is a product of several functions, we first note that regularity and
smoothness of the solution demands that the functions $H_i$ never vanish.
In particular, they must approach a non-zero constant as $r\to0$.
This in turn keeps $\Delta$ finite and non-zero.  As a result, we
conclude that $y=0$ only when $r = 0$.

Since $y$ is an implicit function of the three complex coordinates
\beq
z_i = \rho_i(r^2) \, \mut_i \, \Bigl[\cos\phi_i e^{\varphi_i/2}+i\sin\phi_i
e^{-\varphi_i/2}\Bigr]\,
\eeq
defined in (\ref{eq:3ccc}), the algebraic condition $y=0$ (or
equivalently $r=0$) imposes a single real constraint on the $z_i$
coordinates, yielding a five real dimensional surface embedded in
$\mathbb C^3$.  To examine the shape of this surface, we first use
\bea
\mut_i \cos\phi_i &=& \Re\Bigl(\frac{z_i}{\rho_i}e^{-\varphi_i/2}\Bigr) ,
\nn \\
\mut_i \sin\phi_i &=& \Im\Bigl(\frac{z_i}{\rho_i}e^{\varphi_i/2}\Bigr) ,
\eea
to find
\beq
\mut_i^2=\frac{e^{-\varphi_i}}{\rho_i^2}\Bigl(\frac{z_i+\zb_i}{2}\Bigr)^2-
\frac{e^{\varphi_i}}{\rho_i^2}\Bigl(\frac{z_i-\zb_i}{2}\Bigr)^2 .
\eeq
Finally, using the constraint
\beq
\sum_{i=1}^3 \mut_i^2 =1,
\eeq
we obtain
\beq
\sum_{i=1}^3 \, \Bigl[\frac{\cosh\varphi_i}{\rho_i^2} |z_i|^2 -\frac{\sinh\varphi_i}{\rho_i^2}
\Bigl(\frac{z_i^2+\zb_i^2}{2}\Bigr)
\Bigr]=1.
\eeq
The degeneration surface that we are interested in lies at $r=0$.
Since the functions $\rho_i$ and $\varphi_i$ given above are functions
of $r$, we define
\beq
\bar{\rho}_i \equiv \rho_i(r=0), \qquad \bar{\varphi}_i \equiv
\varphi_i(r=0),
\eeq
to be their boundary values.  Regularity of the three-charge solution
ensures that these values are all non-vanishing.  In this case, the
five-dimensional surface is given simply by
\beq
\label{3Qsurface}
\sum_{i=1}^3 \Bigl[\frac{\cosh\bar{\varphi}_i}{\bar{\rho}_i^2} |z_i|^2 -\frac{\sinh\bar{\varphi}_i}{\bar{\rho}_i^2}
\Bigl(\frac{z_i^2+\zb_i^2}{2}\Bigr) \Bigr]=1 .
\eeq
This is an ellipsoid, as can be seen more clearly by writing it in terms
of real and imaginary parts $z_i = x_i + i y_i$:
\beq
\sum_{i=1}^3 \frac{1}{\bar{\rho}_i^2}\Bigl[e^{-\bar{\varphi}_i} \, x_i^2 + e^{\bar{\varphi}_i} \, y_i^2
\Bigr]=1.
\label{eq:elipeqn}
\eeq

This ellipsoid may be considered to be a deformation of the round
sphere corresponding to the AdS ground state discussed above in
Section~\ref{sec:adsbc}.  One way to see this is to note that turning
off the deformation scalars, $\bar\varphi_i\to0$, forces $H_i\to1$
(to avoid the potential singularity at $r=0$).  In this case, the
three-charge solution reduces to the AdS$_5\times S^5$ vacuum, and
(\ref{eq:rhoieqn}) is trivially integrated to give $\rho_i^2=1+r^2$.
This in turn gives $\bar\rho_i=1$, in which case (\ref{eq:elipeqn})
reduces to the equation for a sphere of unit radius
\begin{equation}
\sum_{i=1}^3(x_i^2+y_i^2)=1,
\end{equation}
corresponding to the ground state `Fermi surface' which yields
the AdS$_5\times S^5$ vacuum.

As we noted for the AdS$_5 \times S^5$ vacuum, only the outside
of the ellipsoid (\ref{eq:elipeqn}) is allowed.  To see
this, it is enough to show that both $\rho_i^2e^{\varphi_i}$
and $\rho_i^2e^{-\varphi_i}$ are monotonically increasing functions of $r^2$.
Using $\cosh\varphi_i = (RH_i)'$ and the equation of motion (\ref{Heq}),
we find
\beq
\partial_R \, \varphi_i = \frac{H_1 H_2 H_3}{H_i f}\,(-\sinh\varphi_i).
\eeq
This may be combined with the expression for $\partial_R\log\rho_i^2$
from (\ref{eq:rhoieqn}) to obtain
\bea
\partial_R  \Bigl(\frac{\rho_i^2}{e^{\varphi_i}}\Bigr)
&=& \frac{\partial_R \rho_i^2 - \rho_i^2 \partial_R\,\varphi_i}{e^{\varphi_i}}
=\frac{\rho_i^{2}\,H_1 H_2 H_3}{e^{\varphi_i} H_i \,f}
(\cosh\varphi_i+\sinh\varphi_i)
=\frac{\rho_i^2 \, H_1 H_2 H_3}{H_i \, f} \geq 0, \nn \\
\partial_R  \Bigl(\frac{\rho_i^2}{e^{-\varphi_i}}\Bigr)
&=& \frac{\partial_R \rho_i^2 + \rho_i^2 \partial_R\,\varphi_i}{e^{-\varphi_i}}
=\frac{\rho_i^{2}\,H_1 H_2 H_3}{e^{-\varphi_i} H_i \,f}
(\cosh\varphi_i-\sinh\varphi_i)
=\frac{\rho_i^2 \, H_1 H_2 H_3}{H_i \, f} \geq0 .\quad
\eea
Thus the six axes of the ellipsoid $\rho_ie^{\varphi_i/2}$
and $\rho_ie^{-\varphi_i/2}$ all increase with $r$, which shows
that only the region outside the smallest ellipsoid (given by $r=0$)
is occupied.

Deforming the round ball into an ellipsoid corresponds to turning on
angular momentum two harmonics on $S^5$.  These modes are part of the
standard Kaluza-Klein spectrum \cite{Kim:1985ez}.  Likewise, the
three-charge smooth gravity solution of \cite{Chong:2004ce}, given by
the fields (\ref{5D3Q}), is dual to $\mathcal N=4$ Yang-Mills in a
1/8 BPS sector built on top of a combination of $\Tr(X^2)$, $\Tr(Y^2)$
and $\Tr(Z^2)$.

It is also instructive to consider the superstar (singular $R$-charged
black hole) limit of the above three-charge solution, which is obtained
by taking $\varphi_i=0$ while keeping at least one of the three $R$ charges
turned on.  In this case, from (\ref{3Qsurface}), we can read off the
corresponding five-dimensional degeneration surface
\beq
\sum_{i=1}^3 \frac{|z_i|^2}{\bar{\rho}_i^2}=1.
\label{eq:ssdegen}
\eeq
However, for a complete picture, we also need information on the
values of $\bar\rho_i$ for the superstar.  For three non-vanishing
charges, we may integrate (\ref{eq:ssrhoi}) using (\ref{eq:sshi}) to
arrive at
\begin{equation}
\rho_i^2=\prod_{a=1}^3(R-R_a)^{\lambda^i_a},
\label{eq:ssrhoint}
\end{equation}
where $R_a$ are the three roots of the cubic expression
\begin{equation}
0=R^2f=R^2+\prod_{a=1}^3(R+Q_a).
\label{eq:sscubic}
\end{equation}
Note that, so long as the charges $Q_a$ are non-negative (which we always
assume as a physical condition), then none of the roots $R_a$ can lie on
the positive real axis.  The exponents in (\ref{eq:ssrhoint}) are given
by
\begin{equation}
\lambda^i_a=\fft{(R_a+Q_{i+1})(R_a+Q_{i+2})}
{(R_a-R_{a+1})(R_a-R_{a+2})},
\end{equation}
where the subscripts are to be taken modulo three ({\it i.e.}~to lie
in the range $1,2,3$).  For a fixed $i$, these exponents satisfy
\begin{equation}
\sum_{a=1}^3\lambda_a^i=1,\qquad
\sum_{a=1}^3R_a\lambda_a^i=-Q_i-1.
\end{equation}
As a result, the large $R$ behavior of (\ref{eq:ssrhoint}) is simply
\begin{equation}
\rho_i^2(R)\sim R+1+Q_i+\mathcal O\Bigl(\fft1R\Bigr).
\label{asym_rho}
\end{equation}

We are of course more interested in the fate of the ellipsoid
(\ref{eq:ssdegen}), which is obtained from the minimum values $\bar\rho_i$.
The three non-vanishing charge case is somewhat unusual, in that the
naked singularity is generally reached for $R<0$ \cite{Behrndt:1998ns}.
This occurs at the first zero of the function $R^3H_1 H_2H_3=\prod(R+Q_i)$,
which we may take to be at $R=-Q_3$ by appropriate ordering of the charges
({\it i.e.}~$Q_1\ge Q_2\ge Q_3>0$).  By expanding (\ref{eq:ssrhoint})
near this singularity, we obtain
\begin{eqnarray}
\rho_i^2&=&\bar\rho_i^2\Bigl[1+(R+Q_3)\delta_{i3}\fft{(Q_1-Q_3)(Q_2-Q_3)}{Q_3^2}
\nonumber\\
&&\qquad+\ft12(R+Q_3)^2\left(|\epsilon_{ij3}|\fft{Q_j-Q_3}{Q_3^2}+\delta_{i3}
\fft{2Q_1Q_2-(Q_1+Q_2)Q_3}{Q_3^3}\right)+\cdots\Bigr],
\label{eq:ssrhohor}
\end{eqnarray}
where
\begin{equation}
\bar\rho_i^2(R)=\prod_{a=1}^3(-R_a-Q_3)^{\lambda_a^i}.
\end{equation}
This shows that, despite the presence of the naked singularity, the
ellipsoid defined by (\ref{eq:ssdegen}), and with interior removed,
is still present for the generic three charge superstar.  Here, the
singularity of the solution is rather subtle, and arises not because
of degeneration of the boundary surface, but rather because vanishing
of the linear term for $\rho_1^2$ and $\rho_2^2$ in (\ref{eq:ssrhohor})
results in unwanted singular behavior of the K\"ahler base near the
ellipsoid.  While the curvature of the K\"ahler base for a regular
solution is supposed to blow up as $R\sim -8/y^4$ where $y$ is the
normal to the boundary, here the singularity is apparently of a
different nature.

The above expressions are slightly modified in the case of one or
more vanishing charges.  For $Q_3=0$, the $\rho_i$ are given by
\begin{eqnarray}
\rho_1^2&=&\sqrt{(R-R_+)(R-R_-)}\left(\fft{R-R_-}{R-R_+}\right)^{\fft{1+Q_1-Q_2}
{2\sqrt{(Q_1+Q_2+1)^2-4Q_1Q_2}}},\nonumber\\
\rho_2^2&=&\sqrt{(R-R_+)(R-R_-)}\left(\fft{R-R_-}{R-R_+}\right)^{\fft{1-Q_1+Q_2}
{2\sqrt{(Q_1+Q_2+1)^2-4Q_1Q_2}}},\nonumber\\
\rho_3^2&=&R\left(\fft{R-R_-}{R-R_+}\right)^{\fft1
{\sqrt{(Q_1+Q_2+1)^2-4Q_1Q_2}}},
\end{eqnarray}
where
\begin{equation}
R_\pm=-\ft12[(Q_1+Q_2+1)\mp\sqrt{(Q_1+Q_2+1)^2-4Q_1Q_2}]
\end{equation}
are the two non-zero roots of (\ref{eq:sscubic}).  Note that
$R_-<R_+<0$.  As a result, the naked singularity is reached at $R=0$,
where $\rho_3^2$ vanishes.  This demonstrates that $\bar\rho_3^2=0$
in the two charge case.  Hence in this case the ellipsoid (\ref{eq:ssdegen})
collapses, and the singularity of the solution is manifest.

The one-charge superstar is even more straightforward.  If $Q_1$ is the only
non-vanishing charge, we have
\begin{eqnarray}
\rho_1^2&=&R+Q_1+1,\nonumber\\
\rho_2^2&=&R^{\fft{Q_1}{Q_1+1}}(R+Q_1+1)^{\fft1{Q_1+1}},\nonumber\\
\rho_3^2&=&R^{\fft{Q_1}{Q_1+1}}(R+Q_1+1)^{\fft1{Q_1+1}}.
\end{eqnarray}
Taking $R\to0$, we read off $\bar\rho_2^2=\bar\rho_3^2=0$, and thus
the ellipsoid collapses in two of the three complex directions.  The
remaining direction defines a circle in the $z_1$ plane, corresponding
to the LLM disk with intermediate value of the LLM $Z(z_1,\bar z_1,y)$
function at $y=0$, as originally demonstrated in \cite{Lin:2004nb}.

%%%%%%%%%%%%%%%%%%%%%%%%%%%%%%%%%%%%%%%%
\subsection{LLM}
\label{sec:18llm}

The exploration of the three charge smooth solutions in the previous
subsection has allowed us to gain some intuition on the nature of turning
on Kaluza-Klein excitations, corresponding to smooth deformations of the
Fermi surface.  However, we are also interested in the case of topology
change and the emergent picture of droplets (particle and hole excitations).
While we do not have a particularly constructive way of obtaining complete
1/8 BPS solutions with non-trivial topology, there is in fact a large
class of topologically interesting solutions which we may investigate,
and these are nothing but the LLM ones.  The LLM geometries of course
preserve 1/2 of the supersymmetries, so comprise a very special
subclass of the configurations described by the 1/8 BPS system of
(\ref{eq:7met2}) and (\ref{eq:7cond2}).

The 1/2 BPS LLM solution (\ref{eq:bubblemets}) has the form \cite{Lin:2004nb}
\begin{equation}
ds_{10}^2=-h^{-2}(dt+V)^2+h^2(|dz_1|^2+dy^2)+ye^Gd\Omega_3^2+ye^{-G}d\widetilde
\Omega_3^2,
\label{eq:llmeqn}
\end{equation}
where $V=V_zdz_1+V_{\overline z}d\overline z_1$ satisfies the relations
\begin{equation}
y\partial_y V_z=i\partial_{z_1}Z,\qquad y\partial_y V_{\overline
z} =-i\partial_{\overline z_1}Z,\qquad 2iy(\partial_{\overline
z_1}V_z -\partial_{z_1}V_{\overline z})=\partial_y Z,
\label{eq:7vzrel}
\end{equation}
and
\begin{equation}
Z=\ft12\tanh G,\qquad h^{-2}=2y\cosh G.
\end{equation}
Here we have deliberately chosen to follow the LLM notation \cite{Lin:2004nb}
so as to avoid confusion with the corresponding quantities in the 1/8 BPS
system.  Furthermore, here we reserve $y$ to only refer to the $y$
coordinate of LLM, and not to the $y$ variable used in (\ref{eq:7met2}).
In particular, the 1/8 BPS metric will be taken in the form
\begin{equation}
ds_{10}^2=-e^{2\alpha}(dt+\omega)^2+e^{-2\alpha}h_{ij}dx^idx^j
+e^{2\alpha}d\Omega_3^2.
\label{eq:1/8eqn}
\end{equation}

By identifying the two three-spheres defined by $d\Omega_3$ in
(\ref{eq:llmeqn}) and (\ref{eq:1/8eqn}), we see that
\begin{equation}
e^{2\alpha}=ye^G.
\label{eq:e2aiden}
\end{equation}
The remaining seven-dimensional metric then has the form
\begin{equation}
ds_7^2=-(e^{2\alpha}+y^2e^{-2\alpha})(dt+V)^2+e^{-2\alpha}
(Z+\ft12)(dy^2+|dz_1|^2)+y^2e^{-2\alpha}d\widetilde\Omega_3^2.
\end{equation}
We again wish to shift the angular coordinates on $d\widetilde\Omega_3$.
This may be done by writing
\begin{equation}
d\widetilde\Omega_3^2=d\theta^2+\cos^2\theta d\phi_1^2+\sin^2\theta d\phi_2^2,
\end{equation}
and then shifting
\begin{equation}
\phi_1=\psi_1-t,\qquad\phi_2=\psi_2-t.
\end{equation}
Performing this shift and completing the square in $dt$ now yields
\begin{eqnarray}
ds_7^2&=&-e^{2\alpha}(dt+\omega)^2
+e^{-2\alpha}\biggl[\fft{y^2}{Z+\fft12}(V^2+2V(\cos^2\theta d\psi_1
+\sin^2\theta d\psi_2))\nonumber\\
&&+y^2\fft{1-2Z}{1+2Z}(\cos^2\theta d\psi_1+\sin^2\theta
d\psi_2)^2+y^2d\widetilde\Omega_3^2+(Z+\ft12)(dy^2+|dz_1|^2)\biggr],
\end{eqnarray}
where
\begin{equation}
\omega=\fft1{Z+\fft12}V+\fft{1-2Z}{1+2Z}(\cos^2\theta d\psi_1
+\sin^2\theta d\psi_2).
\end{equation}
As a result, the metric on the six-dimensional base can be read off
from the terms inside the square brackets above.

To show that this metric is K\"ahler, and to read off the K\"ahler
potential, we introduce complex coordinates
\begin{equation}
z_2=r\cos\theta e^{i\psi_1},\qquad
z_3=r\sin\theta e^{i\psi_2},
\end{equation}
so that
\begin{eqnarray}
|dz_i|^2&=&dr^2+r^2d\widetilde\Omega_3^2,\nonumber\\
|\overline z_idz_i|^2&=&r^2dr^2+r^4(\cos^2\theta d\psi_1+\sin^2\theta
d\psi_2)^2,\nonumber\\
\Im(\overline z_idz_i)&=&r^2(\cos^2\theta d\psi_1+\sin^2\theta d\psi_2),
\end{eqnarray}
where here $i=2,3$ only.  In this case, the metric on the six-dimensional base
becomes
\begin{eqnarray}
ds_6^2&=&(Z+\ft12)(dy^2+|dz_1|^2)-\fft{y^2}{Z+\fft12}\fft{dr^2}{r^2}
+\fft{y^2}{r^2}|dz_i|^2+\fft{y^2}{r^4}\fft{1-2Z}{1+2Z}|\overline z_idz_i|^2
\nonumber\\
&&+\fft{y^2}{Z+\fft12}(V_zdz_1+V_{\overline z}d\overline z_1)^2
+\fft{2y^2}{r^2(Z+\fft12)}(V_zdz_1+V_{\overline z}d\overline z_1)
\Im(\overline z_idz_i).
\label{eq:7met61}
\end{eqnarray}
Note that $r^2=|z_i|^2$.  Since the original LLM coordinate $y$ is
somehow out of place, we need to find a transformation relating $y$
with the complex coordinates $z_1,z_2,z_3$.  To obtain this transformation,
we take a hint from the $dr$ and $dy$ sector of the metric
\begin{equation}
ds_6^2=\fft{y^2}{Z+\fft12}\left((Z+\fft12)^2\fft{dy^2}{y^2}
-\fft{dr^2}{r^2}\right)+\cdots.
\end{equation}
This suggests that we take
\begin{equation}
r^2(z_1,\overline z_1,y)
=\exp\int^{y^2}\left(Z(z_1,\overline z_1,y')+\ft12\right)\fft{d(y'^2)}{y'^2},
\label{eq:7rydef}
\end{equation}
where we are somewhat sloppy about the limits of the indefinite integral.
Because of the $z_1,\overline z_1$ dependence on the right hand side,
this relation is somewhat subtle to manipulate.  For example
\begin{eqnarray}
\fft{dr}r&=&\left(\int^y\partial_{z_1}Z\fft{dy'}{y'}\right)dz_1
+\left(\int^y\partial_{\overline z_1}Z\fft{dy'}{y'}\right)d\overline z_1
+\fft{Z+\fft12}ydy\nonumber\\
&=&-i\left[\left(\int^y\partial_{y'}V_zdy'\right)dz_1
-\left(\int^y\partial_{y'}V_{\overline z}dy'\right)d\overline z_1
\right]+\fft{Z+\fft12}ydy\nonumber\\
&=&-i(V_zdz_1-V_{\overline z}d\overline z_1)
+\fft{Z+\fft12}ydy,
\end{eqnarray}
where we have used (\ref{eq:7vzrel}).  Here we assume that the integration
in (\ref{eq:7rydef}) may be defined so that this differential relation
holds.  Inserting this relation into (\ref{eq:7met61}) finally gives
the complex Hermitian metric
\begin{eqnarray}
ds_6^2&=&\left((Z+\ft12)+\fft{4y^2}{Z+\fft12}V_zV_{\overline z}\right)
|dz_1|^2+\fft{y^2}{r^2}\left(|dz_2|^2+|dz_3|^2\right)
+\fft{y^2}{r^4}\fft{1-2Z}{1+2Z}|\overline z_2dz_2+\overline z_3dz_3|^2
\nonumber\\
&&-\fft{4y^2}{r^2(Z+\fft12)}\Re\left(iV_{\overline z}(\overline z_2dz_2
+\overline z_3dz_3)d\overline z_1\right),
\label{eq:6dhbase}
\end{eqnarray}
where $r^2=|z_2|^2+|z_3|^2$, and $y$ is implicitly defined from
(\ref{eq:7rydef}).

In order to show that the above metric is K\"ahler, we may directly
obtain the
K\"ahler potential $K(z_1,\overline z_1,r^2)$ by integrating the
differential relations
\begin{eqnarray}
&&\partial_{r^2}K=\fft{y^2}{2r^2},\qquad\partial_{r^2}\partial_{r^2}K
=\fft{y^2}{2r^4}\fft{1-2Z}{1+2Z},\qquad
\partial_{z_1}\partial_{\overline z_1}K=\ft12(Z+\ft12)+\fft{2y^2}{Z+\fft12}
V_zV_{\overline z},\nonumber\\
&&\partial_{z_1}\partial_{r^2}K=-\fft{y^2}{2r^2(Z+\fft12)}iV_{\overline z},
\qquad\partial_{\overline z_1}\partial_{r^2}K=\fft{y^2}{2r^2(Z+\fft12)}iV_z.
\end{eqnarray}
The result is particularly simple
\begin{equation}
K(z_1,\overline z_1,y^2)=\ft12\int^{y^2}(Z(z_1,\overline z_1,y')+\ft12)d(y'^2).
\label{eq:7kpotint}
\end{equation}
Of course, $y^2$ has to be rewritten in terms of $z_1$, $\overline z_1$
and $r^2$ using (\ref{eq:7rydef}).  In order to verify that this is
correct, we need the chain rule expressions
\begin{eqnarray}
\partial_{r} f(z_1,\overline z_1,r)&=&\fft1{\partial r/\partial y}
\partial_{y} f(z_1,\overline z_1,y)
=\fft{y}{r(Z+\fft12)}\partial_y f(z_1,\overline z_1,y),\nonumber\\
\partial_{z_1} f(z_1,\overline z_1,r)&=&\left(\partial_{z_1}
-\fft{\partial r/\partial z_1}{\partial r/\partial y}\partial_y\right)
f(z_1,\overline z_1,y)=\left(\partial_{z_1}
+\fft{iyV_z}{Z+\fft12}\partial_y\right)f(z_1,\overline z_1,y),\quad
\end{eqnarray}
where $r=r(z_1,\overline z_1,y)$.

Linearity of the LLM Laplacian (\ref{eq:4llmlap}) allows a Green's function
solution for $Z$ of the form \cite{Lin:2004nb}
\begin{equation}
Z(z_1,\overline z_1,y)=\fft12-\fft{y^2}\pi\int_D\fft{dx_1'dx_2'}
{[|z_1-z_1'|^2+y^2]^2},
\label{eq:llmgreens}
\end{equation}
where the integral is only over the areas of the two-dimensional droplets
($Z=-1/2$) sitting in the $Z=1/2$ background.  This allows us to rewrite
(\ref{eq:7rydef}) as
\begin{equation}
\log(r^2)=\log(y^2)+\fft1\pi\int_D\fft{dx_1'dx_2'}
{|z_1-z_1'|^2+y^2},
\label{eq:7rygreens}
\end{equation}
at least up to an unimportant $y$-independent function arising from the
indefinite $y$ integral in (\ref{eq:7rydef}).
As $y$ approaches $0$, there are two cases to consider: i) $z_1
\in D$ and ii) $z_1 \notin D$. In the first case $r^2\big|_{y=0}$ is finite
and (\ref{eq:7rygreens}) defines a five-dimensional surface,
whereas in the latter $r^2=y^2 + {\cal O}(y^4)$.

In addition, substituting (\ref{eq:llmgreens}) into (\ref{eq:7kpotint})
while ensuring proper asymptotic behavior gives an expression for
the K\"ahler potential
\begin{equation}
K=\ft12y^2+\ft12|z_1|^2+\fft1{2\pi}\int_D\left(\fft{y^2}{|z_1-z_1'|^2+y^2}
-\log[|z_1-z_1'|^2+y^2]\right)dx_1'dx_2'.
\label{eq:7kgreens}
\end{equation}

\subsubsection{The LLM vacuum}

As a simple example, we may consider the AdS$_5\times S^5$ vacuum,
which is specified by a circular disk in the LLM plane.  Taking this
disk to have radius $L$, the Green's function integral (\ref{eq:llmgreens})
gives \cite{Lin:2004nb}
\begin{equation}
Z=\fft{|z_1|^2+y^2-L^2}{2\sqrt{(|z_1|^2+y^2-L^2)^2+4y^2L^2}}.
\label{eq:7adszfunc}
\end{equation}
Before working out the K\"ahler potential, we may use (\ref{eq:7rygreens})
to determine
\begin{equation}
r^2=\fft12\left(L^2+y^2-|z_1|^2+\sqrt{(|z_1|^2+y^2-L^2)^2+4y^2L^2}\right),
\label{eq:7rfy}
\end{equation}
which in turn may be inverted to yield
\begin{equation}
y^2=r^2\left(1-\fft{L^2}{r^2+|z_1|^2}\right).
\end{equation}
The $y=0$ surface reduces to $r^2+|z_1|^2=L^2$ for $|z_1|<L$ and to
$r=0$ for $|z_1|>L$, corresponding to the cases i) and ii) mentioned
in the previous section, after (\ref{eq:7rygreens}).

The K\"ahler potential itself is obtained from (\ref{eq:7kgreens}):
\begin{eqnarray}
K&=&\fft14\biggl[|z_1|^2+y^2+L^2+\sqrt{(|z_1|^2+y^2-L^2)^2+4y^2L^2}\nonumber\\
&&-2L^2\log\left(\fft12\left(|z_1|^2+y^2+L^2+\sqrt{(|z_1|^2+y^2-L^2)^2+4y^2L^2}
\right)\right)\biggr].
\label{eq:7kpotads}
\end{eqnarray}
Using (\ref{eq:7rfy}), this may be rewritten as
\begin{equation}
K=\ft12(|z_1|^2+|z_2|^2+|z_3|^2)-\ft12L^2\log(|z_1|^2+|z_2|^2+|z_3|^2),
\end{equation}
where we have used $r^2=|z_2|^2+|z_3|^2$.  This of course
recovers the symmetrical AdS$_5\times S^5$ K\"ahler potential
(\ref{eq:7ads5kpot}), but this time with the AdS radius $L$ restored.

\subsubsection{Multi-disk configurations}
\label{multi_disk}

Given the vacuum solution corresponding to a single LLM disk,
there is in fact a natural procedure for building up topologically
non-trivial configurations through linear superposition.  Suppose
we have $n$ disks, each with radius $b_i$, centered at the complex
position $a_i$ in the $z_1$ plane.  So long as the disks are
non-overlapping, the function $Z$ obtained by (\ref{eq:llmgreens})
has a superposition solution of the form
\begin{eqnarray}
Z&=&\fft12+\sum_{i=1}^{n}\left(\frac{|z_{1}-a_{i}|^{2}+y^{2}-b_{i}^{2}}
{2\sqrt{(|z_{1}-a_{i}|^{2}+y^{2}-b_{i}^{2})^{2}+4y^{2}b_{i}^{2}}}
-\fft12\right)\nonumber\\
&=&\fft{1-n}2+\sum_{i=1}^{n}\frac{|z_{1}-a_{i}|^{2}+y^{2}-b_{i}^{2}}
{2\sqrt{(|z_{1}-a_{i}|^{2}+y^{2}-b_{i}^{2})^{2}+4y^{2}b_{i}^{2}}}.
\end{eqnarray}
In addition, the form of the integral (\ref{eq:7rygreens}) relating $r^2$
with $y^2$ indicates that $r^2$ may be obtained by superposing $n$
individual terms, each of the form given by (\ref{eq:7rfy})
\begin{eqnarray}
r^2&=&y^2\prod_{i=1}^n\fft1{2y^2}
\left[b_{i}^{2}+y^{2}-|z_{1}-a_{i}|^{2}
+\sqrt{(|z_{1}-a_{i}|^{2}+y^{2}-b_{i}^{2})^{2}+4y^{2}b_{i}^{2}}\right]
\nonumber\\
&=&\fft{y^{2(1-n)}}{2^n}\prod_{i=1}^n\left[b_{i}^{2}+y^{2}-|z_{1}-a_{i}|^{2}
+\sqrt{(|z_{1}-a_{i}|^{2}+y^{2}-b_{i}^{2})^{2}+4y^{2}b_{i}^{2}}\right].
\label{r_yz}
\end{eqnarray}
Similarly, the K\"ahler potential may be obtained by superposing
individual terms of the form (\ref{eq:7kpotads})
\begin{eqnarray}
K&=&\ft12y^2+\ft12|z_1|^2+\sum_{i=1}^n\fft14\biggl[b_i^2-y^2-|z_1-a_i|^2
+\sqrt{(|z_1-a_i|^2+y^2-b_i^2)^2+4y^2b_i^2}\nonumber\\
&&\kern4em-2b_i^2\log\left(\fft12\left(b_i^2+y^2+|z_1-a_i|^2
+\sqrt{(|z_1-a_i|^2+y^2-b_i^2)^2+4y^2b_i^2}\right)\right)\biggr].\kern3em
\label{eq:kdisks}
\end{eqnarray}

In principle, (\ref{r_yz}) ought to be inverted to give $y^2$ as
a function of $z_1$, $\overline z_1$ and $r^2$.  In turn, this could then
be inserted into (\ref{eq:kdisks}) to obtain the final expression for the
K\"ahler potential.  Unfortunately, however, (\ref{r_yz}) is a rather
unwieldy function to invert.  Nevertheless, we can learn a fair bit about
the boundary conditions even without an explicit form of the K\"ahler
potential.

Our main interest is to examine the degeneration surface when $e^{2\alpha}
\to0$ ({\it i.e.}~when the $S^3$ inside AdS$_5$ shrinks).  From
(\ref{eq:e2aiden}), this requires that $y\rightarrow 0$ (along with
some possible requirement on $e^G$, which we are not so concerned about).
Recalling that $r^2=|z_2|^2+|z_3|^2$ in our notation, setting $y=0$
in (\ref{r_yz}) then defines a five-dimensional degeneration surface
through a real algebraic equation in $\mathbb C^3$.  Actually, because
of the $y^{2(1-n)}$ prefactor, some care must be taken before we can
let $y=0$ in (\ref{r_yz}).  To proceed, we may start with the small $y$
expansion of (\ref{r_yz}), and then subsequently take $y\rightarrow 0$.

Because of the square root expressions, this small $y$ expansion is
dependent on our location in the $z_{1}$ plane.  In particular, as
$y\to0$, we have
\begin{eqnarray}
&&\left[ b_{i}^{2}+y^{2}-|z_{1}-a_{i}|^{2}
+\sqrt{(|z_{1}-a_{i}|^{2}+y^{2}-b_{i}^{2})^{2}+4y^{2}b_{i}^{2}}\right]
\nonumber\\
&&\kern14em=\begin{cases}\displaystyle
\frac{|z_{1}-a_{i}|^{2}}{|z_{1}-a_{i}|^{2}-b_{i}^{2}}(2y^2)+\mathcal O(y^4),
&|z_1-a_i|>|b_i|;\\
2(b_{i}^{2}-|z_{1}-a_{i}|^{2})+\mathcal O(y^{2}),
&|z_1-a_i|<|b_i|.
\end{cases}\qquad\qquad
\label{eq:yzlimits}
\end{eqnarray}
The first case corresponds to $z_1$ outside the $i$-th disk, and the
second to $z_1$ inside.  Because of the non-overlapping condition,
$z_1$ can fall inside a single disk, at most.  Suppose we look at the
region inside the $j$-th disk.  In this case, the expression for
$r^2$ in (\ref{r_yz}) receives $n-1$ contributions of the first
type (when $i\ne j$), and a single contribution of the second type.
This combination of expansions introduces a $y^{2(n-1)}$ factor in
the product, canceling the $y^{2(1-n)}$ factor in (\ref{r_yz}).
So the result for this $j$-th region is
\begin{equation}
r^2\equiv|z_2|^2+|z_3|^2=\left(b_{j}^{2}-|z_{1}-a_{j}|^{2}\right)
\prod_{i\neq j}\frac{|z_{1}-a_{i}|^{2}}{|z_{1}-a_{i}|^{2}-b_{i}^{2}}.
\label{eq:sufeqn}
\end{equation}
Note that this equation is exact, even though we had to expand in $y=0$
in order to obtain it.

We recall that this equation defines a five-dimensional surface inside
$\mathbb C^3$ where the $S^3$ inside AdS$_5$ shrinks to a point.  To
understand the implication of this equation better, we may consider the
single-disk limit, when the other $n-1$ finite disks are very
far away from the $j$th disk.  In this case, $|z_{1}-a_{i}|\gg ~b_{i}$
for $i\ne j$, and we get the simplified expression
\begin{equation}
|z_2|^2+|z_3|^2=b_{j}^{2}-|z_{1}-a_{j}|^{2}.
\end{equation}
This describes a round five-sphere centered at $z_{1}=a_{j}$, with radius
$b_{j}$.  When the disks are not so well separated, the additional factors in
(\ref{eq:sufeqn}) lead to a distortion of the five-sphere.  Nevertheless,
the picture that emerges is clear.  The interior of each LLM disk gets
mapped into a (possibly distorted) five-sphere degeneration surface inside
$\mathbb C^3$.  Equation (\ref{eq:sufeqn}) simply describes the $j$-th
disconnected component of the complete five-dimensional degeneration surface.

We have now shown that non-trivial LLM topology has a natural generalization
in the 1/8 BPS system.  In particular, individual LLM droplets (with disk
topology) map directly into degeneration surfaces which are topologically
five-spheres, and which may be considered as canonical 1/8 BPS droplets.
Since the interior of each droplet is not present, the 1/8 BPS system
can be described using a set of coordinates spanning $\mathbb C^3$, but
with various regions removed.  In the LLM picture, a large disk surrounded
by small droplets corresponds to a collection of dual giant gravitons,
all expanding in AdS$_5$ \cite{Lin:2004nb}.  Each droplet modifies the
topology, and may be considered as a backreacted version of a giant
graviton.  In the general 1/8 BPS description, this has a corresponding
picture as a large spherical void at the center of $\mathbb C^3$ surrounded
by a set of five-sphere `bubbles', each bubble being one of the dual giant
gravitons.

Given this understanding of dual giants in the 1/8 BPS context, there is
still one remaining question, and that is how giant gravitons expanding
on $S^5$ fit in  the above framework.  In terms of the LLM picture,
turning on these giant gravitons corresponds to introducing holes in
the AdS disk itself.  Before we consider the effect of holes, however,
we first consider the $y\to0$ behavior of (\ref{r_yz}) in the case that
$z_1$ lies outside all of the disks.  In this case, all $n$ expressions
in (\ref{r_yz}) are of the form of the top line in (\ref{eq:yzlimits}),
and we thus end up with
\begin{equation}
r^2\approx y^2\prod_{i=1}^{n}\frac{|z_1-a_i|^2}{|z_1-a_i|^2-b_i^2},
\end{equation}
as $y\to0$.  The extra $y^2$ factor then ensures that $r\to0$ as $y\to0$,
so long as $z_1$ lies outside the disks.  Recalling that $r^2=|z_2|^2+|z_3|^2$,
this limit corresponding to shrinking $S^3$ inside $S^5$, which of course
agrees with the 1/2 BPS bubbling picture of \cite{Lin:2004nb}.

%%%%%%%%%%%%%%%%%%%%%%%%%%%%%%%%%%%%%%%%
\begin{figure}[t]
\begin{center}
\includegraphics{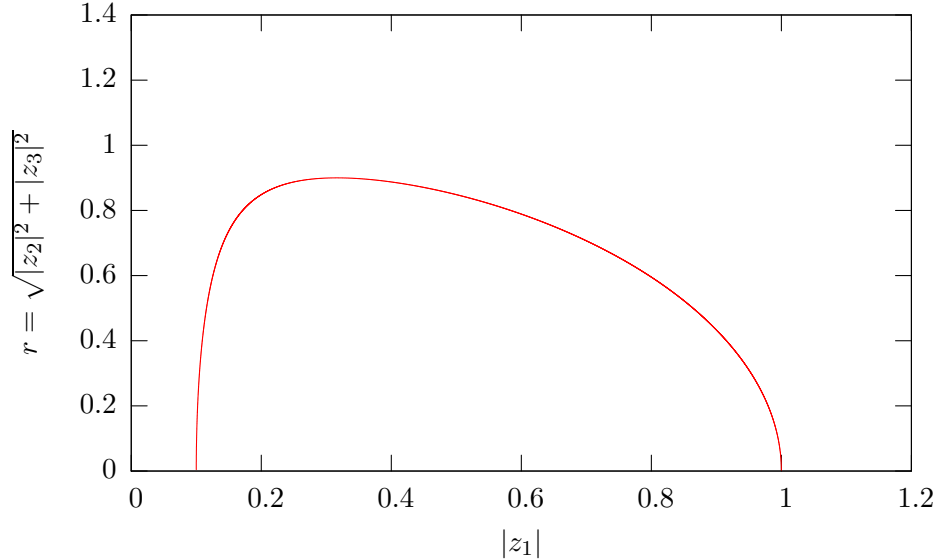}%
\end{center}
\caption{Profile of $r$ versus $|z_1|$ for the configuration corresponding
to a single hole of radius $0.1$ centered at the origin of the AdS disk (of
unit radius).  This picture corresponds to a maximal giant graviton
expanding on $S^5$.}
\label{fig:onegg}
\end{figure}
%%%%%%%%%%%%%%%%%%%%%%%%%%%%%%%%%%%%%%%%

We are now in a position to consider adding holes (giant gravitons expanding
in $S^5$) to the above multi-disk configuration.  Again, because of
linear superposition, we may consider holes as simply `negative' regions
inside a disk (provided, of course, that they are entirely contained within
the corresponding disk).  In this case, for $n$ disks as above along with
$m$ circular holes (each with radius $\tilde b_i$ and centered at
$\tilde a_i$), the generalization of (\ref{r_yz}) is simply
\begin{eqnarray}
r^2&=&y^2\prod_{i=1}^n\fft1{2y^2}
\left[b_{i}^{2}+y^{2}-|z_{1}-a_{i}|^{2}
+\sqrt{(|z_{1}-a_{i}|^{2}+y^{2}-b_{i}^{2})^{2}+4y^{2}b_{i}^{2}}\right]
\nonumber\\
&&\times\prod_{i=1}^m2y^2\left[\tilde b_i^2+y^2-|z_1-\tilde a_i|^2
+\sqrt{(|z_1-\tilde a_i|^2+y^2-\tilde b_i^2)^2+4y^2\tilde b_i^2}\right]^{-1}.
\label{eq:llmbubhole}
\end{eqnarray}
For a single hole inside the AdS disk, the degeneration surface can be
obtained by taking the $y\to0$ limit of this expression for the case
where $z_1$ lies in the disk, but not the hole.  The resulting surface
is described by
\begin{equation}
r^2\equiv|z_2|^2+|z_3|^2=\fft{(L^2-|z_1|^2)(|z_1-\tilde a|^2-\tilde b^2)}
{|z_1-\tilde a|^2},
\label{eq:ggsurf}
\end{equation}
where we have taken the AdS disk to be centered at the origin and to have
radius $L$.  The hole is centered at $\tilde a$, and has radius $\tilde b$.
This describes a five-dimensional surface of topology $S^4\times S^1$,
which was in fact already noticed in \cite{Lin:2004nb} when fibering
$\tilde S^3$ over an annulus in the LLM plane.  As an example, we plot the
profile of the surface given by (\ref{eq:ggsurf}) in Fig.~\ref{fig:onegg}.

%%%%%%%%%%%%%%%%%%%%%%%%%%%%%%%%%%%%%%%%
\begin{figure}[t]
\begin{center}
\includegraphics[width=8cm]{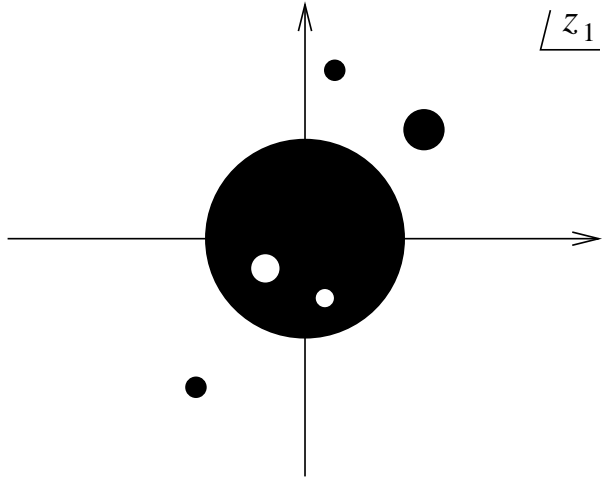}
\end{center}
\caption{An LLM configuration with three droplets and two holes.}
\label{fig:llmbub}
\end{figure}
%%%%%%%%%%%%%%%%%%%%%%%%%%%%%%%%%%%%%%%%

On the gauge theory side of the duality, the picture shown in
Fig.~\ref{fig:onegg} presumably corresponds to the numerical eigenvalue
distribution studied recently in \cite{Berenstein:2007wz} for the one
hole state.  We note that, at least in this coordinate system, the change
of $r$ is very steep near the central hole of the giant graviton.  This
may account for the failure of the numerical eigenvalue distribution to
close on this hole observed in \cite{Berenstein:2007wz}.  However, it
remains to be seen whether or not the present coordinate system is in
fact the one which is preferred when matching to the eigenvalue distribution.

%%%%%%%%%%%%%%%%%%%%%%%%%%%%%%%%%%%%%%%%
\begin{figure}[t]
\begin{center}
\includegraphics[width=12cm]{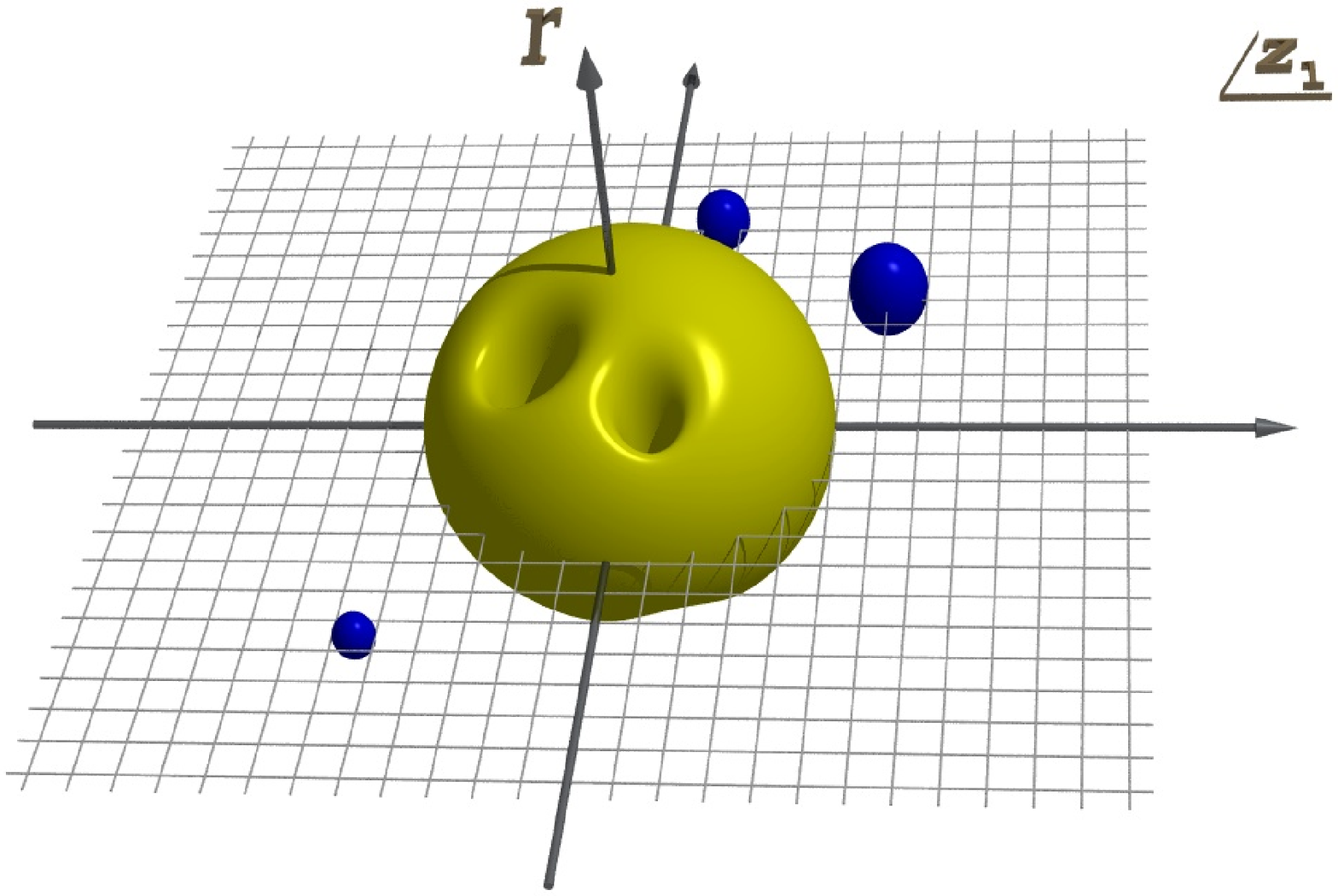}
\end{center}
\caption{The LLM configuration of Fig.~\ref{fig:llmbub} shown as droplets
in the six-dimensional base given by (\ref{eq:6dhbase}).  Here $r^2=|z_2|^2
+|z_3|^2$, and the additional $S^3$ directions are suppressed.  Note that
the physical space is comprised of the region outside of the droplets only.}
\label{fig:llm3dbub}
\end{figure}
%%%%%%%%%%%%%%%%%%%%%%%%%%%%%%%%%%%%%%%%

As more holes are introduced into the AdS disk, more and more non-trivial
topology is generated; adding $m$ holes gives rise to a corresponding
five-dimensional surface which may be described as $S^3$ fibered over the
disk with $m$ holes.  Thus the
five-dimensional boundary surface has a very physical interpretation as the
distortion of the original five-sphere of the AdS$_5\times S^5$ background.
A complete 1/2 BPS bubbling geometry, with both giant gravitons and dual
giants, thus involves an AdS disk along with both particle and hole
excitations.  The holes in the AdS disk change the topology of the
original five-sphere, while the particles outside the disk give rise to
additional degeneration surfaces.  Consider, for example, the LLM
geometry specified in Fig.~\ref{fig:llmbub}, corresponding to the excitation
of two giant gravitons and three dual giants.  When written in the 1/8 BPS
framework, the resulting degeneration surfaces, as given by
(\ref{eq:llmbubhole}), take on the form shown in Fig.~\ref{fig:llm3dbub}.
More complicated geometries, corresponding to non-circular droplets, are of
course possible.  However, for 1/2 BPS states, the boundary surfaces always
contain an additional unbroken $S^3$ isometry related to the angular
directions not indicated in Fig~\ref{fig:llm3dbub}.  This isometry would
not be present for more generic 1/4 and 1/8 BPS bubbles.  Nevertheless, even
in such cases, the overall picture of droplets as removed volumes of
$\mathbb R^6$ remains valid.

%%%%%%%%%%%%%%%%%%%%%%%%%%%%%%%%%%%%%%%%
\section{Examples fitting into the 1/4 BPS case}
\label{QuarterBPSsection}
%%%%%%%%%%%%%%%%%%%%%%%%%%%%%%%%%%%%%%%%

After having studied the general 1/8 BPS case, we now turn to explicit
solutions for the case of 1/4 BPS configurations.  These backgrounds
have an additional $S^1$ isometry compared with the generic 1/8 BPS
backgrounds, and have a ten-dimensional metric of the form
\bea
\label{GenericQuarterBPS}
&&ds_{10}^2= -h^{-2}(dt+\omega)^2+h^2((Z+\ft12)^{-1}2h_{i\bar j}
dz^i d\bar z^{\bar j}+dy^2)+y(e^G d\Omega_3^2+e^{-G}(d\psi+{\mathcal A})^2),
\nonumber\\
&&h^{-2}=2y\cosh G, \qquad h_{i\bar j}=\partial_i\partial_{\bar j}K.
\eea
We have also defined
\beq
Z\equiv\frac{1}{2}{\rm
\tanh}G=-\frac{1}{2}y\partial_y\frac{1}{y}\partial_y K \,,
\label{ZKrel}
\eeq
which is the 1/4 BPS version of the LLM function.  The four-dimensional
base metric $h_{i\bar j}$ is K\"ahler, and is further constrained by
a Monge-Amp\`ere type equation (\ref{eq:sol_MA}), along with auxiliary
condition (\ref{eq:Deqn2})
\begin{eqnarray}
&&\log\det h_{i\bar j}=\log(Z+\frac 12)+n\eta \log y+\frac 1{y}(2-n\eta)
\partial_y K + D(z_i,\bar z_{\bar j}),\nonumber\\
&&(1+*_4)\partial\bar\partial D= \frac 4{y^2}(1-n\eta)
\partial\bar\partial K.
\label{eq:MA2}
\end{eqnarray}
Since we are mainly
interested in the form of the K\"ahler metric on the base, we do not
repeat here the expressions for the two-forms $\mathcal F=d\mathcal A$
and $d\omega$, nor for the IIB self-dual five-form.  These expressions,
along with details of the analysis, may be found above in
Sections~\ref{sec:d=6} and \ref{sec:d=6susy}.  We do note, however, that
the two-form $\mathcal F$ must satisfy the additional constraint given
in (\ref{eq:6fwf2}).

Since the construction of arbitrary new backgrounds by solving
the Monge-Amp\`ere equation (\ref{eq:MA2}) is a rather challenging
task, we instead look at several classes of existing solutions and
see how they may be transformed into the 1/4 BPS form
(\ref{GenericQuarterBPS}).  In this way, we are able to deduce the
generic 1/4 BPS bubbling picture without having to turn directly to
the construction of explicit solutions.

Note, however, that (\ref{eq:MA2}) becomes much simpler to analyze in
certain special cases, such as when the complex two-dimensional base
decomposes into a direct product of two Riemann surfaces.  We will
study this case at the end of this section and show its connection to
the embedding of the LLM solution into the gauged ansatz.

%%%%%%%%%%%%%%%%%%%%%%%%%%%%%%%%%%%%%%%%
\subsection{AdS$_5\times S^5$}

Before expanding on the 1/4 BPS droplet picture, we start with the
embedding of the AdS$_5\times S^5$ ground state into the framework
given by (\ref{GenericQuarterBPS}).  We will then move on to more
complicated geometries.

As in Section~\ref{ads5xs5}, we take global AdS$_5\times S^5$ written as:
\begin{eqnarray}
ds^2&=&-\cosh^2\rho \, dt^2+d\rho^2+\sinh^2\rho \,
d\Omega_3^2+d\Omega_5^2\nonumber\\
&=&-\cosh^2\rho \, dt^2+d\rho^2+\sinh^2\rho \,
d\Omega_3^2\nonumber\\
&&\qquad+\sin^2\theta d\psi^2
+d\theta^2+\cos^2\theta[\cos^2\alpha
d\phi_1^2+d\alpha^2+\sin^2\alpha d\phi_2^2],
\label{eq:6ads}
\end{eqnarray}
where in the second line we have chosen an explicit parameterization of
the five-sphere metric.  In order to embed this into the 1/4 BPS system
of (\ref{GenericQuarterBPS}), we must identify the appropriate $S^3
\times S^1$ isometry for the embedding.  While the $S^3$ factor inside
AdS$_5$ is obvious, there are several possible choices for the circle
factor inside $S^5$.  By writing the five-sphere metric as above, we
have chosen to follow the ungauged 1/4 BPS ansatz, where we set
$\mathcal A=0$ from the start. Then, after comparing with
(\ref{GenericQuarterBPS}), we choose to identify the time
coordinate $t$, as well as the $S^3\times S^1$ factors
$d\Omega_3$ and $d\psi$.  (Another possibility, which we do not
pursue, would be to write $S^5$ as $U(1)$ bundled over $CP^2$ as
in (\ref{eq:7u1bundle}), and then to follow the gauged 1/4 BPS ansatz.)

The above identification allows us to deduce
\begin{eqnarray}
&&ye^G=\sinh^2\rho,\kern5.4em ye^{-G}=\sin^2\theta, \nonumber\\
%\Rightarrow
 && h^{-2}=\sinh^2\rho+\sin^2\theta, \qquad
y=\sinh\rho\sin\theta.
\label{eq:f12f22}
\end{eqnarray}
Thus the $y$ coordinate is easily given in terms of the original global
AdS$_5\times S^5$ variables.  In fact, these expressions are identical
to their 1/2 BPS LLM counterparts.  This suggests that we simply use the
LLM coordinate transformation
\begin{equation}
y=\sinh\rho\sin\theta,\qquad r=\cosh\rho\cos\theta,
\label{eq:ydefn}
\end{equation}
to map between $(\rho,\theta)$ and $(r,y)$ coordinates.  In particular,
this yields
\begin{equation}
dy^2+dr^2=h^{-2}(d\rho^2+d\theta^2).
\end{equation}
For the remaining coordinates, we note, just as in the 1/8 BPS case of
(\ref{eq:7shiftpsi}), that the azimuthal angles $\psi_1$ and $\psi_2$ need
to be shifted
\begin{equation}
\phi_1=\psi_1-t,\qquad\phi_2=\psi_2-t.
\end{equation}
After completing the square in $dt$, and comparing with
(\ref{GenericQuarterBPS}), we now obtain the one-form
\begin{equation}
\omega=h^2\cos^2\theta(\cos^2\alpha d\psi_1+\sin^2\alpha d\psi_2),
\end{equation}
as well as the four-dimensional K\"ahler metric
\begin{eqnarray}
ds_4^2&=&h^2\sinh^2\rho
\bigl[dr^2+h^{-2}\cos^2\theta(\cos^2\alpha d\psi_1^2+d\alpha^2
+\sin^2\alpha d\psi_2^2)\nonumber\\
&&+\cos^4\theta(\cos^2\alpha d\psi_1+\sin^2\alpha d\psi_2)^2\bigr]\nonumber\\
&\equiv& A \,dr^2+B \, d\Omega_3^2+C (\cos^2\alpha d\psi_1+\sin^2\alpha
d\psi_2)^2\, .
\label{eq:kahler}
\end{eqnarray}

In order to demonstrate that (\ref{eq:kahler}) is indeed K\"ahler,
we identify the K\"ahler potential $K$. To do so, we
first write the metric entirely in terms of the
coordinates $(r,\alpha,\psi_1,\psi_2)$. This may be done by inverting
(\ref{eq:ydefn}) to obtain
\begin{eqnarray}
\sinh^2\rho&=&
\ft12(r^2+y^2-1)+\sqrt{\ft14(r^2+y^2-1)^2+y^2},\nonumber\\
\sin^2\theta &=& -\ft12(r^2+y^2-1)+\sqrt{\ft14(r^2+y^2-1)^2+y^2},
\end{eqnarray}
which gives us expressions for $A$, $B$ and $C$ in terms of
$(r,y)$ only.  We now introduce complex coordinates $z_1$, $z_2$ and,
based on symmetry, assume that the K\"ahler potential is only a function of
$r^2=|z_1|^2+|z_2|^2$ and $y$, namely $K=K(r^2,y)$. We then find that the
metric takes the form
\begin{equation}
ds^2=2K'(|dz_1|^2+|dz_2|^2)+2K''|\overline z_1dz_1+\overline z_2dz_2|^2,
\label{eq:rsqkahler}
\end{equation}
where primes are derivatives with respect to $r^2$.  To make
contact with (\ref{eq:kahler}), we choose a parameterization of
$z_1$ and $z_2$ as
\begin{equation}
z_1=r\cos\alpha e^{i\psi_1},\qquad z_2=r\sin\alpha e^{i\psi_2}.
\end{equation}
Using
\begin{eqnarray}
|dz_1|^2+|dz_2|^2&=&dr^2+r^2d\Omega_3^2,\nonumber\\
|\overline z_1dz_1+\overline z_2dz_2|^2&=&r^2dr^2+r^4(\cos^2\alpha d\psi_1
+\sin^2\alpha d\psi_2)^2,
\end{eqnarray}
the K\"ahler metric (\ref{eq:rsqkahler}) becomes
\begin{equation}
ds^2=2(K'+r^2K'')dr^2+2r^2K'd\Omega_3^2+2r^4K''(\cos^2\alpha d\psi_1
+\sin^2\alpha d\psi_2)^2.
\label{eq:fkahler}
\end{equation}
Comparing (\ref{eq:fkahler}) with (\ref{eq:kahler}) gives the
identifications
\begin{equation}
K'+r^2K''=\ft12A,\qquad r^2K'=\ft12B,\qquad
r^4K''=\ft12C.
\label{eq:abcsys}
\end{equation}
Notice that this system is overdetermined, since the function
$K(r^2)$ is determined by three equations.  However, we
may verify that $B+C=r^2A$ and $A=B'$.  As a result, the three
equations are redundant, and we are left with only $K'=B/2r^2$,
which may be integrated to give the K\"ahler potential
\begin{equation}
K(r^2,y)=\fft12\int^{r^2}\fft{B(r^2,y)}{r^2}d(r^2) .
\end{equation}
Although it is not particularly illuminating, we can perform the integral
explicitly. Using the expression for $B$,
\beq
B=\ft12(r^2-y^2-1)+\sqrt{\ft14(r^2+y^2-1)^2+y^2} ,
\eeq
we find that the K\"ahler potential is
\begin{eqnarray}
K&=&\ft12\left(\ft12(r^2+y^2+1)+\sqrt{\ft14(r^2+y^2-1)^2+y^2}\right)\nonumber\\
&&-\ft12\log\left(\ft12(r^2+y^2+1)+\sqrt{\ft14(r^2+y^2-1)^2+y^2}\right)
\nonumber\\
&&-\ft12y^2\log\left(\ft12(-r^2+y^2+1)+\sqrt{\ft14(r^2+y^2-1)^2+y^2}\right)
+\ft12 y^2\log(y).
\label{eq:kpot}
\end{eqnarray}
The final function of $y$ ensures that $K$ satisfies the relation (\ref{ZKrel}).

\subsubsection{Boundary conditions}

In analogy with the $1/8$ BPS embedding of $AdS_5 \times S^5$ as well
as the 1/2 BPS LLM embedding, we expect to find that boundary conditions
at $y=0$ will give us a spherical surface. To make this apparent, we start
by pointing out that our complex coordinates are such that
\begin{equation}
|z_1|^2+|z_2|^2=r^2.
\end{equation}
The coordinate $y=\sinh\rho \, \sin\theta$ vanishes in two cases,
either when $\rho=0$ or $\theta=0$. The $\rho=0$ case,
corresponding to the $S^3$ shrinking to zero size, tells us from
(\ref{eq:ydefn}) that $r\leq 1$. In turn, this translates into the
interior of a spherical (unit radius) droplet:
\begin{equation}
|z_1|^2+|z_2|^2 \leq 1.
\end{equation}
On the other hand, the $\theta=0$ limit, which describes collapse
of the $S^1$, corresponds to the outside of the spherical droplet,
\begin{equation}
|z_1|^2+|z_2|^2 \geq 1.
\end{equation}
Thus the two regions are separated by a three-dimensional sphere
of unit radius%
\footnote{The reason for the unit radius is that we have taken
the AdS$_5$ radius to be one.}.
This may be viewed as a higher-dimensional realization
of the unit LLM circle, which describes the 1/2 BPS embedding of AdS$_5
\times S^5$, as well as a lower-dimensional realization of the five-sphere
which describes the 1/8 BPS embedding.

%%%%%%%%%%%%%%%%%%%%%%%%%%%%%%%%%%%%%%%%
\subsection{Two-charge smooth solutions}
\label{sec:2Q}

Starting from the round three-sphere, which describes the AdS$_5\times S^5$
ground state, we now move on to less trivial backgrounds.  In particular,
we now turn to the case of the smooth, two-charge ($1/4$ BPS) solutions which
can be obtained from the more general three-charge case
(\ref{3Qmetric}) by setting one of the charges to zero.  To be specific,
we choose to set $H_1=1$ and the corresponding scalar field $\varphi_1=0$ in
(\ref{3Qmetric}).

Using the explicit expressions (\ref{mus}) for $\mu_1$ and $\mu_2$, the
metric takes the form
\bea
ds^2_{10}&=& -\frac{f\sqrt{\Delta}}{(H_2 H_3)^{2/3}}
\, dt^2+ \sqrt{\Delta}(H_2 H_3)^{1/3} \, f^{-1}\, dr^2 +
r^2 \, \sqrt{\Delta}(H_2 H_3)^{1/3}\, d\Omega_3^2\nn \\
&&+ \frac{1}{\sqrt{\Delta}(H_2 H_3)^{1/3}}\, \Bigl[\cos^2\theta d\theta^2 + \sin^2\theta(d\phi_1-dt)^2\Bigr]\nn \\
&&+ \frac{1}{\sqrt{\Delta}} \Biggl\{ \frac{H_2^{2/3}}{H_3^{1/3}}
\Bigl[ e^{\varphi_2}\Bigl(d\mu_3-\,H_2^{-1}\,\mu_4 \, dt\Bigr)^2
+ e^{-\varphi_2}\Bigl( d\mu_4+\,H_2^{-1}\,\mu_3 \, dt \Bigr)^2\Bigr]\nn \\
&&\kern3em+\frac{H_3^{2/3}}{H_2^{1/3}} \Bigl[e^{\varphi_3}
\Bigl(d\mu_5-\,H_3^{-1}\,\mu_6 \, dt\Bigr)^2
+ e^{-\varphi_3}\Bigl(d\mu_6+\,H_3^{-1}\,\mu_5 \, dt\Bigr)^2
\Bigr]\Biggr\}.
\label{eq:6met2ch}
\eea
If we let $d\psi=d\phi_1-dt$, we can think of $\psi$ as parameterizing
the $S^1$ direction of the 1/4 BPS ansatz.  In particular, this suggests
the ungauged ansatz, as $d\psi$ is trivially fibered over the remaining
directions of the metric (\ref{eq:6met2ch}).  However, for convenience
in subsequent manipulations, we will formally allow $\mathcal A\ne0$
for the moment. Along with $S^1$, the $S^3$ is also clearly visible,
which brings us to the following identifications:
\beq
y \, e^G = r^2 (H_2 H_3)^{1/3} \sqrt{\Delta}, \qquad
y \, e^{-G} =
\frac{\sin^2\theta}{\sqrt{\Delta}(H_2 H_3)^{1/3}},
\label{S13radius}
\eeq
with
\bea
\Delta &=& (H_2 H_3)^{1/3}\sin^2\theta +
\frac{H_3^{1/3}}{H_2^{2/3}}\cos^2\theta
\sin^2\alpha(e^{-\varphi_2}\sin^2\phi_2+e^{\varphi_2}\cos^2\phi_2)\nn\\
&&+ \frac{H_2^{1/3}}{H_3^{2/3}}\cos^2\theta
\cos^2\alpha(e^{-\varphi_3}\sin^2\phi_3+e^{\varphi_3}\cos^2\phi_3).
\label{Delta}
\eea
Thus, we find
\bea
&&y=r\,\sin\theta , \qquad
e^G=\frac{\sqrt{\Delta}\,(H_2 H_3)^{1/3}\,r}{\sin\theta} , \nn \\
&&h^{-2}=y \, e^G + y \, e^{-G} = \sqrt{\Delta}(H_2 H_3)^{1/3}
\left(r^2+\frac{\sin^2\theta}{\Delta (H_2 H_3)^{2/3}}\right) .
\label{y2Q}
\eea
We will come back to these relations when we discuss
boundary conditions.

To show that this solution fits into the $1/4$ BPS ansatz
(\ref{GenericQuarterBPS}), we could of course try to embed it
directly, by first identifying the four-dimensional base, and
expressing it in terms of complex coordinates. However, for the
case of non-vanishing scalar fields $\varphi_i$, this calculation
turns out to be particularly cumbersome. We will instead make
use of the $1/8$ BPS embedding of the three-charge solution given
in Section~\ref{sec:3Q}, and require that the solution has an
additional $U(1)$ isometry. Note that this is the same strategy
that was employed in the general $1/4$ BPS discussion of
Section~\ref{sec:d=6susy}.
%This more indirect way to obtain the embedding
%will turn out to shed some light on the issue of boundary
%conditions in the $1/4$ BPS case.

To impose an additional $U(1)$, we take the K\"{a}hler
potential of the $1/8$ BPS solution to be of the form
\beq
K=K(|z_1|^2, z_i,\zb_i), \qquad i=2,3,
\eeq
where $z_1=\tilde{r}\, e^{i\psi}$. This clearly corresponds to setting
one of the scalar fields to zero, $\varphi_1=0$ (and also $H_1=1$). The
six-dimensional base of the $1/8$ BPS ansatz (\ref{Gen1/8Met})
then becomes
\bea
ds_6^2 = h_{mn}dx^m \, dx^n &=& 2\Bigl[\partial_i
\bar{\partial}_j K - \frac{\tilde{r}^2}{(\tilde{r}^2
K')'}\partial_i K' \, \bar{\partial}_j K' \Bigr]dz_id\zb_j
+\frac{d(\tilde{r}^2 K')^2}{2\tilde{r}^2(\tilde{r}^2 K')'} \nn \\
&&+ 2\tilde{r}^2(\tilde{r}^2 K')'\Bigl[d\psi
+\frac{\Im(\partial_iK' dz_i)}{(\tilde{r}^2 K')'}\Bigr]^2 ,
\eea
where now a prime denotes derivatives with respect to $|z_1|^2=\tilde{r}^2$.
Next, we would like to make the somewhat natural identification
\beq
y^2=2\tilde{r}^2 K',
\label{yK}
\eeq
which allows us to rewrite the base as
\beq
ds_6^2 = 2\Bigl[\partial_i \bar{\partial}_j K
- \frac{\tilde{r}^2}{(\tilde{r}^2 K')'}\partial_i K' \,
\bar{\partial}_j K' \Bigr]dz_id\zb_j +\frac{K'}{(\tilde{r}^2K')'}dy^2
+2\tilde{r}^2(\tilde{r}^2 K')'
\Bigl[d\psi+\frac{\Im(\partial_i K' dz_i)}{(\tilde{r}^2K')'}\Bigr]^2.
\eeq
The ten-dimensional metric then becomes
\bea
ds^2_{10}&=& e^{2\alpha} d\Omega_3^2 + 2e^{-2\alpha}\Bigl[\partial_i \bar{\partial}_j K - \frac{\tilde{r}^2}{(\tilde{r}^2 K')'}
\partial_i K' \, \bar{\partial}_j K' \Bigr]dz_id\zb_j + e^{-2\alpha}\frac{K'}{(\tilde{r}^2 K')'}dy^2 \nn \\
&&+2e^{-2\alpha}\tilde{r}^2(\tilde{r}^2K')'\Bigl[d\psi
+\frac{\Im(\partial_i K' dz_i)}{(\tilde{r}^2K')'}\Bigr]^2
-e^{2\alpha} (dt+\omega_{1/8})^2,
\eea
where we are adopting the notation $\omega_{1/8}$ for the three-charge
($1/8$ BPS) solution, so as to avoid confusion with the $1/4$ BPS $\omega$.
Clearly, the condition (\ref{yK}), if general, might shed
some light on the meaning of the $y$ coordinate inside of the
$1/8$ BPS ansatz. Specifically, it is natural to ask whether
$K'=0$ plays a crucial role in determining boundary conditions on
the $y=0$ plane.

A first check of whether we have identified the $y$ coordinate
correctly is to show that the $g_{yy}$ component of the metric
takes the expected form, $h^2$. To do so, we will use the explicit
relations for the K\"{a}hler potential of the three-charge
solution. Recall that, in the notation of Section~\ref{sec:3Q}, we
had $z_1=\rho_1(r^2)\, \sin\theta \, e^{i\phi_1}$.  Setting $\phi_1=\psi$,
and using (\ref{KDEQ14}), we then find that
\bea
K' &=& \partial_{z_1}\partial_{\zb_1}K=\frac{r^2}{2\rho_1^2}=
\frac{r^2 \sin^2\theta}{2\tilde{r}^2}, \nn\\
(\tilde{r}^2 K')'&=& \frac{\sin^2\theta}{2\tilde{r}^2}
\frac{h^{2}}{\sqrt{\Delta} (H_2H_3)^{1/3}},
\eea
which allows us to show that
\beq
g_{yy}= e^{-2\alpha}\frac{K'}{(\tilde{r}^2 K')'} =h^2.
\eeq
Using $e^{2\alpha}=ye^G$ and $e^{-2\alpha}\,\tilde{r}^2(2\tilde{r}^2 K')'
=h^{-2}e^{-2G}$, we find that the ten-dimensional metric becomes
\beq
ds^2_{10}= ye^G d\Omega_3^2 + \frac{1}{ye^G}ds_4^2 + h^2 \, dy^2+
h^{-2}e^{-2G}\Bigl[d\psi+\frac{\Im(\partial_i K'
dz_i)}{(\tilde{r}^2 K')'}\Bigr]^2- e^{2\alpha} (dt+\omega_{1/8})^2,
\eeq
where we have defined
\beq ds_4^2 =
2\Bigl[\partial_i \bar{\partial}_j K -
\frac{\tilde{r}^2}{(\tilde{r}^2 K')'}
\partial_i K' \, \bar{\partial}_j K' \Bigr] dz_{i} d {\bar z}_{j}.
\label{4dbase}
\eeq
Notice that the $g_{tt}$ component is still
not in the appropriate $1/4$ BPS form. To obtain the correct
form of $g_{tt}$, it is enough to let $\psi=\tilde{\psi}-t$.
After using the following decomposition,
\beq
\omega_{1/8}=\omega_\psi\, d\psi + \tilde{\omega},
\eeq
we shift the angle and find
\bea
ds^2_{10}&=&  ye^G d\Omega_3^2 + \frac{1}{ye^G}ds_4^2 + h^2 \, dy^2 \nn \\
&&+ h^{-2} e^{-2G} \Bigl[d\psit - dt + \frac{\Im(\partial_i K'
dz_i)}{(\tilde{r}^2 K')'}\Bigr]^2- e^{2\alpha}
(dt(1-\omega_\psi)+\omega_\psi d\psit +\tilde{\omega})^2.
\eea
Furthermore, using (\ref{omegaphi}), we find that $\omega_\psi=
-e^{-2G}$. It is then easy to show that the $g_{tt}$ and $g_{\psit\psit}$
terms take the expected form:
\bea
g_{tt} &=& h^{-2} e^{-2G}-y e^G (1-\omega_\psi)^2
=-h^{-2}\, , \nn \\
g_{\psit \psit} &=& h^{-2} e^{-2G} -
e^{2\alpha}\omega_\psi^2=ye^{-G},
\eea
so that the metric becomes
\bea
ds^2_{10}&=&  ye^G d\Omega_3^2 + \frac{1}{ye^G}ds_4^2 + h^2 \,
dy^2 + y e^{-G}d\psit^2
-h^{-2} (dt+\omega)^2 +  h^{-2} \omega^2 \nn \\
&+&  h^{-2} e^{-2G} \Bigl[2 d\psit \, \frac{\Im(\partial_i K'
dz_i)}{(\tilde{r}^2 K')'} + \Bigl(\frac{\Im(\partial_i K'
dz_i)}{(\tilde{r}^2 K')'}\Bigr)^2 \Bigr] -
e^{2\alpha}(\tilde{\omega}^2 + 2 \omega_\psi \, d\psit \,
\tilde{\omega})\, ,
\eea
where
\beq
\omega =\tilde{\omega}+e^{-2G}\, \frac{\Im(\partial_i K'
dz_i)}{(\tilde{r}^2 K')'}\,.
\eeq

We now deal with a possible $U(1)$ gauging by completing the square
in $d\psit$.  In particular, by defining
\beq
\mathcal{A}=\tilde{\omega}+\frac{h^{-2}}{y \, e^G} \,
\frac{\Im(\partial_i K' dz_i)}{(\tilde{r}^2 K')'}\, ,
\eeq
the metric can then be put into precisely the gauged form of the 1/4 BPS
ansatz:
\beq
ds^2_{10} =  ye^G
d\Omega_3^2 + \frac{1}{ye^G}\, ds_4^2 + h^2 \, dy^2
-h^{-2}(dt+\omega)^2 + y e^{-G}(d\psit+\mathcal{A})^2.
\label{2Qgauged}
\eeq
where we have used the fact that
\beq -ye^{-G} \mathcal{A}^2 +  h^{-2}  \Bigl[e^{-2G}
\frac{\Im(\partial_i K' dz_i)}{(\tilde{r}^2
K')'}+\tilde{\omega}\Bigr]^2 + h^{-2} e^{-2G}
\Bigl(\frac{\Im(\partial_i K' dz_i)}{(\tilde{r}^2 K')'}\Bigr)^2
-e^{2\alpha}\tilde{\omega}^2 =0.
\eeq

As indicated by the form of the initial metric (\ref{eq:6met2ch}),
where the circle defined by $d\psi=d\phi_1-dt$ is trivially fibered
over the base, it is surprising to see that the gauged form of the
1/4 BPS ansatz has now turned up.  However, we still have not used
the explicit relations (\ref{KDEQ14})
for the K\"{a}hler potential to simplify $\mathcal{A}$. Once we do
this, we indeed find $\mathcal{A}=0$, which brings the solution to the
ungauged form:
\beq
ds^2_{10} =  ye^G d\Omega_3^2 +
\frac{1}{ye^G}\, ds_4^2 + h^2 \, dy^2 -h^{-2}(dt+\omega)^2 + y
e^{-G}d\psit^2,
\eeq
in agreement with initial expectations.  Furthermore, we can use the
condition $\mathcal{A}=0$ to express $\omega$ in terms of the $1/8$ BPS
one-form $\omega_{1/8}$:
\beq
\omega=-\frac{\Im(\partial_i K'
dz_i)}{(\tilde{r}^2 K')'}=y\, h^2 \Bigl[e^G \,
\omega_{1/8}+e^{-G}\,d\psi\Bigr].
\eeq

The final step is to find the explicit expression for the four-dimensional
K\"{a}hler metric (\ref{4dbase}).  Using the following expressions for
derivatives of the K\"{a}hler potential
\bea
\partial_i
K^\prime &=& \frac{\bar{z}_i  \cosh\varphi_i - z_i \sinh\varphi_i}
{2\Lambda H_2 H_3 \rho_i^2}, \nn\\
\bar\partial_{j}K^\prime &=&\frac{z_j\cosh\varphi_j-\bar{z}_j\sinh\varphi_j}
{2\Lambda H_2 H_3  \rho_j^2}, \nn\\
(\tilde{r}^2 K')' &=& \frac{r^2 \, \Delta \, (H_2 H_3)^{2/3}+ \tilde{r}^2}
{2\Delta \, (H_2 H_3)^{2/3}},\nn\\
\partial_i \bar{\partial}_j K dz_i d\zb_j &=&
\sum_i \frac{r^2 H_i}{2\rho_i^2}dz_id\zb_i\nn \\
&&+ \sum_{i,j} \frac{(\zb_i\cosh\varphi_i-z_i\sinh\varphi_i)
(z_j\cosh\varphi_j-\zb_j\sinh\varphi_j)} {2\Lambda H_2 H_3 \,
\rho_i^2 \, \rho_j^2}dz_id\zb_j ,
\eea
where
\beq
\Lambda=\frac{\Delta}{(H_2 \, H_3)^{1/3}} ,
\eeq
we finally obtain
\beq
ds_4^2 = \frac{r^2 H_i}{\rho_i^2}|dz_i|^2 + \frac{(\bar{z}_i
\cosh\varphi_i-z_i\sinh\varphi_i) (z_j
\cosh\varphi_j-\bar{z}_j\sinh\varphi_j)} {\rho_i^2 \rho_j^2 \bigl[
\Delta (H_2 H_3)^{2/3} +{y^2}/{r^4} \bigr] } dz_i d\zb_j .
\eeq

Note that one can obtain the special case of the singular
two-charge black hole (superstar) from the expressions above by
setting $\varphi_i=0$. This is very similar to the three-charge
extremal black hole solution (superstar) which we described with
the $1/8$ BPS examples.  In this case,
the harmonic functions are given by
\beq
H_i = 1+\frac{Q_i}{r^2},
\eeq
where $Q_i$ label the black hole charges.
The expression for $\Delta$ now simplifies
\beq
\Delta = (H_2 H_3)^{1/3}\sin^2\theta + \frac{H_3^{1/3}}{H_2^{2/3}}\cos^2\theta
\sin^2\alpha + \frac{H_2^{1/3}}{H_3^{2/3}}\cos^2\theta \cos^2\alpha,
\eeq
and so do the derivatives of $K$:
\bea
\partial_i K^\prime &=& \frac{\bar{z}_i}{2\Lambda H_2 H_3 \rho_i^2}, \nn\\
(\tilde{r}^2 K')' &=& \frac{r^2 \, \Delta \, (H_2 H_3)^{2/3}+ \tilde{r}^2}
{2\Delta\,(H_2 H_3)^{2/3}},\nn\\
\partial_i \bar{\partial}_j K dz_id\zb_j &=&
\sum_i\frac{r^2 H_i}{2\rho_i^2}dz_id\zb_i + \sum_{i,j} \frac{\zb_i \, z_j}
{2\Lambda H_2 H_3 \,\rho_i^2 \, \rho_j^2}dz_id\zb_j .
\eea
The final metric then becomes
\beq
ds_4^2 = \frac{r^2 H_i}{\rho_i^2}|dz_i|^2 +
\frac{\bar{z}_i\, z_j} {\rho_i^2 \rho_j^2 \bigl[ \Delta
(H_2H_3)^{2/3} +{y^2}/{r^4} \bigr] } dz_i d\zb_j  .
\eeq
We note that, for the specific case of two equal charges,
we were also able to embed this solution directly, without
resorting to the $1/8$ reduction, and found agreement.

\subsubsection{Boundary conditions}

We would like to emphasize again that for the LLM 1/2 BPS picture of
\cite{Lin:2004nb}, the boundary value (on the $y=0$ plane) of the function
$Z$ gave the black and white coloring of all the solutions, and was a
crucial element in the development of the droplet picture.  In this respect,
the 1/4 BPS system is more similar to the LLM case than to the 1/8 BPS
case, as it also involves a $y=0$ boundary plane and a binary choice of
either the $S^3$ or the $S^1$ collapsing.  As in the LLM case, this
boundary condition is encoded in the behavior of $Z$ [defined in the
usual manner according to (\ref{ZKrel})] as the $y$ coordinate
vanishes.  We now investigate this for the two charge bubble solutions.

We first combine the expressions (\ref{S13radius}) and (\ref{Delta})
above to find $Z$ for the smooth two-charge solutions:
\bea
Z = \frac{1}{2}\tanh{G}
&=& \frac{1}{2}\frac{r^2 \Delta (H_2 H_3)^{2/3} - \sin^2\theta}
{r^2 \Delta (H_2 H_3)^{2/3} +\sin^2\theta}\nn \\
&=& \frac{1}{2}-\frac{\sin^2\theta}{r^2 \Delta (H_2 H_3)^{2/3} +
\sin^2\theta} .
\eea
Since $y=r\sin\theta$ from (\ref{y2Q}), the $y\to0$ boundary is reached
when either $r\to0$ or $\theta\to0$.  Looking at the non-trivial denominator
of the expression above,
\bea
r^2\Delta (H_2H_3)^{2/3}+\sin^2\theta&=& \sin^2\theta(1+r^2H_2H_3) \nn \\
&&+ r^2\,H_3 \,\cos^2\theta \sin^2\alpha (\sin^2\phi_2 e^{-\varphi_2}
+\cos^2\phi_2 e^{\varphi_2}) \nn \\
&&+ r^2 \ H_2 \,\cos^2\theta \cos^2\alpha (\sin^2\phi_3 e^{-\varphi_3}
+\cos^2\phi_3e^{\varphi_3}) ,
\eea
one finds that
\beq
Z(\theta\rightarrow 0) = + \half .
\eeq
The case of $r \rightarrow 0$ with deformations turned on is more delicate,
especially since explicit solutions for $H_{2,3}$ are not known. However, we
note that as long as $H_2$ and $H_3$ approach a constant (and even in the case
$H_{2,3}\sim1/r$) as $r\rightarrow 0$, we find
\beq
Z(r \rightarrow 0) = - \half \qquad (\text{for}\quad \varphi_{2,3}\neq 0).
\eeq

Next, we would like to ask whether the boundary conditions found above
translate into the presence of a three-dimensional surface embedded in four
dimensions.  From (\ref{y2Q}) we know that the $y$ coordinate of the
two-charge solution was identified to be
\beq
y= r\sin\theta = r\mut_1 \, .
\eeq
Clearly $y$ vanishes when either $r=0$ or when $\theta=0$.  Using
$\sum_i \mut_i^2=1$ and the definition of our complex coordinates, we find
\bea
y^2 &=& r^2 (1-\mut_2^2-\mut_3^2) \nn \\
&=& r^2 \Bigl[1-\sum_{i=2,3}\frac{1}{4\rho_i^2}
\Bigl(e^{-\varphi_i}(z_i+\zb_i)^2-e^{\varphi_i}(z_i-\zb_i)^2 \Bigr)\Bigr]\, .
\label{2Qbc}
\eea
Thus, we see that the $\theta=0$ condition guaranteeing $y=0$
corresponds to the surface
\beq
\Bigl[1-\sum_{i=2,3}\frac{1}{4\bar{\rho}_i^2}
\Bigl(e^{-\bar{\varphi}_i}(z_i+\zb_i)^2-e^{\bar{\varphi}_i}(z_i-\zb_i)^2   \Bigr)\Bigr]=0,
\eeq
where $\bar{\rho}_i \equiv \rho_i(r=0)$ and
$\bar{\varphi}_i \equiv \varphi_i(r=0)$.
The surface denotes the boundary between regions where the $S^3$
shrinks ($r\rightarrow 0$) and regions where the $S^1$
shrinks ($\theta \rightarrow 0$). To see more explicitly that this
surface is in fact an ellipsoid embedded in four dimensions, we can
rewrite it using $z_i=x_i+iy_i$ in the following way:
\beq
\sum_{i=2,3} \Bigl[x_i^2 \,
\frac{e^{-\bar{\varphi}_i}}{\bar{\rho}_i^2}+ y_i^2 \,
\frac{e^{\bar{\varphi}_i}}{\bar{\rho}_i^2} \Bigr] =1 .
\label{QuarterEllips}
\eeq

We would like to make a few simple comments about the relation between
the ellipsoid above and the five-dimensional one (\ref{eq:elipeqn})
obtained in the 1/8 BPS case.  The $1/4$ BPS ellipsoid (\ref{QuarterEllips})
can be thought of as the $\varphi_1=0$, $\bar{\rho}_1=1$ limit of the
1/8 BPS ellipsoid (\ref{eq:elipeqn}), with the $S^1$ which rotates $x_1$
and $x_2$ shrinking to zero.  Furthermore, we can consider the 1/2 BPS
limit of (\ref{QuarterEllips}) by setting another charge to zero (say
$Q_2=0$, or $\bar{\varphi_2}=0$), and looking at the subspace where
$x_2^2+y_2^2=0$.  By doing so, we find a simpler one-dimensional surface
described by
\beq
x_3^2 \, \frac{e^{-\bar{\varphi}_3}}{\bar{\rho}_3^2}+
y_3^2 \, \frac{e^{\bar{\varphi}_3}}{\bar{\rho}_3^2} = 1 ,
\eeq
which is an ellipse in the two-dimensional (LLM) droplet plane.
This corresponds to a horizon-free, smoothed-out solution for the
1/2 BPS singular black hole.
One can alternatively arrive at this one-dimensional ellipse by
considering another limit of the $1/8$ BPS ellipsoid
(\ref{eq:elipeqn}), in which
$\varphi_1=\varphi_2=0$, $\bar{\rho}_1=\bar{\rho}_2=1$, and the
$S^3$ rotating the $x_1,x_2,x_3$ and $x_4$ coordinates is
shrinking to zero.

Let us now turn to the two-charge singular black hole (superstar) case,
which is obtained by turning off the deformations, {\it i.e.}~by setting
$\varphi_i=0$. Recalling that $H_i=1+{Q_i}/{r^2}$, we find that the
function $Z$ becomes
\beq
\label{2Qz}
Z=\half -  \frac{1}{1+Q_2+Q_3+r^2+{Q_2 \, Q_3}/{r^2}+
\cot^2\theta[(r^2+Q_3)\sin^2\alpha+(r^2+Q_2)\cos^2\alpha]}.
\eeq
We can now see that $Z$ approaches the same constant value
independently of how $y$ is going to zero,
\bea
\label{ztheta}
Z(\theta\rightarrow 0) &=& + \half , \\
\label{zr}
Z(r \rightarrow 0) &=& + \half,
\eea
provided neither charge vanishes.  In particular, it is the
${Q_2 \, Q_3}/{r^2}$ factor in the denominator of $Z$ which causes
$Z\rightarrow + \half$ even when $r\rightarrow 0$. This is consistent
with what we find if we look at what happens to the radii of $S^3$
and $S^1$ as $r \rightarrow 0$:
\bea
&& r(S^3) \rightarrow \sqrt{Q_2 Q_3} \sin\theta, \nn \\
&& r(S^1) \rightarrow 0.
\eea
On the other hand, when $\theta \rightarrow 0$, one recovers the
usual result, with the $S^3$ staying finite and the $S^1$
shrinking to zero. It is precisely the fact that $S^1 \rightarrow 0$
in both limits which makes $Z=1/2$ all the time.

Clearly, if we take one of the two charges in (\ref{2Qz}) to
vanish, our result should be comparable to the one-charge
superstar configuration. In that case it was found that, as $r
\rightarrow 0$, the function $Z$ approached a $Q$-dependent factor
\cite{Lin:2004nb}%
\footnote{Studies have shown that this distribution corresponds to
``typical states" given by approximately triangular Young
diagrams \cite{Balasubramanian:2005mg}.}
\beq
\label{zsuperstar}
Z\rightarrow \half \frac{Q-1}{Q+1}.
\eeq
Indeed, if we take, for example, $H_1=H_2=1$ and
$H_3=1+Q/r^2$, we find that $Z$ becomes
\beq
Z=\half -
\frac{1}{1+r^2+Q+\cot^2\theta(r^2+Q \sin^2\alpha)} \longrightarrow
\half \frac{Q(1+\cot^2\theta \sin^2\alpha)-1}{Q(1+\cot^2\theta
\sin^2\alpha)+1} \quad\text{as} \quad r\rightarrow 0 \nn \, ,
\eeq
a result that is similar to (\ref{zsuperstar}), except for some additional
angular dependence.  To conclude, we would like to note that the
$r\rightarrow 0$ behavior (\ref{zr}) of $Z$ for the two-charge black hole
is due to the additional presence of flux, forcing the second term in
(\ref{2Qz}) to approach zero.

Finally, we would like to identify, for the superstar, the regions in
the four-dimensional subspace where $y=0$. We can take the smooth
two-charge solution result (\ref{2Qbc}) and set $\varphi_i=0$.  We then
see that $y$ vanishes either when $r=0$ or on the ellipsoidal surface
\beq
\sum_{i=2,3} \frac{|z_i|^2}{\bar{\rho}_i^2}=1 .
\eeq
Similarly to the three-charge black hole case, if the charges are the same
the surface degenerates into a sphere.

%%%%%%%%%%%%%%%%%%%%%%%%%%%%%%%%%%%%%%%%
\subsection{LLM}
\label{sec:14llm}

We now turn to the embedding of configurations which preserve $1/2$ of
the available supersymmetries, namely the LLM solutions. These are
clearly a subset of the $1/4$ BPS states.
Recall the general form of the LLM metric, which is given in (\ref{eq:llmeqn}),
and which we repeat here for convenience
\begin{equation}
ds_{10}^2=-\hat h^{-2}(d\hat t+V)^2+\hat h^2(|dz_1|^2+d\hat y^2)
+\hat y(e^{\hat G}d\Omega_3^2+e^{-\hat G}d\widetilde\Omega_3^2).
\label{eq:6llmhat}
\end{equation}
Note that we have added a hat over LLM quantities to distinguish them
from their 1/4 BPS counterparts.
The most straightforward way to embed this into the 1/4 BPS ansatz
(\ref{GenericQuarterBPS}) is to write the second three-sphere $\tilde{S}^3$
of (\ref{eq:6llmhat}) as the Hopf fibration of U(1) bundled over $CP^1$,
and then to proceed with the gauged form of the 1/4 BPS ansatz.  This
is done by grouping $z_1$ with the complex coordinate on $CP^1$ to form
a four-dimensional K\"ahler base
\begin{eqnarray}
ds_{10}^{2} &=&-\hat{h}^{-2}(d\hat{t}+V_{z_{1}}dz_{1}
+V_{\overline{z}_1}d\overline{z}_1)^2+\hat{h}^{2}d\hat{y}^{2}
+\left[\hat{h}^2dz_1d\overline{z}_1+\hat{y}e^{-\hat{G}}ds^2(CP^1)\right]\nn\\
&&+\hat{y}e^{\widehat{G}}d\Omega_3^2+\hat{y}e^{-\hat{G}}
(d\hat\psi +\hat{\mathcal A})^2,
\label{14bpsllmans}
\end{eqnarray}
where $d\hat{\mathcal A}=2\hat J$ and $\hat J$ is the K\"ahler form on $CP^1$.

A direct comparison of the above with the 1/4 BPS form of the metric
(\ref{GenericQuarterBPS}) allows us to make the identifications:
\begin{eqnarray}
&&h=\hat{h}=(2\hat y\cosh\hat G)^{-\frac 12},\qquad
t=\hat{t},\qquad y=\hat{y},\qquad
e^{G}=e^{\hat{G}},\qquad\hat\psi=\psi,\nn\\
&&\omega=V_{z_{1}}dz_{1}+V_{\overline{z}_{1}}d\overline{z}_{1},\kern2.8em
{\mathcal A}=\hat{\mathcal A},\kern5.9em
{\mathcal{F}}=d{\mathcal A}=2\hat J.
\label{eq:6liftid}
\end{eqnarray}
The field strength ${\mathcal{F}}$ has flux through $CP^1$ and is quantized.
We also infer that the four-dimensional subspace is given by:
\begin{eqnarray}
{h}_{i\bar j}dz^{i}d\bar z^{\bar j} &=&ye^{G}\left[h^2dz_1
d\overline{z}_1+ye^{-G}\fft{dz_2d\overline z_2}{(1+|z_2|^2)^2}\right] \nn\\
&=&(Z+\frac{1}{2})dz_{1}d\overline{z}_{1}+{y}^{2}
\fft{dz_2d\overline z_2}{(1+|z_2|^2)^2},
\label{14BPSLLMh}
\end{eqnarray}
where we have written out the explicit metric on $CP^1$.
Here ${Z}={Z}(z_{1},\overline{z}_{1},{y})= \frac 12\text{tanh}G$ is just
the LLM harmonic function introduced in \cite{Lin:2004nb} and satisfying
(\ref{eq:4llmlap})
\begin{equation}
4\partial_{1}\partial_{\bar{1}}{ Z}+{y}\partial_{{y}}
(\frac{1}{{y}}\partial_{{y}}{ Z})=0,
\label{harm}
\end{equation}
where we have used that $z_1=x_1+i x_2$ and have rewritten the
two-dimensional Laplacian in terms of complex derivatives.

It is now clear that the four-dimensional base with metric (\ref{14BPSLLMh})
decomposes into a direct product of two complex subspaces, the first
being related to the two-dimensional LLM base and the second being simply
$CP^1$ warped by $y^2$.  To be explicit, we may write out the K\"ahler
potential yielding (\ref{14BPSLLMh}) as a sum of two terms
\begin{equation}
K=\ft12y^2\log(1+|z_2|^2)+\ft12\int\!\!\int^{z_1, {\bar z_1}}
({Z}(z_{1}^\prime,\bar{z}_{1}^\prime,{y})+\ft12)
dz_{1}^\prime d\bar{z}_{1}^\prime,
\label{14BPSLLMK}
\end{equation}
where the above integral is an indefinite integral (which allows for
K\"ahler transformations). Lastly, we observe that the harmonic function
$Z$ obeys the 1/4 BPS constraint (\ref{ZKrel}), $Z=-(y/2)\partial_y y^{-1}
\partial_y K$.  (Note that this condition removes the freedom to perform
K\"ahler transformations on $K$.)  To see this, it is useful to act on
both sides with  $\partial_1\bar\partial_1$, substitute (\ref{14BPSLLMK}),
and notice that the ensuing equation is nothing but the harmonic equation
(\ref{harm}).

To ensure that we really have a valid embedding, we would like to verify
that the non-linear Monge-Amp\`{e}re equation (\ref{eq:MA2}) is satisfied
as well:
\begin{equation}
\det {h_{i\bar j}}=\fft{y^2(Z+\fft12)}{(1+|z_2|^2)^2}
=e^D(Z+\ft12) y^{n\eta}\exp \left( \frac{1}{y}(2-n\eta)
\partial_y {K}
\right),
\label{eq:detgeqn}
\end{equation}
where we have used the explicit form of the four-dimensional metric
(\ref{14BPSLLMh}).  We now see that the $y$-dependence matches, provided
that we identify the $U(1)$ charge of the Killing spinor with
\beq
n\eta=2\,.
\label{eq:neta=2}
\eeq
In this case, the final term in (\ref{eq:detgeqn}) becomes trivial,
and we are left with the identification
\beq
e^D=\fft1{(1+|z_2|^2)^2},
\label{Dequation}
\eeq
which must be compatible with (\ref{eq:MA2}), which constrains $D$.
Since $D=-2\log(1+|z_2|^2)$, we see that $\partial\bar\partial D=4iJ_2$
where $J_2$ is the K\"ahler form on $CP^1$.  In this case, it is easy
to verify that
\begin{equation}
(1+*_4)\partial\bar\partial D=\fft{4i}{y^2}J_4,
\end{equation}
where $J_4=i\partial\bar\partial K$ is the K\"ahler
form on the full base metric (\ref{14BPSLLMh}).
This verifies that the constraint (\ref{eq:MA2})
is indeed satisfied.

\subsubsection{Boundary conditions}

Finally, we are interested in the lifting of the LLM boundary conditions
into the gauged 1/4 BPS ansatz.  Here, we notice from (\ref{eq:6liftid})
that, since both $y=\hat y$ and $G=\hat G$, the 1/4 BPS function $Z$ is
identified with the corresponding LLM one
\begin{equation}
Z(z_1,z_2,\bar z_1,\bar z_2,y)=Z_{\rm LLM}(z_1,\bar z_1,y).
\label{eq:6llmliftz}
\end{equation}
As usual, the boundary conditions are imposed on the $y=0$ subspace where
either $S^3$ or $S^1$ (inside $\tilde S^3$) shrinks to zero size.
The LLM solutions are regular if either $Z=-1/2$, which corresponds to
shrinking $S^3$, or if $Z=1/2$, which corresponds to shrinking $\tilde S^3$.

When lifted to the gauged 1/4 BPS ansatz, the boundary surfaces
implied by (\ref{eq:6llmliftz}) are $z_2$ independent.  This indicates
that the 1/2 BPS LLM droplets lift into four-dimensional
droplets which are simply the direct product of of the two-dimensional
droplet in the $(z_1,\bar z_1)$ plane with the $CP^1$ formed by
$(z_2,\bar z_2)$.  The boundaries of these droplets are then three-real
dimensional surfaces formed from the direct product of the boundary
lines of the LLM droplets with $CP^1$.

Unlike the above two examples of the AdS$_5\times S^5$ sphere and
the ellipsoidal deformations of the two-charge BPS bubble solution,
here the shapes of the droplets are different.  The reason for this
is because we have used a different choice of embedding for the LLM
system, corresponding to the gauged ansatz, instead of the ungauged
ansatz which was used above. In fact, because the LLM configurations
preserve the full $\tilde S^3$ isometry, and since the gauged ansatz
has an explicit $S^1$ fiber, the three dimensional boundary surfaces
necessarily have a $CP^1$ invariance (so that $S^1$ fibered over $CP^1$
forms the round $\tilde S^3$).  Therefore these surfaces must be of
the form of a direct product of a real curve in the LLM plane with
$CP^1$.  (The $CP^1$ is determined from the solution for the function
$D$ in (\ref{Dequation}).)

Note that, unlike in the case of the $(z_1,\bar z_1)$ LLM plane, which has
a regular flat metric, here the four-dimensional $y=0$ subspace given in
(\ref{14BPSLLMh}) has non-trivial geometry; it is in fact singular since
the $CP^1$ metric vanishes as $y\to0$.  (In general, the behavior of the
base may be different for the two separate cases $Z\to\fft12$ and
$Z\to-\fft12$.)
This singularity as $y\to0$ is reminiscent of the 1/8 BPS case, where the
six-dimensional base also develops a curvature singularity as the 1/8 BPS
$y$ variable approaches zero.  Although the full ten-dimensional metric is
non-singular, this nevertheless complicates the issue of making any
direct comparison of the four-dimensional boundary subspace with any
corresponding phase space in the dual gauge theory.

To make a closer comparison with the 1/8 BPS lifting of
Section~\ref{sec:18llm}, it may be advantageous to turn instead to
an ungauged embedding of LLM into the 1/4 BPS ansatz.  This is
perhaps most straightforwardly accomplished by reducing the 1/8
BPS lift of Section~\ref{sec:18llm} on a circle according to
either (\ref{eq:6reds1}) or some variation thereof. However, since
the result of doing so would only yield a modified interpretation
of the 1/8 BPS picture considered in Section~\ref{sec:18llm}, we
will not pursue this here.

%%%%%%%%%%%%%%%%%%%%%%%%%%%%%%%%%%%%%%%%
\subsection{General analysis with a decomposable four-dimensional base}

The above LLM embedding in the 1/4 BPS ansatz was facilitated by taking
the four-dimensional base to be a warped product of the LLM plane with $CP^1$.
In this subsection, we address the question of whether new classes of 1/4 BPS
solutions may be obtained where the four-dimensional base, parameterized by
the complex coordinates $z_1, z_2$, is a direct product of two Riemann
surfaces.  In particular, if the base is factorizable, then the
K\"{a}hler potential would be given by the sum
\beq
K=K_1 (z_1,\bar z_1,y) + K_2(z_2,\bar z_2,y).
\eeq
Following the general outline of the LLM embedding, we shall also assume that
\beq
Z=Z(z_1,\bar z_1, y),\qquad D=D(z_2,\bar z_2,y).
\eeq
Since $Z$ is related to $K$ by (\ref{ZKrel}), the requirement that
$Z$ is independent of $z_2,\bar z_2$ translates into
\beq
\partial_2 \partial_y(\frac 1y \partial_y K)=0,\qquad
\partial_{\bar 2}\partial_y(\frac 1y \partial_y K)=0.
\eeq
Therefore, we find that the $y$-dependence of $K_2$ is fixed:
\beq
K_2=y^2 k_2(z_2,\bar z_2) + \tilde k_2 (z_2,\bar z_2\,).
\label{k2y}
\eeq
Also, from the second equation in (\ref{eq:MA2}), we find that
\beq
D=\frac{4}{y^2}(1-n\eta) K_2 + d(z_2,y) + \bar d(\bar z_2, y)\,.
\label{dy}
\eeq

The immediate advantage of the assumptions we have made is that
the non-linear Monge-Amp\`{e}re equation factorizes. Under these
conditions, the first equation in (\ref{eq:MA2}) is replaced by
the following two equations:
\bea
&&\partial_1\partial_{\bar 1} K_1 = \fft{k(y)}2(Z+\ft12)
\exp\bigg(\frac{1}{y}(2-n\eta)\partial_y K_1\bigg),\nn\\
&&\partial_2\partial_{\bar 2} K_2 = \frac{1}{2k(y)} y^{n\eta}
\exp\bigg(\frac{1}{y}(2-n\eta)\partial_y K_2\bigg)e^D,
\label{14BPSKfact}
\eea
where $k(y)$ is an arbitrary function.
Substituting (\ref{k2y}) and (\ref{dy}) into (\ref{14BPSKfact}), we find
\beq
y^2 \partial_2\partial_{\bar 2} k_2 + \partial_2\partial_{\bar 2}\tilde k_2
=\frac{1}{2k(y)} y^{n\eta}\exp\bigl(2(2-n\eta) k_2\bigr)
\exp\bigg(\frac{4}{y^2}(1-n\eta)(y^2k_2 + \tilde k_2)+d+\bar d\bigg).
\label{k2eqn1}
\eeq
Since $k_2$, and $\tilde k_2$ are $y$-independent, matching the
$y$-dependence on both sides of the previous equation requires that
\beq
4(1-n\eta)\frac 1{y^2}\tilde k_2+d+\bar d=0,
\eeq
and
\beq
y^2=\frac{1}{k(y)} y^{n\eta}.
\eeq
Here we used the fact that the left-hand side of (\ref{k2eqn1}) is
a polynomial of degree two in $y$ to infer that the infinite series
in $y$ on the right-hand side must truncate.
After the $y$-dependence has been factored out, we are left with
\beq
\partial_2\partial_{\bar 2} k_2=\ft12\exp\bigl(2(4-3n\eta)k_2\bigr).
\eeq
Alternatively, we can rewrite this as a Liouville equation for $D$:
\beq
\frac{1}{1-n\eta}\partial_2\partial_{\bar 2}D
=2\exp\bigg(\frac{4-3n\eta}{2(1-n\eta)}D\bigg).
\eeq
The $z_1$ dependence of the four-dimensional K\"{a}hler base is
dictated by the remaining equation:
\beq
\partial_1\partial_{\bar 1} K_1 = \ft12y^{n\eta -2}(Z+\ft12)\exp\bigg
(\frac{1}{y}(2-n\eta)\partial_y K_1\bigg).
\label{k1eqn}
\eeq
A further restriction, namely
\beq
n\eta=2,
\label{cond1}
\eeq
which is identical to the LLM embedding case (\ref{eq:neta=2}), then allows
us to find explicit solutions.

Using (\ref{cond1}), the Liouville equation for $D$ becomes
\beq
\partial_2\partial_{\bar 2} D+2e^D=0,
\eeq
whose solutions are expressed in terms of an arbitrary holomorphic function
${\mathcal D}(z_2)$:
\beq
e^D=\frac{\left|\partial_2 {\mathcal D}(z_2)\right|^2}
{(1+|{\mathcal D}(z_2)|^2)^2}.
\label{liou}
\eeq
The choice of Killing spinor $U(1)$ charge according to (\ref{cond1})
leads to a drastic simplification of (\ref{k1eqn})
\beq
\partial_1\partial_{\bar 1} K_1 = \ft12(Z+\ft12),
\eeq
which can be easily integrated. Of course, $Z$ is constrained by
(\ref{ZKrel}). The compatibility of these two equations yields
\beq
4\partial_1\partial_{\bar 1}Z+y\partial_y \frac 1y\partial_y Z=0\,,
\eeq
which is the harmonic equation that we encountered before in the
context of LLM solutions.

Given the above, it is now easy to see that the base has a metric of
the form
\begin{eqnarray}
ds_4^2&=&(Z+\ft12)dz_1d\bar z_1+y^2e^Ddz_2d\bar z_2\nonumber\\
&=&(Z+\ft12)dz_1d\bar z_1+y^2\fft{|\partial_2\mathcal D(z_2)|^2}
{(1+|\mathcal D(z_2)|^2)^2}dz_2d\bar z_2.
\end{eqnarray}
A change of variables $z_2\to w\equiv\mathcal D(z_2)$ then results in
\begin{equation}
ds_4^2=(Z+\ft12)dz_1d\bar z_2+y^2\fft{dw d\bar w}{(1+|w|^2)^2},
\end{equation}
which is identical in form to that of (\ref{14BPSLLMh}).  This
demonstrates that the LLM lift examined in Section~\ref{sec:14llm}
is essentially the unique configuration corresponding to a decomposable
base.  Additional possibilities may exist, however, where the Killing
spinors carry a different $U(1)$ charge, $n\eta\ne2$.

%%%%%%%%%%%%%%%%%%%%%%%%%%%%%%%%%%%%%%%%
\subsection{Flux quantization}

Until now, we have focused on developing a droplet picture by examining
the loci of shrinking surfaces ($S^3$ or $S^1$) while ignoring flux issues.
However, we conclude this section by considering the IIB five-form flux
integral near $y=0$, with the goal of obtaining a flux quantization condition.
To obtain explicit results, we limit the following analysis to the LLM
embedding, where the four-dimensional base is decomposable.  In this
case, the ten-dimensional metric and flux take the form
\begin{eqnarray}
ds_{10}^{2} &=&-h^{-2}(dt+\omega )^{2}+h^{2}dy^{2}
+ye^{G}d\Omega_{3}^{2}+ye^{-G}(d\psi +\mathcal{A})^{2}\notag \\
&&+\frac{1}{y e^{G}}\left[ (\ft12+Z)|dz_{1}|^{2}
+y^{2}e^{D}|dz_{2}|^{2}\right] , \\
F_{(5)} &=& (1+\ast _{10})\Bigl(d[y^{2}e^{2G}(dt+\omega )]
+y^{2}(d\omega -d\mathcal{A})  \notag \\
&&-i\,[(\ft12+Z)dz_{1}\wedge d\bar{z}_{1}
+y^{2}e^{D}dz_{2}\wedge d\bar{z}_{2}]\Bigr)\wedge \Omega_{3},
\label{1/4 flux}
\end{eqnarray}
where we used (\ref{eq:6fss}) to obtain the components of the five-form.
Note that here we have explicitly set $\eta=-1$%
\footnote{In general, taking the period of $\psi$ to be $2\pi$, choosing
$\eta=1$ or $-1$ corresponds to choosing chirality $(1,2)$ or $(2,1)$ under
$SU(2)_{L}\times SU(2)_{R}$ for the Killing spinors on $S^{3}$ in
(\ref{s3spinor}).}.

We first want to consider the flux that is orthogonal to the
$(dt+\omega)\wedge dy\wedge\Omega_{3}$ directions.  This flux component
is easy to identify using (\ref{1/4 flux}) and (\ref{eq:6fss}).
The integral of its Hodge dual is given by:
\begin{eqnarray}
\int_{Z=-\fft12}\ast _{10}F_{(5)}
&=&\int_{y=0}\ast_{10}[\partial
_{y}(y^{2}e^{2G})dy\wedge (dt+\omega )\wedge \Omega _{3}]  \notag \\
&=&\int_{y=0}\left[ 2(\ft12-Z)+\frac{y\partial_{y}Z}
{\fft12+Z} \right] _{y=0}e^{D}\frac i2 dz_{1}\wedge
d\bar{z}_{1}\wedge
\frac i2 dz_{2}\wedge d\bar{z}_{2} \wedge(d\psi +\mathcal{A}).\nn\\
\label{flux1}
\end{eqnarray}
It can be seen from (\ref{eq:6fss}) that $\mathcal{A}$ has components along
the coordinates on the four-dimensional K\"{a}hler base only, so
$dz_{1}\wedge d\bar{z}_{1}\wedge dz_{2}\wedge d\bar{z}_{2}\wedge
\mathcal{A}=0$.  We can perform the flux integral by first integrating
out $d\psi$, and then reducing it to an integral over the four-dimensional
K\"{a}hler base (at $y=0$). Notice that (always assuming we are at $y=0$)
when $Z=-\frac{1}{2}$, which corresponds to the $S^3$ collapsing to zero
size, we have
\beq
\left[2(\ft12-Z)+\frac{y\partial_{y}Z}{\fft12+Z}\right]_{y=0}=4.
\eeq
Thus, the flux integral reduces to
\begin{eqnarray}
\int_{Z=-\fft12}\ast _{10}F_{(5)}&=&\int_{y=0}(2\pi)4e^{D}\big|_{y=0}
\fft{i}2 dz_1\wedge d\bar{z}_1\fft{i}2\wedge dz_2\wedge d\bar{z}_2\notag\\
&=&4~\mbox{Vol}\big(\Sigma_{3}\big|_{y=0}\big)\int_{Z=-\fft12}
\fft{i}2 dz_1\wedge d\bar{z}_1\sim N_{Z=-\fft12},
\end{eqnarray}
where $\mbox{Vol}\big(\Sigma _{3}\big|_{y=0}\big)=\int_{y=0}e^{D}\big|_{y=0}
\frac i2dz_{2}\wedge d\bar{z}_{2} \wedge (d\psi +\mathcal{A})$ is the volume
of a three dimensional surface at $y=0$.  This corresponds to the case of
D3-branes originally wrapping the $S^3$ in AdS$_5$ being replaced by five-form
fluxes through dual five-cycles ({\it i.e.}~$\Sigma_3\big|_{y=0}$ fibered
over the $Z=-\frac{1}{2}$ region of the $z_1$ plane).

Next, we consider the self-dual five-form with component along
$dz_{1}\wedge d\bar{z}_{1} \wedge \Omega _{3}$, and evaluate its
flux integral:
\begin{eqnarray}
\int_{Z=\frac{1}{2}}\ast _{10}F_{(5)} &=&\int_{y=0}[-i (\ft12+Z)_{y=0}
dz_{1}\wedge d\bar{z}_{1}\wedge \Omega _{3}
+\left(y^{2}e^{2G}\fft1y\partial_{y}J\right)_{y=0}\wedge\Omega_3] \notag\\
&=&\int_{y=0}\left[2(\ft12+Z)-\frac{y\partial_yZ}{\fft12-Z}\right]_{y=0}
\frac{-i}{2} dz_{1}\wedge d\bar{z}_{1 }\wedge \Omega_{3}.
\end{eqnarray}
The second term in the first line comes from the
$y^{2}e^{2G}d\omega \wedge \Omega _{3}$ term in the flux near $y=0$ in
expression (\ref{eq:6fss}).  Notice that, similarly to what happened in
(\ref{flux1}), when $Z=+\frac{1}{2}$ (corresponding to the three-cycle
$\Sigma_3\big|_{y=0}$ collapsing), we have
\beq
\left[2(\ft12+Z)-\frac{y\partial_{y}Z}{\fft12-Z}\right]_{y=0}=4.
\eeq
Thus, the flux integral reduces to
\begin{equation}
-\int_{Z=\frac{1}{2}} \ast _{10}F_{(5)}=4~\mbox{Vol}(S^{3})
\int_{Z=\frac{1}{2}}\frac i2 dz_{1} \wedge d\bar{z}_{1}\sim N_{Z=\frac{1}{2}}.
\end{equation}
Once again, this corresponds to the case of D3-branes, originally wrapping
the $\Sigma_3\big|_{y=0}$ in $S^5$, being replaced by five-form fluxes
through dual five-cycles ({\it i.e.}~$S^3$ fibered over the
$Z=\frac{1}{2}$ region of the $z_1$ plane).

%%%%%%%%%%%%%%%%%%%%%%%%%%%%%%%%%%%%%%%%
\section{Regularity conditions for 1/8 BPS configurations}
\label{regular}
%%%%%%%%%%%%%%%%%%%%%%%%%%%%%%%%%%%%%%%%

In the previous few sections, we have been concerned with developing a
droplet description of generic 1/8 and 1/4 BPS smooth solutions of type
IIB supergravity, corresponding to bubbling AdS configurations. These
configurations have either an $S^3$ isometry, or an $S^3\times S^1$ isometry.
The only non-trivial ten-dimensional fields are the self-dual five-form field
strength and the metric. We have also studied in detail several
classes of explicit solutions, and investigated their corresponding boundary
conditions at $y=0$.  It should be noted, however, that by starting with
known regular solutions (such as the three-charge smooth solutions of
\cite{Chong:2004ce} or the original 1/2 BPS LLM solutions \cite{Lin:2004nb}),
we are necessarily guaranteed to obtain regular examples of 1/4 and 1/8 BPS
embeddings.

It would be desirable, of course, to explore both regularity conditions
as well as boundary conditions on the BPS geometries directly, without
prior knowledge of explicit solutions.  What we mean here by boundary
conditions are the conditions specifying the droplets, {\it i.e.}~the
one or three-dimensional droplet boundaries on the $y=0$ subspaces for
the cases of 1/2 and 1/4 BPS solutions, or the five-dimensional droplet
boundaries for the 1/8 BPS case.  For 1/2 BPS LLM solutions, the uniqueness
of the Green's function solution to (\ref{eq:4llmlap}) ensures that each
droplet picture corresponds to a unique geometry%
\footnote{Note also that boundary conditions at $y\to\infty$ are encoded
in the Green's function.  These are necessary to ensure a proper asymptotic
AdS$_5$ geometry.}.
Furthermore, in the absence of cusps or other pathologies in the droplets,
all such 1/2 BPS solutions are regular.  Hence no additional regularity
conditions need to be imposed, at least for generic smooth droplets.

Because of the nonlinear equations underlying the supersymmetry analysis,
however, the regularity situation for 1/4 and 1/8 BPS configurations  is
less clear.  In principle, just as in the LLM case, it appears that droplets
can have any arbitrary shape or configuration; we simply choose any
desired three or five-dimensional boundary surface inside $\mathbb R^4$ or
$\mathbb R^6$, respectively, for the 1/4 and 1/8 BPS cases.  However, it
is not obvious that an arbitrary choice would always lead to a regular
smooth geometry in the full ten-dimensional sense.  After all, it is the
nature of non-linear equations that they do not always admit well behaved
solutions throughout their entire parameter range.  Furthermore, even if
a regular geometry exists, its uniqueness could be questioned.

For the droplet picture that we have presented to be useful, each droplet
configuration ought to give rise to a unique geometry.  Based on the LLM
experience, it certainly seems to be the case that droplet collections
would be unique, so long as we demand the geometry to be asymptotically
AdS$_5\times S^5$.  We are, however, unable to prove such uniqueness.
Nevertheless, we will motivate this statement by examining the approach
to AdS$_5\times S^5$ in the asymptotic regime.  Before doing so, however,
we first examine conditions on the regularity of the geometry near the
$y=0$ boundary.

For concreteness, we focus our attention on the 1/8 BPS
configurations.  (This also encompasses 1/4 and 1/2 BPS configurations
as special cases.\footnote{In Appendix D we  perform a regularity analysis directly on the 1/4 BPS solutions with an ungauged $S^3\times S^1$ isometry.} )  These solutions can be viewed as $\mathbb{R}
\times S^{3}$ fibrations over a six-dimensional K\"{a}hler base
which ends, as $y\rightarrow 0$, on (generally disconnected) five-dimensional
surfaces, where the $S^{3}$ fiber shrinks to zero size.  We are
interested in understanding the necessary conditions which ensure the
regularity of such solutions as $y\to0$.  These conditions then allow us
to understand the behavior of the K\"{a}hler potential near the
five-dimensional droplet boundaries, and will provide additional
insight into the moduli space of droplets in reduced supersymmetry
configurations.

%%%%%%%%%%%%%%%%%%%%%%%%%%%%%%%%%%%%%%%%
\subsection{Regular boundary conditions near $y=0$}

Focusing on 1/8 BPS configurations, we recall from (\ref{eq:bubblemets})
that the full ten-dimensional metric is of the form
\begin{equation}
ds_{10}^{2}=-y^{2}(dt+\omega )^{2}+\frac{2}{y^{2}}\partial_{i}
\partial_{\overline j}Kdz^{i}d\bar{z}^{\bar{j}}+y^{2}d\Omega
_{3}^{2},
\end{equation}
where the radial direction $y$ corresponds to the size of the $S^{3}$
\begin{equation}
y^2=e^{2\alpha }(z^{i},\bar{z}^{\bar{j}}).
\end{equation}
If the scalar field $\alpha $ is constant (as in the case of the
AdS$_{3}\times S^{3}\times T^{4}$ solution), then the only regularity
condition which must be enforced is on the six-dimensional K\"{a}hler metric
$h_{i\bar{j}}$. Otherwise, $y=0$ corresponds to a potentially singular locus,
with the three-sphere $d\Omega _{3}^{2}$ shrinking to zero size. To avoid
this singularity, the ten-dimensional metric must take the form
\begin{equation}
ds_{10}^{2}=-y^{2}(dt+\omega )^{2}+\frac{1}{y^{2}}\bigg(y^{2}dy^{2}+y^{2}d%
\Sigma _{4}^{2}+\mathcal{N}_{\psi }^{2}(d\psi +A)^{2}\bigg)+y^{2}d\Omega
_{3}^{2},\qquad y\ll 1
\label{y=0 10d metric}.
\end{equation}
As long as the four-dimensional subspace $d\Sigma_{4}^{2}$ is $y$-independent,
then the $y^{2}d\Omega_{3}^{2}+dy^{2}$ line element yields a regular (locally
flat) four-dimensional component of the ten-dimensional geometry. The
four-dimensional component $d\Sigma_{4}^{2}$ is similarly regular (at least
in terms of taking the $y\rightarrow 0$ limit).  Here $\mathcal{N}_{\psi }$
is a function of the coordinates on $d\Sigma_{4}^{2}$, and is finite at
$y=0$. However, the remaining two-dimensional component involving $t$ and
$\psi$ is still potentially singular, as $g_{tt}\rightarrow 0$ and
$g_{\psi \psi}\rightarrow \infty$.

To completely elucidate the $y\rightarrow 0$ behavior of the 1/8 BPS
solution, we first turn to the requirement that the six-dimensional base
is K\"{a}hler. In this case, we may take the three complex coordinates to
be given by
\begin{equation}
z_{j}=r_{j}e^{i\phi _{j}},\qquad j=1,2,3.
\end{equation}
Furthermore, the metric is determined by the K\"{a}hler potential
$K(z^{i},\bar{z}^{\bar{j}})$
\begin{equation}
ds_{6}^{2}=h_{mn}dx^mdx^n=2h_{i\bar{j}}dz^{i}d\bar{z}^{\bar{j}}
=2\partial_{i}\partial_{\bar{j}}Kdz^{i}d\bar{z}^{\bar{j}},
\end{equation}
where $m$ and $n$ are real indices and $i$ and $\overline{j}$ are complex
indices.  Assuming toric geometry, we now introduce a new real function
$F$, defined in the following way:
\begin{equation}
y^{2}\equiv F(r_{1}^{2},r_{2}^{2},r_{3}^{2}).
\label{y_eq}
\end{equation}
We are looking for a K\"{a}hler potential which will give us,
in the region near $y=0$, a metric of the form
\begin{equation}
ds_6^2=y^2dy^2+y^2d\Sigma_4^2+\mathcal{N}_{\psi }^{2}(d\psi+A)^{2}.
\label{y0_metric}
\end{equation}
Henceforth our analysis will refer strictly to the $y\ll 1$ region.
A K\"{a}hler potential satisfying our requirement is
\begin{equation}
K(z^{i},z^{\bar{j}})=\frac{1}{4}y^{4}+\mathcal{O}(y^{6}),
\label{kahler_y}
\end{equation}
up to an irrelevant constant.
Given the definition of $y$ in (\ref{y_eq}), it follows that for $y\ll 1$,
the six-dimensional base is toric, with a $U(1)^{3}$ isometry. This may be
too strong a requirement, but it allows us to consider a rather large
class of solutions ($F$ is only required to be a smooth non-singular
function of $r_{1}^{2}$, $r_{2}^{2}$, $r_{3}^{2}$), and at the same time
to be very specific.  Then, using the chain rule
\begin{equation}
ydy=F_{1}r_{1}dr_{1}+F_{2}r_{2}dr_{2}+F_{3}r_{3}dr_{3},\qquad
F_{i}=\frac{\partial F}{\partial {r_{i}^{2}}},
\end{equation}
we find
\begin{equation}
ds_6^2=\sum_{a=1}^3\left[F_a^2r_a^2+y^2(F_{aa}r_a^2+F_a)\right]
\left(dr_{a}^{2}+r_{a}^{2}d\phi _{a}^{2}\right)
+2\sum_{a<b}^3\left[F_aF_b+y^2F_{ab}\right]r_a^2r_b^2
\left(\frac{dr_{a}dr_{b}}{r_{a}r_{b}}+d\phi _{a}d\phi _{b}\right),
\label{kahler_metric}
\end{equation}
where
\begin{equation}
F_{ij}=\frac{\partial ^{2}F}{\partial r_{i}^{2}\partial r_{j}^{2}}.
\end{equation}
{}From $y^{2}=F(r_{1}^{2},r_{2}^{2},r_{3}^{2})$, by eliminating, say,
$r_{1}$ in favor of $y$,
\begin{equation}
r_{1}^{2}=f(y^{2},r_{2}^{2},r_{3}^{2}),
\label{r1f}
\end{equation}
we can express the metric in terms of the $\{y,\phi_{1},r_{2},\phi_{2},
r_{3},\phi _{3}\}$ coordinates.

The leading order terms of the six-dimensional metric are
\begin{eqnarray}
ds_{6}^{2} &=&dy^{2}\,y^{2}+\frac{1}{f_{y}^{2}}(fd\phi
_{1}-f_{2}r_{2}^{2}d\phi _{2}-f_{3}r_{3}^{2}d\phi _{3})^{2} \notag \\
&&+ \biggl[ dr_{2}^{2}\;\frac{y^{2}}{f_{y}}\left(-f_{2}-f_{22}r_{2}^{2}
+\frac{f_{2}^{2}r_{2}^{2}}{f}\right)+dr_{3}^{2}\;\frac{y^{2}}{f_{y}}
\left(-f_{3}-f_{33}r_{3}^{2}+\frac{f_{3}^{2}r_{3}^{2}}{f}\right)\notag \\
&&+2dr_2dr_3\;\fft{y^2}{f_y}r_2r_3\left(-f_{23}+\fft{f_2f_3}{f}\right)\notag\\
&&+d\phi_{2}^{2}\;\frac{y^{2}r_{2}^{2}}{f_{y}^{2}}
\left(2r_{2}^{2}f_{2y}f_{2}-f_{2}f_{y}-r_{2}^{2}f_{y}f_{22}
-\frac{r_{2}^{2}f_{2}^{2}f_{yy}}{f_{y}}\right)  \notag \\
&&+d\phi _{3}^{2}\;\frac{y^{2}r_{3}^{2}}{f_{y}^{2}}
\left(2r_{3}^{2}f_{3y}f_{3}-f_{3}f_{y}-r_{3}^{2}f_{y}f_{33}
-\frac{r_{3}^{2}f_{3}^{2}f_{yy}}{f_{y}}\right)  \notag \\
&&+2d\phi _{2}d\phi _{3}\;\frac{y^{2}r_{2}^{2}r_{3}^{2}}{f_{y}}
(-f_{23}+f_{2y}f_{3}+f_{3y}f_{2}-f_{2}f_{3}f_{yy}) \biggr],
\label{y_metric}
\end{eqnarray}
where
\begin{equation}
f_{y}=\frac{\partial f}{\partial y^{2}},\qquad
f_{2}=\frac{\partial f}{\partial r_{2}^{2}},\qquad
f_{2y}=\frac{\partial ^{2}f}{\partial r_{2}^{2}\partial y^{2}},\qquad
\mbox{etc}\ldots.
\end{equation}
The subleading terms in this metric are given in
Appendix~\ref{regularappendix}. A direct comparison of (\ref{y_metric})
and (\ref{y0_metric}) shows that in the $y\ll 1$ region they are identical,
provided that we identify $\psi \equiv \phi_{1}$. Therefore, as anticipated,
the K\"{a}hler potential $K=y^{4}/4$ yields a six-dimensional metric which
is of the desired form, as in (\ref{y0_metric}).

We now have all the necessary ingredients to study the regularity of the
ten-dimensional metric. As discussed above, any potentially singular
behavior as $y\rightarrow 0$ would come from the following two-dimensional
part of the ten-dimensional metric
\begin{equation}
ds_{2}^{2}=-y^{2}(dt+\omega )^{2}+\frac{1}{y^{2}f_{y}^{2}}
(fd\phi_{1}-f_{2}r_{2}^{2}d\phi _{2}-f_{3}r_{3}^{2}d\phi _{3})^{2}.
\label{2d_metric}
\end{equation}
We now recall that the one-form $\omega $ is determined by the
K\"{a}hler potential of the six-dimensional base
\begin{equation}
2\eta d\omega =\mathcal{R},
\end{equation}
where $\mathcal{R}$ is the Ricci form of the base. Noting that
\begin{equation}
\mathcal{R}=iR_{i\bar{j}}dz^{i}\wedge d\bar{z}^{\bar{j}}
=\frac{i}{2}\partial_{i}\partial _{\bar{j}}\log (\det h_{mn})dz^{i}\wedge
d\bar{z}^{\bar{j}}
\end{equation}
is a $(1,1)$ form and that $d=\partial +\bar{\partial}$, we have
\begin{equation}
\omega =\omega _{i}dz^{i}+\bar{\omega}_{\bar{j}}d\bar{z}^{\bar{j}},\qquad
\omega _{i}=-\frac{i\eta }{8}\partial _{i}\log (\det h_{mn}),\qquad
\bar{\omega}_{\bar{j}}=(\omega _{j})^{\ast }.
\label{omega}
\end{equation}
Since $\sqrt{\det h_{mn}}$ is a scalar density, this means that
$\omega$ in (\ref{omega}) is locally defined.  From (\ref{kahler_metric})
we find that $\det h_{mn}=\mathcal{O}(y^{8})$ in the coordinate system of
$\{r_i, \phi_i \}$. Thus, the leading order term in $\omega $ is
\begin{eqnarray}
\omega &=&\frac{\eta }{8}\sum_{a=1}^{3}\partial _{r_{a}}\log (\det
h_{mn})\;r_{a}d\phi _{a}  \notag \\
&=&\frac{\eta }{8}\frac{8}{y^{2}}(F_{1}r_{1}^{2}d\phi
_{1}+F_{2}r_{2}^{2}d\phi _{2}+F_{3}r_{3}^{2}d\phi
_{3})+\mathcal{O}(y^{0}),
\end{eqnarray}
where, on the second line, we have used the chain rule to evaluate
$(dz_{1}\partial _{1}-d\bar{z}_{1}\partial _{\bar{1}})\log (y^{8})$ {\it etc}.
To leading order in $y$, we find that
\begin{equation}
\omega =\frac{1}{y^{2}f_{y}}(fd\phi _{1}-f_{2}r_{2}^{2}
d\phi_{2}-f_{3}r_{3}^{2}d\phi _{3})+\mathcal{O}(y^{0}).
\end{equation}
Plugging this expression back into (\ref{2d_metric}), the potentially
singular terms cancel, and we arrive at
\begin{equation}
ds_{2}^{2}=-\frac{2}{f_{y}}dt(fd\phi _{1}-f_{2}r_{2}^{2}d\phi
_{2}-f_{3}r_{3}^{2}d\phi _{3})+\mathcal{O}(y^{2}),
\end{equation}
which is regular.

To summarize, we have investigated the region of the 1/8 BPS solutions
near $y=0$. Assuming a toric base, we have seen that the $y=0$ locus is
a five-dimensional surface $\Sigma _{5}$ specified by
\begin{equation}
F(r_{1}^{2},r_{2}^{2},r_{3}^{2})=0.
\end{equation}
Furthermore, the $y$ coordinate is orthogonal to $\Sigma _{5}$.
The complete ten-dimensional solution is generated by choosing an
arbitrary smooth (generally disconnected) five-dimensional surface
embedded in the six-dimensional K\"{a}hler base.  Then the ten-dimensional
solution will be non-singular provided that, in the vicinity of the
$\Sigma _{5}$ surface,
\begin{eqnarray}
ds_{10}^{2} &=&-g_{tt}\big|_{y=0}~dt^{2}+\left(\frac{f}{f_{y}}+9a
\frac{f^{2}}{f_{y}^{2}}-\frac{3f^{2}f_{yy}}{f_{y}^{3}}\right)
\left(d\phi_1-\frac{f_2r_2^2}{f}d\phi _{2}
-\frac{f_{3}r_{3}^{2}}{f}d\phi_{3}-w_{t}dt\right)^{2}\bigg|_{y=0}\notag \\
&&+\frac{2f}{f_{y}}\left(d\phi _{1}-\frac{f_{2y}r_{2}^{2}}{f_{y}}d\phi _{2}
-\frac{f_{3y}r_{3}^{2}}{f_{y}}d\phi _{3}\right)^{2}\bigg|_{y=0}
+d\widetilde{\Sigma}_{4}{}^{2}(r_{2},\phi _{2},r_{3},\phi _{3})
+d\mathbb R_{4}{}^{2},
\label{10d_metric_partial}
\end{eqnarray}
where $g_{tt}\big|_{y=0}$ is finite and $d\widetilde{\Sigma}_{4}{}^{2}
(r_{2},\phi _{2},r_{3},\phi_{3})$ is the metric of a four dimensional
surface, and $d\mathbb R_{4}{}^{2}=dy^2 +y^2 d\Omega_3^2$. More
details of the intermediate steps are presented in
Appendix~\ref{regularappendix}.

The cancellation of the leading order $\mathcal{O}(y^{-2})$ terms in
$d\phi _{1}^{2}$, which was necessary to ensure the regularity of the
solution at $y=0$, forces us to keep the subleading
$\mathcal{O}(y^{2})$ terms from (\ref{y_metric}).  As we show in
Appendix~\ref{regularappendix}, we also take into account the leading
order terms generated from the \emph{correction} to the K\"{a}hler
potential, $\delta K=ay^{6}$. For the ten-dimensional metric, all the
terms collected in (\ref{10d_metric_partial}) are of the same order,
namely $\mathcal{O}(y^{0})$.  Note that, for regularity, one must also
require that $(f/{f_{y}})\big|_{y=0}$ is finite as a function of
$r_{2}^{2}$, $r_{3}^{2}$.  We remind the reader that the function $f$
is defined through $r_{1}^{2}=f(y^{2},r_{2}^{2},r_{3}^{2})$, so the
five-dimensional surface at $y=0$ is given by the constraint
$r_{1}^{2}=f(0,r_{2}^{2},r_{3}^{2})$.

The full K\"{a}hler potential is obtained by evolving the approximate
$K=y^{4}/4+\mathcal{O}(y^{6})$ according to (\ref{eq:7cond})
\begin{equation}
\Box_{6}R=-R_{mn}R^{mn}+\frac{1}{2}R^{2},
\label{box_eq}
\end{equation}
where $R$ is the Ricci scalar of the six-dimensional K\"{a}hler base, and
$m,n=1,\ldots,6$ are real indices.

For completeness we shall also verify two consistency conditions. Since we
have identified the three-sphere warp factor $e^{2\alpha }$ with $y^{2}$,
and since $y=(-8/R)^{1/4}$, we must check that indeed $R=-8/y^{4}$ to
leading order for $y\ll1 $. From the expression of the Ricci tensor on a
K\"{a}hler space
\begin{equation}
R_{i\bar{j}}=\partial _{i}\partial _{\bar{j}}\sqrt{\det h_{mn}},
\end{equation}
we find that, to leading order in $y$,
\begin{equation}
R_{i\bar{j}}=-2\frac{F_{i}F_{j}r_{i}r_{j}}{y^{4}}+\mathcal{O}(y^{-2}),\qquad
F_{i}=\partial_{r_{i}^{2}}F,\qquad\mbox{etc}\ldots.
\end{equation}
Hence
\begin{equation}
R_{r_ar_b}=-4\frac{F_{a}F_{b}r_{a}r_{b}}{y^{4}}+\mathcal{O}(y^{-2}),\qquad
R_{\phi_{a}\phi _{b}}=-4\frac{F_aF_br_a^2r_b^2}{y^4}+\mathcal{O}(y^{-2}).
\end{equation}
By inverting the K\"{a}hler metric (\ref{kahler_metric}) we can evaluate the
Ricci scalar
\begin{equation}
R=R_{mn}h^{mn}=-\frac{8}{y^{4}},
\end{equation}
as anticipated. The corrections to the K\"{a}hler potential are expected to
cancel any potential contributions to order $y^{-2}$ from the K\"{a}hler
metric (\ref{kahler_metric}).  The second check we perform on the
K\"{a}hler potential is that, to leading order in $y$, the equation
(\ref{box_eq}) is satisfied. Indeed this is so, since
\begin{equation}
R_{mn}R^{mn}=\frac{96}{y^{8}}+\mathcal{O}(y^{-6}),\qquad
R^{2}=\frac{64}{y^{8}}+\mathcal{O}(y^{-6}),\qquad
\Box_{6}R=-\frac{64}{y^{8}}+\mathcal{O}(y^{-6}).
\end{equation}

We now turn to a discussion of the fluxes.  Near each
disconnected component of the five dimensional surface, we may
perform an integral of $F_{5}$ over the five-surfaces and measure
the number of flux-quanta threading it. We use the $y\ll 1$
metric (\ref{y=0 10d metric}) and the flux
\begin{equation}
F_{5}=\ast _{10}F_{5}=(1+\ast _{10})(d[y^{4}(dt+\omega )]-2\eta
J^{(6)})\wedge \Omega _{3}.
\end{equation}
The component of $F_{5}$ which is needed contains
$(dt+\omega )\wedge dy\wedge \Omega_{3}$. We consider its Hodge dual,
\begin{equation}
\ast _{10}F_{5}=4y^{3}\frac{1}{y^{3}}\mathcal{N}_{\psi }(d\psi +A)\wedge
\mbox{Vol}_{\Sigma _{4}}+\cdots,
\end{equation}
and see that the $y$ dependence cancels nicely, which is a
consequence of the regularity of the expression (\ref{y=0 10d metric})
for the metric.  Hence the integral of the five-form flux through the $i$-th
disconnected piece of the 5d surface $\Sigma _{5}^{(i)}$ is
\begin{equation}
\int\limits_{\Sigma _{5}^{(i)}}\ast _{10}F_{5}=\int 4\mathcal{N}_{\psi}
(d\psi +A)\wedge \mbox{Vol}_{\Sigma _{4}}=N_{i},
\end{equation}
which is expected to be quantized.  The total D3 brane flux quanta $N$ of
the solution is the sum of the flux quanta threading each disconnected
component of the surfaces, {\it i.e.}~$N=\sum_{i}N_{i}$.

We have thus seen that, in order for the ten-dimensional 1/8 BPS
configurations to be regular, we need to specify the following boundary
conditions.  We begin with defining a five-dimensional surface via the
algebraic constraint $y^2\equiv F(r_1^2,r_2^2,r_3^2)=0$ (for generic
non-toric geometries, we should allow for a dependence on the three
angular coordinates as well, even though we have not done so here).
Then we require that the K\"{a}hler potential behaves (up to an irrelevant
constant) as $y^{4}/4$, to leading order in $y$ for $y\ll 1$.
This guarantees that the $dy^2+y^2 d\Omega_3^{2}$ part of the ten-dimensional
metric will be regular, and it ensures that, at leading order,
there will be no mixing between $y$ and the remaining coordinates.
Further requiring that the remaining part of the metric be regular imposes
additional constraints on the function $F(r_{1}^2,r_{2}^2,r_{3}^2)$.
In particular, a necessary condition for regularity is that
$(f/{f_{y}})\big|_{y=0}$ is finite, where $f$ was defined in (\ref{r1f}).
So, in the end, the six-dimensional K\"{a}hler base is allowed to end
only on {\it smooth} five-dimensional surfaces.

Other than for this smoothness condition, we have shown (at least locally
near $y=0$) that arbitrary droplet configurations are allowed by
regularity.  Of course, it remains to be seen whether this conclusion
holds globally as well.  Proving this appears to be highly non-trivial,
although there are no obvious obstructions to the existence of global
solutions starting from arbitrary droplet data.

%%%%%%%%%%%%%%%%%%%%%%%%%%%%%%%%%%%%%%%%
\subsection{Asymptotic conditions at large $y$}

Finally, while we do not address the uniqueness of solutions directly,
we now turn to an examination of the asymptotic boundary conditions.
In addition to addressing regularity and uniqueness issues, these
asymptotic conditions are also useful for identifying the 1/8 BPS
$\mathcal{N}=4$ SYM states that are dual to this class of regular
supergravity solutions.  (Other asymptotic boundary conditions could
correspond to 1/4 BPS or 1/2 BPS states of $\mathcal{N}=2$ or
$\mathcal{N}=1$ gauge theories arising from D3 branes.)

As we have seen earlier, demanding that the asymptotic geometry approaches
AdS$_5\times S^5$ gives rise to a leading K\"ahler potential of the
form (\ref{eq:7ads5kpot})
\begin{equation}
K=\ft12|z_i|^2-\ft12\log(|z_i|^2)+\cdots.
\label{eq:kpasy}
\end{equation}
Since the small $y$ K\"ahler potential behaves as (\ref{kahler_y})
\begin{equation}
K=\fft14y^4(z_i,\bar z_i)+\cdots,
\label{eq:kpy0}
\end{equation}
a complete solution would interpolate between (\ref{eq:kpasy}) in the
asymptotic region and various expressions behaving as (\ref{eq:kpy0}),
one for each disconnected component of the $y=0$ boundary.  The question
of uniqueness is then whether the 1/8 BPS condition (\ref{box_eq})
admits a unique solution with these boundary conditions.

As a preliminary step, we may consider the asymptotic expansion of $K$,
and in particular the form of the correction terms in (\ref{eq:kpasy}).
Recall that a general 1/8 BPS droplet configuration can be described by
excising regions from $\mathbb C^3$, coordinatized by $z_1$, $z_2$ and
$z_3$.  Near asymptotic infinity, the geometry of these excised regions
may then be encoded by generalized multipole moments.  This then allows
a multipole expansion of the K\"ahler potential at infinity.
Instead of developing the general multipole expansion, we give as an
example the next-to-leading expression of $K$ for 1/8 BPS solutions
with three $U(1)$ $R$-charges $(J_{1},J_{2},J_{3}) \propto
(Q_{1},Q_{2},Q_{3})$ turned on.  It suffices to obtain this term from
the asymptotic expression of the 1/8 BPS smooth configuration given by
the elliptic surface in (\ref{eq:elipeqn}).

Since we need the next-to-leading terms, we can start from the
expression in (\ref{asym_rho}) and keep leading terms in
$1/(R+1)$ or $1/R$ and linear in $Q_{i}$:
\begin{equation}
\rho_{i}^{2}\simeq (R+1)(1+\frac{Q_{i}}{R+1}).
\end{equation}
Note that if $Q_{i}=0$, we find $\rho _{i}^{2}=(R+1)$, corresponding to the
AdS$_{5}\times S^{5}$ vacuum.  We want to solve for $R$ in terms of
$|z_{i}|^{2}.$ We have the constraint equation
\begin{equation}
\sum_{i}\frac{\ |z_{i}|^{2}}{\rho_{i}^{2}}\simeq
\sum_{i}\frac{\ |z_{i}|^{2}}{R+1}(1-\frac{Q_{i}}{R+1})\simeq 1,
\end{equation}
which then gives
\begin{equation}
R+1\simeq \sum_{i}|z_{i}|^{2}-\frac{\sum_{i}Q_{i}|z_{i}|^{2}}
{\sum_{i}|z_{i}|^{2}}.
\end{equation}
We also checked that the above expression satisfies (\ref{r2_deriv})
by plugging in (\ref{eqn_lambda}):
\begin{equation}
\Lambda =\sum_{i}\frac{\ |z_{i}|^{2}}{\rho _{i}^{2}H_{i}}\simeq
1-\frac{\sum_{i}Q_{i}|z_{i}|^{2}}{(\sum_{i}|z_{i}|^{2})^{2}}.
\end{equation}

In the asymptotic region, the leading and next-to-leading terms in the
K\"{a}hler potential are expected to be a function of $|z_{i}|^{2}$,
$i=1,2,3,$
\begin{equation}
K=K(|z_{i}|^{2}).
\end{equation}
Note that the derivatives of $K$ are known, since they were evaluated in
(\ref{k_2nd_deriv})
\begin{equation}
\partial _{|z_{j}|^{2}}\partial _{|z_{i}|^{2}}K
=\frac{1}{2\Lambda H_{1}H_{2}H_{3}\rho _{j}^{2}\rho _{i}^{2}}
\simeq \frac{1}{2(\sum_{i}|z_{i}|^{2})^{2}}
+\frac{3\sum_{i}Q_{i}|z_{i}|^{2}}{2(\sum_{i}|z_{i}|^{2})^{4}}
-\frac{Q_{j}+Q_{i}}{2(\sum_{i}|z_{i}|^{2})^{3}}
-\frac{\sum_{i}Q_{i}}{2(\sum_{i}|z_{i}|^{2})^{3}}.
\end{equation}
After integrating $\int d|z_{j}|^{2}\int d|z_{i}|^{2}$ we get
\begin{equation}
K\simeq \ft12\sum_{i}|z_{i}|^{2}-\ft12\log\Bigl(\sum_{i}|z_{i}|^{2}\Bigr)
+\frac{1}{4}\frac{\sum_{i}Q_{i}|z_{i}|^{2}}{(\sum_{i}|z_{i}|^{2})^{2}}
-\frac{1}{8}\frac{\sum_{i}Q_{i}}{(\sum_{i}|z_{i}|^{2})}.
\end{equation}
The first two terms provide the leading  AdS$_{5}\times S^{5}$ behavior
of (\ref{eq:kpasy}), while the latter two terms give the first order
deviations from the AdS$_5\times S^5$ vacuum that are linear in the 
$R$-charges, which characterize the solutions.

In principle, this expansion can be carried out to higher orders, and
with more general multipole distributions.  In this case, individual
complex components $z_i$ and $\bar z_i$ would also begin to enter into
the expansion of $K$.  Nevertheless, since any
arbitrary distribution of droplets in $\mathbb C^3$ may be fully
characterized by their (infinite set of) multipole moments, and since
the multipole expansion of $K$ appears to be unique (although we have
not proven this), this provides evidence that the droplet description
of bubbling AdS is well defined in the sense that there is a one-to-one
mapping between droplets and geometries.

%%%%%%%%%%%%%%%%%%%%%%%%%%%%%%%%%%%%%%%%
\section{Conclusions}
\label{sec:conclusion}
%%%%%%%%%%%%%%%%%%%%%%%%%%%%%%%%%%%%%%%%

In this paper we investigated the supergravity duals of BPS states
in ${\mathcal N}=4$ super Yang-Mills. We found evidence for a
universal bubbling AdS picture for all 1/2, 1/4 and 1/8 BPS
geometries in IIB supergravity for these states. This picture
emerges from a careful consideration of the necessary conditions
which ensure the regularity of these supergravity solutions.

In the case of generic 1/8 BPS solutions, which have an $S^3$
isometry and are time-fibered over a six-real dimensional K\"ahler
base, regularity is enforced when the radius of $S^3$ (denoted by $y$)
vanishes: $y=0$. Since $y$ is a function of all the base
coordinates, $y=y(x^i)$, $i=1,\ldots,6$, the geometric locus where the
$S^3$ shrinks to zero size is a generally disconnected
five-dimensional boundary surface. We have found that regular 1/8 BPS
geometries are determined by the following boundary data: the
general smooth five-dimensional surfaces located at $y=0$ and the
six-dimensional K\"ahler potential $K=\frac 14 y^4 +{\cal O}(y^6)$
near $y=0$. The interior of these five-dimensional surfaces is
excised from the six-dimensional base, since the base ends at
$y=0$. Each regular solution is thus associated with a smooth
five-dimensional surface. For example, the boundary data for the
AdS$_5\times S^5$ ground state is a five-dimensional round sphere,
whose interior, {\it i.e.}~a round ball, is removed from the
six-dimensional base. A generic 1/8 BPS state is then characterized by
a combination of topologically trivial deformations of the $S^5$ (gravitons),
topologically non-trivial ones (giant gravitons), and/or excisions
of other six-dimensional droplets from the base (dual giant gravitons).
One may view these surfaces as the locus where the matrix eigenvalues of the
three complex scalars in the dual theory are distributed. In order
for these configurations to be dual to ${\mathcal N}=4$ super
Yang-Mills states, we must impose additional conditions such
that asymptotically one recovers an AdS$_5\times S^5$ geometry.

In the case of 1/4 BPS solutions, which have an $S^3\times S^1$
isometry, we have identified a four-dimensional K\"ahler base
where the regularity conditions must be imposed. The droplets are
four-dimensional regions of shrinking $S^3$ inside a background where
the $S^1$ shrinks to zero size. This is a natural extension of the LLM
droplet picture of 1/2 BPS states, which was obtained by
specifying the two-dimensional regions inside a two-dimensional
phase-space where the $S^3$ inside AdS$_5$ collapses.
Therefore the 1/4 BPS regular solutions are characterized by three
dimensional surfaces separating the regions where either the $S^3$
or the $S^1$ collapses. For example, in the ungauged 1/4 BPS case,
the AdS$_5\times S^5$ ground state corresponds to a round
three-sphere in the four-dimensional base space, and a generic 1/4
BPS state is given by a deformation and/or topologically
non-trivial distortion of the round three-sphere.

%%%%%%%%%%%%%%%%%%%%%%%%%%%%%%%%%%%%%%%%
\begin{figure}[t]
\begin{center}
\includegraphics[width=12cm]{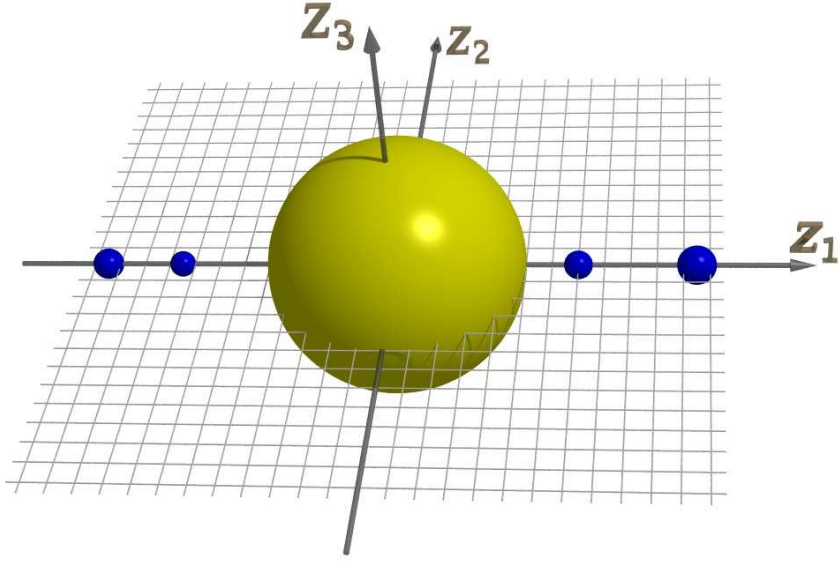}
\end{center}
\caption{Schematic picture of a 1/2 BPS configuration corresponding to
four dual giant gravitons excited on top of the AdS vacuum (central sphere).
Giant gravitons expanding on $S^5$ are not pictured, but would correspond
to giving the AdS sphere a non-trivial topology.  These 1/2 BPS configurations
always preserve an $\tilde S^3$ invariance corresponding to rotations in the
$z_2$-$z_3$ planes.}
\label{fig:droplet12}
\end{figure}
%%%%%%%%%%%%%%%%%%%%%%%%%%%%%%%%%%%%%%%%

We discussed several examples to better illustrate the
universality of the `bubbling AdS' picture in the 1/2, 1/4 and 1/8
BPS sectors. Given the non-linearity of the equations which
determine the explicit form of the 1/4 and 1/8 BPS solutions, our
regularity analysis focused on the small $y$ region of the ten
dimensional geometry and our analysis of the boundary behavior of
the K\"ahler potential is perturbative in small $y$; the boundary
conditions ensure the regularity of the ten-dimensional solution
in a neighborhood patch near $y=0$.  Although we have given
plausibility arguments, we have not rigorously shown that the
solutions which are generated after specifying the boundary data
are unique, nor can we say whether the perturbative analysis near
the $y=0$ region is sufficient to guarantee the regularity in the
whole space at arbitrary non-zero $y$. Clearly such questions
deserve a more thorough investigation. Although the differential
equations determining the whole geometry are non-linear, the
mapping between the topology of the boundary surfaces in the
K\"ahler base and the topology of the eigenvalue distributions of
the complex scalars in the dual ${\cal N}=4$ gauge theory should
be quite straightforward and robust.

\begin{figure}[t]
\begin{center}
\includegraphics[width=12cm]{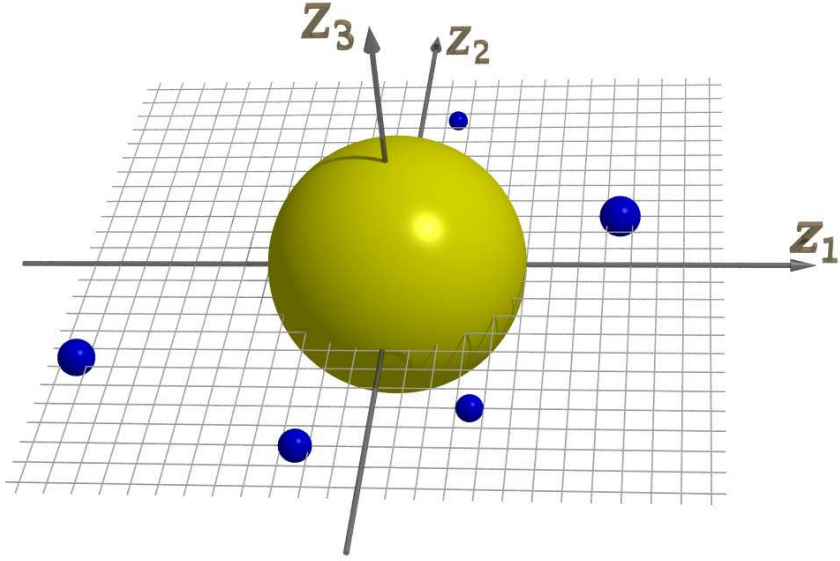}
\end{center}
\caption{Picture of a 1/4 BPS configuration with five dual giant gravitons.
The configuration is symmetric under $S^1$ rotations in the $z_3$ plane
(which, however, cannot be directly visualized since the imaginary
components of the axes are suppressed).}
\label{fig:droplet14}
\end{figure}

\begin{figure}[t]
\begin{center}
\includegraphics[width=12cm]{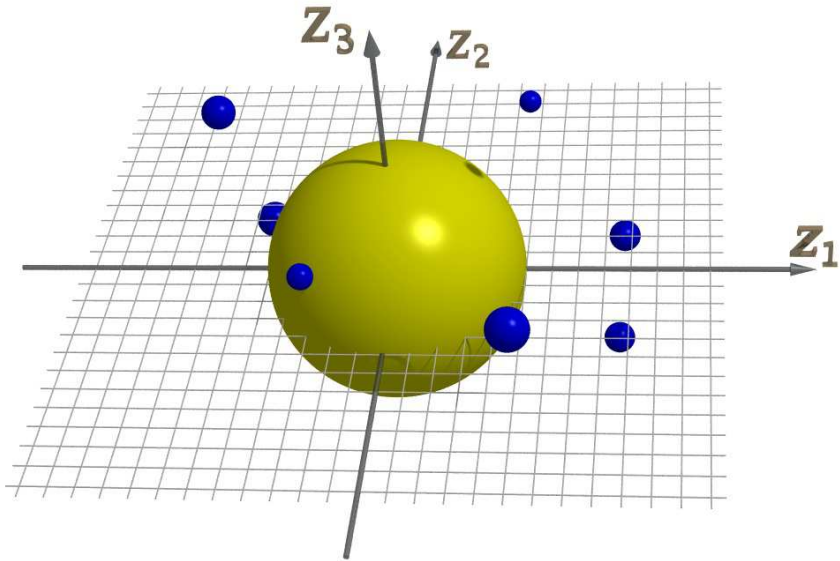}
\end{center}
\caption{Schematic picture of a 1/8 BPS configuration with
seven dual giant gravitons.  In general, 1/8 BPS droplets may
have any topology and geometry allowed by regularity.}
\label{fig:droplet18}
\end{figure}

The family of 1/2, 1/4 and 1/8 BPS geometries may be summarized using
the generic 1/8 BPS picture, where the droplets live on $\mathbb C^3$,
the coordinate space of the six-real dimensional K\"ahler base.  As
shown in Section~\ref{sec:18llm}, 1/2 BPS ({\it i.e.}~lifted LLM)
configurations are described by $\tilde S^3$ invariant droplets
in the $z_1$ plane.  Such configurations are shown schematically in
Figure~\ref{fig:droplet12}.  Moving to 1/4 BPS geometries entails
generalizing the droplets to lie anywhere in the $z_1$-$z_2$ planes,
but to maintain an $S^1$ invariance corresponding to rotations in the
$z_3$ plane.  This is shown in Figure~\ref{fig:droplet14}.  Finally,
generic 1/8 BPS droplets may lie anywhere in $\mathbb C^3$, as indicated
in Figure~\ref{fig:droplet18}.

%%%%%%%%%%%%%%%%%%%%%%%%%%%%%%%%%%%%%%%%
%%%%%%%%%%%%%%%%%%%%%%%%%%%%%%%%%%%%%%%%

It is interesting to note that the droplets which comprise the boundary
data for $1/2^n$ BPS geometries belong to a $2n$ ($n=1,2,3$)
real-dimensional K\"ahler space, which is naturally endowed with a
symplectic form, and therefore admits a phase-space interpretation.
It is also endowed with a complex structure, which is naturally
related to the existence of the $n$ complex scalars in the dual
theory. This observation should be sharpened after quantizing the
1/4 and 1/8 BPS classical solutions discussed here. The
five-dimensional surfaces that we observe are expected to become
non-commutative after the quantization.

It is expected that the 1/2 BPS droplets of Figure~\ref{fig:droplet12}
are non-interacting (as they admit a dual free-fermion description).
This is supported by the linearity of the LLM harmonic function
equation (\ref{eq:4llmlap}).  Furthermore, the complex $z_1$ plane is
unaffected by the presence of the droplets, and hence remains flat
regardless of the details of the 1/2 BPS configuration.  This is no
longer true in the reduced supersymmetry cases.  In particular, note
that Figures~\ref{fig:droplet14} and \ref{fig:droplet18} visualize the
1/4 and 1/8 BPS droplet data in coordinate space, given by Euclidean
$\mathbb C^3$.  The K\"ahler metric itself is highly non-trivial, so the
geometry of the K\"ahler base is curved by the droplets themselves; in
fact, the curvature on the 1/8 BPS base blows up ($R\to-\infty$) as one
approaches the boundaries of the droplets.  This suggests that the 1/4
and 1/8 BPS droplets will have non-trivial interactions, as would also
be expected based on reduced supersymmetry.

%%%%%%%%%%%%%%%%%%%%%%%%%%%%%%%%%%%%%%%%
%%%%%%%%%%%%%%%%%%%%%%%%%%%%%%%%%%%%%%%%

Understanding this non-trivial geometry on the K\"ahler base and its
implications for droplet dynamics seems to be essential in constructing
the moduli space of these BPS configurations. Among other things,
this geometry should shed light on the scattering of BPS droplets in a
non-trivial background.

%%%%%%%%%%%%%%%%%%%%%%%%%%%%%%%%%%%%%%%%
\section*{Acknowledgments}

SC would like to thank Joan Simon for critical comments and
suggestions, and Scott Watson for useful discussions. JTL wishes to
thank David Berenstein for his valuable insight, and for discussions
related to \cite{Berenstein:2007wz}.  AD wishes to thank Antal Jevicki.
HL would like to thank Oleg Lunin and Juan Maldacena for helpful 
communications and Harvard University and the City University of
New York for hospitality.  SC and JTL would like to thank the KITP for
hospitality.
The work of SC is supported in part by the Michigan
Society of Fellows.  The work of BC was partially supported by NSFC
Grant No.~10405028, 10535060, NKBRPC (No.~2006CB805905) and the Key Grant
Project of the Chinese Ministry of Education (No.~305001).  FLL and WYW
acknowledge the support of the Taiwan NSC under Grants 94-2112-M-003-014
and 95-2811-M-002-028.  JTL was supported in part by the
Stephen Hawking Chair in Fundamental Physics at the Mitchell Institute
for Fundamental Physics, Texas A\&M University.  This research was
supported in part by the MCTP, the US Department of Energy under Grant
Nos.~DE-FG02-91ER40688-Task~A and DE-FG02-95ER40899 and the National
Science Foundation under Grant No.~PHY99-07949.

%%%%%%%%%%%%%%%%%%%%%%%%%%%%%%%%%%%%%%%%
\appendix
%%%%%%%%%%%%%%%%%%%%%%%%%%%%%%%%%%%%%%%%

%%%%%%%%%%%%%%%%%%%%%%%%%%%%%%%%%%%%%%%%
\section{Differential identities for the $S^3$ reduction}
%%%%%%%%%%%%%%%%%%%%%%%%%%%%%%%%%%%%%%%%

The seven-dimensional system given in Section~\ref{sec:d=7}
comprises a metric, scalar and two-form field strength,
$(g_{\mu\nu},\alpha,F_{(2)})$.  The differential identities are
obtained by taking the supersymmetry variations (\ref{eq:7gto})
and (\ref{eq:7dto}) and contracting on the left with either
$\overline\epsilon$ or $\epsilon^c$ along with a complete set of
Dirac matrices $\{1,\gamma_\mu,\gamma_{\mu\nu},\gamma_{\mu\nu\lambda}\}$.

After appropriate rearrangement, most of the differential identities for
the Dirac bilinears $\{f,K,V,Z\}$ may be written in form notation
\begin{eqnarray}
i_Kd\alpha&=&0,\\
d(e^{-\alpha}f)&=&0,\label{eq:7emaf}\\
d(e^{3\alpha}f)&=&-i_KF,\\
d(e^{2\alpha}K)&=&-e^{-\alpha}fF-2\eta e^\alpha V,\\
d(e^{-2\alpha}K)&=&-e^{-5\alpha}*(F\wedge Z)+2\eta e^{-3\alpha}V,\\
d(e^\alpha V)&=&0,\label{eq:7eay}\\
d(e^{2\alpha}*V)&=&2\eta e^\alpha*K,\\
d(e^{4\alpha}Z)&=&e^\alpha F\wedge V-4\eta e^{3\alpha}*Z,\\
d(e^{3\alpha}*Z)&=&0.
\end{eqnarray}
The remaining identities are of the form
\begin{eqnarray}
0&=&F_{\mu\nu}V^{\mu\nu}+8\eta e^{2\alpha}f,\\
\nabla_{(\mu}K_{\nu)}&=&0,\label{eq:7killing}\\
\nabla_\mu V_{\nu\lambda}&=&\ft14e^{-3\alpha}
(2Z_{\mu[\nu}{}^\rho F_{\lambda]\rho}-Z_{\nu\lambda}{}^\rho F_{\mu\rho}
-g_{\mu[\nu}Z_{\lambda]\rho\sigma}F^{\rho\sigma}),\\
\nabla_\mu Z_{\nu\lambda\rho}&=&\ft14e^{-3\alpha}
(-\ft12\epsilon_{\mu\nu\lambda\rho}{}^{\alpha\beta\gamma}F_{\alpha\beta}
K_\gamma+3F_{\mu[\nu}V_{\lambda\rho]}+3F_{[\nu\lambda}V_{\rho]\mu}
+6g_{\mu[\nu}F_\lambda{}^\sigma V_{\rho]\sigma}).\quad
\end{eqnarray}
Note, in particular, that (\ref{eq:7killing}) demonstrate that $K^\mu$
is a Killing vector.  Although the `dilatino' variation (\ref{eq:7dto})
leads to algebraic expressions on the spinor bilinears, they naturally
combine with the gravitino variation expressions, and this is what we
have done above in writing down a complete set of differential identities
on the Dirac bilinears.

For the Majorana bilinears $\{f^m,Z^m\}$, we find instead
\begin{eqnarray}
\eta f^m&=&0,\label{eq:7fmzero}\\
d(e^{\alpha}f^m)&=&0,\\
d(e^{-3\alpha}f^m)&=&\ft{i}2e^{-\alpha}
(Z^m_{\mu\nu\lambda}F^{\nu\lambda})dx^\mu,\\
d(e^{2\alpha}Z^m)&=&-2\eta e^\alpha*Z^m,\label{eq:7e2azm}\\
d(e^\alpha*Z^m)&=&\ft{i}4e^{-2\alpha}f^m*F,\\
\nabla_\mu Z^m_{\nu\lambda\rho}&=&\ft{i}{16}e^{-3\alpha}
(-\ft23\delta_\mu^\alpha\epsilon_{\nu\lambda\rho}{}^{\beta\gamma\delta\epsilon}
+g_{\mu[\nu}\epsilon_{\lambda\rho]}{}^{\alpha\beta\gamma\delta\epsilon}
-2\delta_{[\nu}^\alpha\epsilon_{\lambda\rho]\mu}{}^{\beta\gamma\delta\epsilon})
F_{\alpha\beta}Z^m_{\gamma\delta\epsilon}.
\end{eqnarray}
Since $\eta=\pm1$ is non-vanishing (for Killing spinors on $S^3$), the first
expression, (\ref{eq:7fmzero}), immediately demonstrates that the Majorana
scalar invariant vanishes, $f^m=0$.  This leads to the identification
of SU(3) structure and a resulting simplification of the above expressions,
as discussed in Section~\ref{sec:d=7susy}.

%%%%%%%%%%%%%%%%%%%%%%%%%%%%%%%%%%%%%%%%
\section{Differential identities for the $S^3\times S^1$ reduction}
%%%%%%%%%%%%%%%%%%%%%%%%%%%%%%%%%%%%%%%%

In Section~\ref{sec:d=6}, we presented the reduction of the bosonic
fields of IIB supergravity on $S^3\times S^1$ along with the
relevant set of supersymmetry variations (\ref{eq:6gto}) and (\ref{eq:6dto}).
Here we present a partial list of differential identities related
to these variations.  However, before doing so, we recall that the
bosonic fields in six dimensions are the metric $g_{\mu\nu}$, two
abelian gauge fields $A_\mu$ and $\mathcal A_\mu$, as well as
two `dilatonic' scalars $\alpha$ and $\beta$ and one `axionic' scalar
$\chi$.  The differential identities serve to related these
fields with each other, as well as the Dirac $\{f_1,f_2,K,L,V,Y,Z\}$
and Majorana $\{f^m,Y^m,Z^m\}$ bilinears given in (\ref{eq:6bil}).

Because of the large number of fields and bilinears, the complete
list of differential identities is rather long.  Here we only list
the more relevant ones to the supersymmetry analysis.  We begin
with the scalar identities
\begin{eqnarray}
0&=&i_Kd\alpha=i_Kd\beta=i_Kd\chi,\\
0&=&F_{\mu\nu}V^{\mu\nu}+2e^{-\beta}L^\mu\partial_\mu\chi+8\eta e^{2\alpha}
f_2,\\
0&=&F_{\mu\nu}Y^{\mu\nu}-8e^{3\alpha}L^\mu\partial_\mu\alpha-8\eta e^{2\alpha}
f_1,\\
0&=&\mathcal F_{\mu\nu}V^{\mu\nu}+4e^{-\beta}L^\mu\partial_\mu(\alpha+\beta)
+4\eta e^{-\alpha-\beta}f_1+4ne^{-2\beta}f_2,\\
0&=&\mathcal F_{\mu\nu}Y^{\mu\nu}+2e^{-3\alpha-2\beta}L^\mu\partial_\mu\chi
+4\eta e^{-\alpha-\beta}f_2-4ne^{-2\beta}f_1,\\
0&=&\eta f^m=nf^m,\label{eq:6fmzero}\\
0&=&F_{\mu\nu}Y^{m\,\mu\nu}=\mathcal F_{\mu\nu}Y^{m\,\mu\nu}.
\end{eqnarray}
Although the $U(1)$ charge $n$ of the Killing spinor may vanish,
the $S^3$ Killing spinor parameter $\eta=\pm1$ cannot vanish.  As
a result, (\ref{eq:6fmzero}) indicates that $f^m=0$.  This vanishing
of the Majorana scalar invariant simplifies the structure analysis
of Section~\ref{sec:d=6susy}, and is needed for the demonstration of
U(2) structure.

After some rearrangement, the one-form identities may be written as
\begin{eqnarray}
d(e^{-\alpha}f_2)&=&0,\label{eq:6emaf2}\\
d(e^{2\alpha+\beta}f_1+e^{-\alpha}f_2\chi)&=&-2\eta e^{\alpha+\beta}L,
\label{eq:6e2abf1}\\
d(e^{3\alpha}f_2)&=&-i_KF+e^{-\beta}f_1d\chi,\label{eq:6e3af2}\\
d(e^{-\beta}f_1)&=&-i_K\mathcal F,\label{eq:6embf1}\\
d(e^{\alpha+2\beta}f_2)&=&-\ft12e^{\alpha+3\beta}*Z_\mu{}^{\nu\lambda}
\mathcal F_{\nu\lambda}dx^\mu+e^{-2\alpha+\beta}f_1d\chi-2ne^{\alpha+\beta}L,\\
d(e^{-2\alpha+\beta}f_1)&=&\ft12e^{-5\alpha+\beta}*Z_\mu{}^{\nu\lambda}
F_{\nu\lambda}dx^\mu+2\eta e^{-3\alpha+\beta}L,\label{eq:6em2abf1}\\
D(e^\alpha f^m)&=&0,\\
D(e^{-3\alpha}f^m)&=&\ft{i}2e^{-6\alpha}Z^m_\mu{}^{\nu\lambda}F_{\nu\lambda}
dx^\mu+ie^{-6\alpha-\beta}Y^m_\mu{}^\nu\partial_\nu\chi dx^\mu,\\
D(e^{-\alpha-2\beta}f^m)&=&\ft12e^{-\alpha-\beta}*Z_\mu^m{}^{\nu\lambda}
\mathcal F_{\nu\lambda}dx^\mu+ie^{-4\alpha-3\beta}Y^m_\mu{}^\nu\partial_\nu\chi
dx^\mu,
\end{eqnarray}
where $D=d+in\mathcal A$ is the $U(1)$ gauge covariant derivative.

Turning to the two-form identities, we have
\begin{eqnarray}
d(e^{2\alpha}K)&=&-e^{-\alpha}f_2F-e^{2\alpha+\beta}f_1\mathcal F-2\eta
e^\alpha V,\\
d(e^{-2\alpha}K)&=&[-\ft14e^{-5\alpha}*Y_{\mu\nu}{}^{\lambda\sigma}
F_{\lambda\sigma}+\ft12e^{-5\alpha-\beta}*Z_{\mu\nu}{}^\lambda
\partial_\lambda\chi]dx^\mu\wedge dx^\nu
-e^{-2\alpha+\beta}f_1\mathcal F+2\eta e^{-3\alpha}V,\nonumber\\
&&\\
d(e^{2\beta}K)&=&[-\ft14e^{-3\alpha+2\beta}*Y_{\mu\nu}{}^{\lambda\sigma}
F_{\lambda\sigma}-\ft14e^{3\beta}
*V_{\mu\nu}{}^{\lambda\sigma}
\mathcal F_{\lambda\sigma}]dx^\mu\wedge dx^\nu-2ne^\beta Y,\\
d(e^{\alpha+\beta}L)&=&0,\\
dL&=&\ft12e^\beta\mathcal F_\mu{}^\lambda V_{\nu\lambda}dx^\mu\wedge dx^\nu
-\ft14e^{-3\alpha-\beta}Z_{\mu\nu}{}^\lambda\partial_\lambda\chi dx^\mu
\wedge dx^\nu,\\
d(e^{2\alpha}L)&=&\ft12e^{-\alpha}F_\mu{}^\lambda Y_{\nu\lambda}dx^\mu
\wedge dx^\nu
+\ft12e^{2\alpha+\beta}\mathcal F_\mu{}^\lambda V_{\nu\lambda} dx^\mu
\wedge dx^\nu.
\end{eqnarray}
In addition
\begin{eqnarray}
\nabla_{(\mu}K_{\nu)}&=&0,\label{eq:6killing}\\
\nabla_{(\mu}L_{\nu)}&=&\ft18e^{-3\alpha}(4F_{(\mu}{}^\lambda Y_{\nu)\lambda}
-g_{\mu\nu}F_{\lambda\sigma}Y^{\lambda\sigma})+\ft12e^\beta
\mathcal F_{(\nu}{}^\lambda V_{\nu)\lambda}.
\end{eqnarray}
In particular, this shows that $K^\mu$ is a Killing vector.  Also, while
$L_\mu$ is not a closed one-form, the combination $e^{\alpha+\beta}L$ is.

For the three-form identities, we have
\begin{eqnarray}
d(e^\alpha V)&=&e^{\alpha+\beta}\mathcal F\wedge L,\label{eq:6eav}\\
d(e^{-\alpha+2\beta}V)&=&\ft12e^{-4\alpha+2\beta}Z_{\mu\nu}{}^\sigma
F_{\lambda\sigma}dx^\mu\wedge dx^\nu\wedge dx^\lambda-e^{-\alpha+3\beta}
i_k*\mathcal F-2ne^{-\alpha+\beta}*Z,\\
d(e^{3\alpha+2\beta}V)&=&-e^{3\alpha+3\beta}i_K*\mathcal F
-e^{\beta}Y\wedge d\chi-2ne^{3\alpha+\beta}*Z,\\
d(e^\beta Y)&=&-e^{-3\alpha+\beta}i_K*F,\\
d(e^\alpha Y)&=&-\ft12e^{-2\alpha}(i_K*F-F\wedge L)+\ft14e^{\alpha+\beta}
Z_{\mu\nu}{}^\sigma\mathcal F_{\lambda\sigma}dx^\mu\wedge dx^\nu\wedge
dx^\lambda+\eta*Z,\\
d(e^{3\alpha}Y)&=&F\wedge L+\ft14e^{3\alpha+\beta}Z_{\mu\nu}{}^\sigma
\mathcal F_{\lambda\sigma}dx^\mu\wedge dx^\nu\wedge dx^\lambda+\ft12e^{-\beta}
V\wedge d\chi+3\eta e^{2\alpha}*Z,\quad\\
D(e^{2\alpha+\beta}Y^m)&=&2\eta e^{\alpha+\beta}*Z^m-ine^{2\alpha}Z^m,
\label{eq:6e2aby}\\
D(e^\beta Y^m)&=&[\ft{i}4e^{-3\alpha+\beta}*Z^m_{\mu\nu}{}^\sigma
F_{\lambda\sigma}+\ft{i}{12}e^{-3\alpha}
*Y^m_{\mu\nu\lambda}{}^\sigma\partial_\sigma\chi]dx^\mu\wedge dx^\nu\wedge
dx^\lambda-inZ^m,\\
D(e^\alpha Y^m)&=&[\ft14e^{\alpha+\beta}Z^m_{\mu\nu}{}^\sigma
\mathcal F_{\lambda\sigma}
+\ft{i}{12}e^{-2\alpha-\beta}*Y^m_{\mu\nu\lambda}{}^\sigma\partial_\sigma\chi]
dx^\mu\wedge dx^\nu\wedge dx^\lambda+\eta*Z^m.
\end{eqnarray}
When the indices are not taken to be fully antisymmetric, we must also
include the identities
\begin{eqnarray}
\nabla_\mu V_{\nu\lambda}&=&-\ft14e^{-3\alpha}(Z_{\nu\lambda}{}^\sigma
F_{\mu\sigma}-2Z_{\mu[\nu}{}^\sigma F_{\lambda]\sigma}+g_{\mu[\nu}
Z_{\lambda]}{}^{\alpha\beta}F_{\alpha\beta})+e^\beta\mathcal F_{\mu[\nu}
L_{\lambda]}\nonumber\\
&&-\ft14e^{-3\alpha-\beta}(Y_{\nu\lambda}\partial_\mu\chi-2Y_{\mu[\nu}
\partial_{\lambda]}\chi+2g_{\mu[\nu}Y_{\lambda]}{}^\sigma
\partial_\sigma\chi),\label{eq:6dmuv}\\
\nabla_\mu Y_{\nu\lambda}&=&\ft14e^{-3\alpha}(*F_{\mu\nu\lambda}{}^\sigma
K_\sigma-F_{\nu\lambda}L_\mu+2F_{\mu[\nu}L_{\lambda]}-2g_{\mu[\nu}
F_{\lambda]}{}^\sigma L_\sigma)+\ft12e^\beta Z_{\nu\lambda}{}^\sigma
\mathcal F_{\mu\sigma}\nonumber\\
&&+\ft14e^{-3\alpha-\beta}(V_{\nu\lambda}\partial_\mu\chi-2V_{\mu[\nu}
\partial_{\lambda]}\chi+2g_{\mu[\nu}V_{\lambda]}{}^\sigma\partial_\sigma\chi),\\
D_\mu Y^m_{\nu\lambda}&=&-\ft{i}4e^{-3\alpha}(*Z^m_{\nu\lambda}{}^\sigma
F_{\mu\sigma}-2*Z^m_{\mu[\nu}{}^\sigma F_{\lambda]\sigma}+g_{\mu[\nu}
*Z^m_{\lambda]\alpha\beta}F^{\alpha\beta})
+\ft12e^\beta Z^m_{\nu\lambda}{}^\sigma\mathcal F_{\mu\sigma}\nonumber\\
&&+\ft{i}4e^{-3\alpha-\beta}(*Y^m_{\mu\nu\lambda}{}^\sigma\partial_\sigma\chi
-2f^mg_{\mu[\nu}\partial_{\lambda]}\chi).
\end{eqnarray}
%

%%%%%%%%%%%%%%%%%%%%%%%%%%%%%%%%%%%%%%%%
\section{Differential identities for the $S^3\times S^3$ reduction}
%%%%%%%%%%%%%%%%%%%%%%%%%%%%%%%%%%%%%%%%

For the round $S^3\times S^3$ reduction, corresponding to the original
LLM system of \cite{Lin:2004nb}, the relevant supersymmetry variations
are given by (\ref{eq:4s3s3susy}).  Many of the differential identities
for this system have been tabulated in Appendix C of \cite{Liu:2004ru}.
We nevertheless give them here again, using our present notation.

Most of the differential identities can be presented in form notation.
For the Dirac bilinears $\{f_1,f_2,K,L,Y\}$, the scalar (or zero-form)
identities are
\begin{eqnarray}
&&0=i_Kd\alpha=i_Kd\beta,\\
&&0=i_Ld(\alpha+\beta)+\eta e^{-\alpha}f_1+\tilde\eta e^{-\beta}f_2,\\
&&0=i_Ld(\alpha-\beta)-\ft14e^{-3\alpha}F_{\mu\nu}Y^{\mu\nu}+\eta e^{-\alpha}f_1-\tilde\eta e^{-\beta}f_2,\\
&&\ft18e^{-3\alpha}F_{\mu\nu}*Y^{\mu\nu}=\tilde\eta e^{-\beta}f_1
=\eta e^{-\alpha}f_2.
\end{eqnarray}
The one-form identities are
\begin{eqnarray}
d(e^{-\beta}f_1)&=&0,\label{eq:4debf1}\\
d(e^\alpha f_1)&=&-\eta L,\\
d(e^{3\beta}f_1)&=&e^{-3\alpha+3\beta}i_K*F,\\
d(e^{-\alpha}f_2)&=&0,\\
d(e^\beta f_2)&=&-\tilde\eta L,\\
d(e^{3\alpha}f_2)&=&-i_KF.\label{eq:4de3af2}
\end{eqnarray}
The identities given here are derived by taking linear combinations of
those in \cite{Liu:2004ru}.  Of course, the particular choice we have made
for which linear combinations to take is not unique.  However, we find
the above choice particularly useful when completing the solution in
Section~\ref{sec:d=4susy}.  Continuing with the two-form identities, we
have
\begin{eqnarray}
dK&=&-\ft12e^{-3\alpha}(f_2F-f_1*F),\label{eq:4dk}\\
d(e^{2\alpha}K)&=&-e^{-\alpha}f_2F+2\eta e^\alpha*Y,\\
d(e^{2\beta}K)&=&-e^{-3\alpha+2\beta}f_1*F-2\tilde\eta e^\beta Y,\\
dL&=&0,\label{eq:4dl}\\
d(e^{\alpha+\beta}L)&=&0,\\
d(e^{2\alpha}L)&=&\ft12e^{-\alpha}F_\mu{}^\lambda Y_{\nu\lambda}
dx^\mu\wedge dx^\nu.
\end{eqnarray}
Finally, we give the three-form identities
\begin{eqnarray}
d(e^\beta Y)&=&0,\label{eq:4eby}\\
d(e^{-\alpha}Y)&=&-\eta e^{-2\alpha}*K,\\
d(e^{-3\beta}Y)&=&-e^{-3\alpha-3\beta}L\wedge F,\\
d(e^\alpha *Y)&=&0,\label{eq:4ea*y}\\
d(e^{-\beta}*Y)&=&-\tilde\eta e^{-2\beta}*L,\\
d(e^{-3\alpha}*Y)&=&e^{-6\alpha}L\wedge *F.
\end{eqnarray}

Additional information is contained in the original (non-form notation)
differential identities obtained from the gravitino variation
\begin{eqnarray}
\nabla_\mu K_\nu&=&-\ft14e^{-3\alpha}(f_2F_{\mu\nu}-f_1*F_{\mu\nu}),\nonumber\\
\nabla_\mu L_\nu&=&\ft14e^{-3\alpha}(2F_{(\mu}{}^\lambda Y_{\nu)\lambda}
-\ft12g_{\mu\nu}F_{\rho\lambda}Y^{\rho\lambda}),\nonumber\\
\nabla_\mu Y_{\nu\lambda}&=&-\ft14e^{-3\alpha}(F_{\nu\lambda}L_\mu
+2g_{\mu[\nu}F_{\lambda]}{}^\sigma L_\sigma-2F_{\mu[\nu}L_{\lambda]}).
\end{eqnarray}
Note that the vector identities may be decomposed into antisymmetric and
symmetric parts.  The former are contained in (\ref{eq:4dk}) and
(\ref{eq:4dl}), while the latter are
\begin{eqnarray}
2\nabla_{(\mu}K_{\nu)}&=&0,\\
2\nabla_{(\mu}L_{\nu)}&=&e^{-3\alpha}(F_{(\mu}{}^\lambda Y_{\nu)\lambda}
-\ft14g_{\mu\nu}F_{\rho\lambda}Y^{\rho\lambda}).
\end{eqnarray}

For the Majorana bilinears $\{K^m,Y^m\}$, we have the gravitino differential
identities
\begin{eqnarray}
\nabla_\mu K_\nu^m&=&\ft18e^{-3\alpha}(\ft12g_{\mu\nu}F_{\rho\sigma}
*Y^{m\,\rho\sigma}-2F_{(\mu}{}^\lambda *Y_{\nu)\lambda}^m),\nonumber\\
\nabla_\mu Y_{\nu\lambda}^m&=&\ft12e^{-3\alpha}(*F_{\mu[\nu}K_{\lambda]}^m
-g_{\mu[\nu}*F_{\lambda]\rho}K^{m\,\rho}-\ft12*F_{\nu\lambda}K_\mu^m),
\end{eqnarray}
as well as the zero-form identities
\begin{eqnarray}
F^{\mu\nu}Y^m_{\mu\nu}&=&0,\\
d(*K^m)&=&0,\\
d(e^{\alpha+\beta}*K^m)&=&0,\\
d(e^{4\alpha}*K^m)&=&-e^\alpha F\wedge Y^m,
\end{eqnarray}
two-form identities
\begin{eqnarray}
dK^m&=&0,\label{eq:4dkm}\\
d(e^\alpha K^m)&=&\ft14e^{-2\alpha}F_\mu{}^\lambda*Y_{\nu\lambda}^mdx^\mu
\wedge dx^\nu-i\eta*Y^m,\\
d(e^\beta K^m)&=&-\ft14e^{-3\alpha+\beta}F_\mu{}^\lambda*Y_{\nu\lambda}^m
dx^\mu\wedge dx^\nu+i\tilde\eta Y^m,
\end{eqnarray}
and three-form identities
\begin{eqnarray}
d(e^{-\beta}Y^m)&=&0,\\
d(e^\alpha Y^m)&=&i\eta*K^m,\\
d(e^{3\beta}Y^m)&=&e^{-3\alpha+3\beta}(*F)\wedge K,\\
d(e^{-\alpha}*Y^m)&=&0,\\
d(e^\beta*Y^m)&=&i\tilde\eta*K^m,\\
d(e^{3\alpha}*Y^m)&=&-F\wedge K^m.
\end{eqnarray}
%

%%%%%%%%%%%%%%%%%%%%%%%%%%%%%%%%%%%%%%%%%%%%%%%%%%%%%%%%%%%%%%%%%%%
\section{Regularity analysis for 1/4 BPS solutions}

As an example of how we uncover the droplet picture for the 1/4 BPS geometries from a regularity analysis, we consider the case when the U(1) charge of the Killing spinor is
\beq
n\eta=1,
\eeq
which corresponds to the ungauged $S^3\times S^1$ reduction, as discussed in Section 3.2. Under this assumption, since $D(z_i,\bar z_{\bar j})$ is constrained by $(1+*_4)\partial\bar\partial D=0$, it follows that $D$ is a harmonic function of the four-dimensional K\"ahler base parametrized by $z_1, z_2$. 

The metric given in (\ref{GenericQuarterBPS}) is potentially singular when $y=0$, i.e. when the radius of either the $S^3$ or $S^1$ shrinks to zero.
To avoid conical singularities at $y=0$, $G$ ought to behave such that 
$e^{\pm G}=y f_\pm(z_1,\bar z_1,z_2,\bar z_2)+{\cal O}(y^2)$ where the $\pm$ sign corresponds to having either the $S^3$ or $S^1$ collapse to zero size.     
Since $Z=\frac 12 {\rm tanh} G$ is also tied to the four-dimensional base K\"ahler potential $Z=-\frac 12 y\partial_y\frac 1y\partial_y K$, this yields
\bea
K&=&\frac 12 y^2 \ln y + f_0+\frac {y^2}2 f_2-\frac{y^4}4 f_4 + \dots\nn\\
&{\rm or}&\nn\\
K&=&-\frac 12 y^2\ln y + g_0+ \frac{y^2}2 g_2 + \frac{y^4}4 g_4 + \dots 
\eea
where $f_{0,2,4}$ and $g_{0,2,4}$ are functions of $z_1,z_2$ and their complex conjugates.  
In the first case $Z\to -\frac 12$ as $y\to 0$ and in the second, $Z\to \frac12$ as $y\to 0$. The $y=0$ four-dimensional base is then decomposed into regions (``droplets'') with $Z\to \pm \frac 12$, similar to the LLM decomposition of the two-dimensional base. The requirement that the asymptotics of the 1/4 BPS solutions be $AdS_5\times S^5$ introduce the additional constraint that the droplet distribution  must be such that, at large $|z_1|^2+|z_2|^2$, one sees a large spherical droplet plus small distortions which can appear as deformations of the large  droplet and/or as additional disconnected small droplets.

To confirm that the complete ten-dimensional geometry is non-singular we first notice that $h^{-2}=2y\cosh G$ is finite at $y=0$. Second, from 
$\ln\det h_{i\bar j}=\ln (Z+\frac 12)+\ln y+\frac 1y \partial_yK+D$ we find that
\beq
\det h_{i\bar j}= y^4 e^{D+\frac 12+f_2} f_4 + \dots,\qquad  {\rm or} \qquad 
\det h_{i\bar j}= y^0 e^{D-\frac 12+g_2}  + \dots
\eeq
The regularity of the full ten-dimensional metric is assured since the K\"ahler subspace, together with its warp factor $(Z+\frac 12)^{-1}$, is 
non-singular. This follows from the evaluation of the volume of this subspace at $y=0$: 
\beq
\det(h^2 (Z+\frac 12)^{-1} h_{i\bar j})={\rm finite\;at}\;y=0.
\eeq

%%%%%%%%%%%%%%%%%%%%%%%%%%%%%%%%%%%%%%%%
\section{Detailed analysis of regularity conditions for 1/8 BPS
configurations}
\label{regularappendix}
%%%%%%%%%%%%%%%%%%%%%%%%%%%%%%%%%%%%%%%%

In this appendix we present the details of the regularity
analysis of 1/8 BPS configurations as discussed in Section~\ref{regular}.

In order to bring the ten-dimensional metric near $y=0~$to the form
(\ref{y=0 10d metric}), the K\"{a}hler potential has a Taylor expansion
of the form
\begin{equation}
K(z^{i}, z^{\bar{j}})=\frac{1}{4}y^{4}+ay^{6} + \cdots,
\end{equation}
up to a unimportant shift via a K\"{a}hler transformation, and
where $a=a(r_{2}^{2},r_{3}^{2})$ at $y=0$.

First we calculate the metric of the six-dimensional base to leading order in
$y$. The leading order (except in the 1st, 2nd, and 4th lines, which also
contain ${\cal O}(y^{2})$ terms) six-dimensional base metric is given by
\begin{eqnarray}
ds_{6}^{2} &=&dy^{2}\,y^{2}\biggl[1+\frac{y^{2}}{f_{y}}
\biggl(f_{y}-\frac{ff_{yy}}{f_{y}}\biggr)\biggr]  \notag \\
&&+2dydr_{2}\;y^{3}r_{2}\biggl(f_{2}-\frac{ff_{2y}}{f_{y}}\biggr)
+2dydr_{3}\;y^{3}r_{3}\biggl(f_{3}-\frac{ff_{3y}}{f_{y}}\biggr)\notag \\
&&+\frac{1}{f_{y}^{2}}(fd\phi _{1}-f_{2}r_{2}^{2}d\phi_{2}
-f_{3}r_{3}^{2}d\phi _{3})^{2}  \notag \\
&&+d\phi_1^2\;\frac{fy^{2}}{f_{y}^{2}}\biggl(f_{y}-\frac{ff_{yy}}{f_{y}}\biggr)
+2d\phi_{1}d\phi_{2}\;\frac{y^{2}fr_{2}^{2}}{f_{y}^{2}}
\biggl(\frac{f_{2}f_{yy}}{f_{y}}-f_{2y}\biggr)
+2d\phi _{1}d\phi _{3}\;\frac{y^{2}fr_{3}^{2}}{f_{y}^{2}}
\biggl(\frac{f_{3}f_{yy}}{f_{y}}-f_{3y}\biggr) \notag \\
&&+dr_{2}^{2}\;\frac{y^{2}}{f_{y}}\biggl(-f_{2}-f_{22}r_{2}^{2}
+\frac{f_{2}^{2}r_{2}^{2}}{f}\biggr)+dr_{3}^{2}\;\frac{y^{2}}{f_{y}}
\biggl(-f_{3}-f_{33}r_{3}^{2}+\frac{f_{3}^{2}r_{3}^{2}}{f}\biggr) \notag \\
&&+2dr_{2}dr_{3}\;\frac{y^{2}}{f_{y}}r_{2}r_{3}
\biggl(-f_{23}+\frac{f_{2}f_{3}}{f}\biggr)  \notag \\
&&+d\phi_{2}^{2}\;\frac{y^{2}r_{2}^{2}}{f_{y}^{2}}
\biggl(2r_{2}^{2}f_{2y}f_{2}-f_{2}f_{y}-r_{2}^{2}f_{y}f_{22}
-\frac{r_{2}^{2}f_{2}^{2}f_{yy}}{f_{y}}\biggr) \notag \\
&&+d\phi_{3}^{2}\;\frac{y^{2}r_{3}^{2}}{f_{y}^{2}}
\biggl(2r_{3}^{2}f_{3y}f_{3}-f_{3}f_{y}-r_{3}^{2}f_{y}f_{33}
-\frac{r_{3}^{2}f_{3}^{2}f_{yy}}{f_{y}}\biggr) \notag \\
&&+2d\phi_{2}d\phi_{3}\;\frac{y^{2}r_{2}^{2}r_{3}^{2}}{f_{y}}
(-f_{23}+f_{2y}f_{3}+f_{3y}f_{2}-f_{2}f_{3}f_{yy}).
\label{y_metric_app}
\end{eqnarray}

The leading metric pertaining to the four-dimensional surface is
\begin{eqnarray}
d\Sigma _{4}^{2} &=&dr_{2}^{2}\;\frac{1}{f_{y}}\biggl(-f_{2}-f_{22}r_{2}^{2}
+\frac{f_{2}^{2}r_{2}^{2}}{f}\biggr)+dr_{3}^{2}\;\frac{1}{f_{y}}
\biggl(-f_{3}-f_{33}r_{3}^{2}+\frac{f_{3}^{2}r_{3}^{2}}{f}\biggr) \notag \\
&&+2dr_{2}dr_{3}\;\frac{1}{f_{y}}r_{2}r_{3}
\biggl(-f_{23}+\frac{f_{2}f_{3}}{f}\biggr)  \notag \\
&&+d\phi_{2}^{2}\;\frac{r_{2}^{2}}{f_{y}^{2}}
\biggl(2r_{2}^{2}f_{2y}f_{2}-f_{2}f_{y}-r_{2}^{2}f_{y}f_{22}
-\frac{r_{2}^{2}f_{2}^{2}f_{yy}}{f_{y}}\biggr) \notag \\
&&+d\phi _{3}^{2}\;\frac{r_{3}^{2}}{f_{y}^{2}}
\biggl(2r_{3}^{2}f_{3y}f_{3}-f_{3}f_{y}-r_{3}^{2}f_{y}f_{33}
-\frac{r_{3}^{2}f_{3}^{2}f_{yy}}{f_{y}}\biggr) \notag \\
&&+2d\phi_{2}d\phi_{3}\;\frac{r_{2}^{2}r_{3}^{2}}{f_{y}}
(-f_{23}+f_{2y}f_{3}+f_{3y}f_{2}-f_{2}f_{3}f_{yy}).
\end{eqnarray}
We also notice that the leading piece of the $(1/y^{2}f_{y}^{2})
(fd\phi_{1}-f_{2}r_{2}^{2}d\phi_{2}-f_{3}r_{3}^{2}d\phi_{3})^{2}$
term cancels with the leading $y^{2}\omega^{2}$ term coming from
$g_{tt}$ in the ten-dimensional metric, as discussed in Section~\ref{regular}.
We therefore need to consider the subleading contributions of these terms
to the ten-dimensional metric.

We focus on the ${\cal O}(y^{2})$ piece of the metric components for
$d\phi_{1}^{2}$,  $2d\phi_{1}d\phi_{2}$ and $2d\phi_{1}d\phi_{3}$ from the
six-dimensional base, and refer to these as a subspace of the ten-dimensional
metric. These components come from the subleading terms of $K=\frac14y^{4}$
and the leading terms of $\delta K =ay^{6}$.

The terms originating from $K=\frac{1}{4}y^{4}$ are
\begin{eqnarray}
ds_{10}^{2}\big|_{\rm subspace}&=&\frac{-2f_{yy}}{f_{y}^{3}}
(fd\phi_{1}-f_{2}r_{2}^{2}d\phi_{2}-f_{3}r_{3}^{2}d\phi_{3})^{2} \notag \\
&&+\frac{2}{f_{y}^{2}}(fd\phi_{1}-f_{2}r_{2}^{2}d\phi_2-f_3r_3^2d\phi_3)
(f_{y}d\phi _{1}-f_{2y}r_{2}^{2}d\phi_{2}-f_{3y}r_{3}^{2}d\phi_{3}) \notag \\
&&+\frac{f}{f_{y}^{2}}\left(f_{y}-\frac{ff_{yy}}{f_y}\right)d\phi_1^2\notag\\
&&+2\frac{fr_{2}^{2}}{f_{y}^{2}}\left(\frac{f_{2}f_{yy}}{f_{y}}-f_{2y}\right)
d\phi_{1}d\phi_{2}+2\frac{fr_{3}^{2}}{f_{y}^{2}}
\left(\frac{f_{3}f_{yy}}{f_{y}}-f_{3y}\right)d\phi_{1}d\phi_{3}.
\end{eqnarray}
Next we calculate the terms from $\delta K=ay^{6}$. We define
\begin{equation}
\widetilde{F}=\sqrt{2a}y^{3}=\sqrt{2a}F^{3/2},\qquad F\equiv y^{2},
\end{equation}
and we have
\begin{equation}
\widetilde{F}_{a}=\sqrt{2a}\frac32F^{1/2}F_{a}=\sqrt{2a}\frac{3}{2}yF_{a}.
\end{equation}
Notice that we can use the general formula (\ref{kahler_metric}):
\begin{equation}
ds_{6}^{2}=2\sum_{a=1}^{3}\left[\widetilde{F}_{a}^{2}r_{a}^{2}
+\widetilde{F}(\widetilde{F}_{aa}r_{a}^{2}+\widetilde{F}_{a})\right]
\left(dr_{a}^{2}+r_{a}^{2}d\phi _{a}^{2}\right)
+4\sum_{a<b}^3\left[\widetilde{F}_{a}\widetilde{F}_{b}
+\widetilde{F}\widetilde{F}_{ab}\right]r_{a}^{2}r_{b}^{2}
\left(\frac{dr_{a}dr_{b}}{r_{a}r_{b}}+d\phi_{a}d\phi_{b}\right).
\end{equation}
It is easy to see that the first terms in each of the two sums are of
order ${\cal O}(y^{2})$, while the second terms
are of order ${\cal O}(y^{4})$ or higher. So we only keep
\begin{eqnarray}
ds_{6}^{2} &=&2\sum_{a=1}^{3}\left[\widetilde{F}_{a}^{2}r_{a}^{2}\right]
\left(dr_{a}^{2}+r_{a}^{2}d\phi _{a}^{2}\right)+4\sum_{a<b}^3
\left[\widetilde{F}_{a}\widetilde{F}_{b}\right]r_{a}^{2}r_{b}^{2}
\left(\frac{dr_{a}dr_{b}}{r_{a}r_{b}}+d\phi_{a}d\phi_{b}\right) \nn\\
&=&{9a}y^{2}\left\{\sum_{a=1}^{3}\left[F_{a}^{2}r_{a}^{2}\right]
\left(dr_{a}^{2}+r_{a}^{2}d\phi _{a}^{2}\right)+2\sum_{a<b}^3
\left[F_{a}F_{b}\right]r_{a}^{2}r_{b}^{2}
\left(\frac{dr_{a}dr_{b}}{r_{a}r_{b}}+d\phi_{a}d\phi_{b}\right)\right\}.
\end{eqnarray}
In other words, the subleading contribution from $\delta K=a y^6$
is actually $9ay^{2}$ times the leading order metric coming from
$\frac{1}{4}y^{4}$.

Focusing on the subspace mentioned above, we find that the contribution from
$\delta K=ay^{6}$ is:
\begin{equation}
ds_{10}^{2}\big|_{\rm subspace}=\frac{9a}{f_{y}^{2}}
(fd\phi_{1}-f_{2}r_{2}^{2}d\phi _{2}-f_{3}r_{3}^{2}d\phi_{3})^{2}.
\end{equation}
The total contribution to the ten dimensional metric in this subspace is
then given by:
\begin{eqnarray}
ds_{10}^{2}\big|_{\rm subspace} &=&\left(\frac{-2f_{yy}}{f_{y}^{3}}
+\frac{9a}{f_{y}^{2}}\right)(fd\phi_{1}-f_{2}r_{2}^{2}d\phi_{2}
-f_{3}r_{3}^{2}d\phi_{3})^{2}  \notag\\
&&+\frac{2}{f_{y}^{2}}(fd\phi_{1}-f_{2}r_{2}^{2}d\phi_2-f_3r_3^2d\phi_{3})
(f_{y}d\phi_{1}-f_{2y}r_{2}^{2}d\phi_{2}-f_{3y}r_{3}^{2}d\phi_{3}) \notag \\
&&+\frac{f}{f_{y}^{2}}\left(f_{y}-\frac{ff_{yy}}{f_{y}}\right)d\phi_1^2\nn\\
&&+2\frac{fr_{2}^{2}}{f_{y}^{2}}\left(\frac{f_{2}f_{yy}}{f_{y}}-f_{2y}\right)
d\phi_{1}d\phi_{2}+2\frac{fr_{3}^{2}}{f_{y}^{2}}
\left(\frac{f_{3}f_{yy}}{f_{y}}-f_{3y}\right)d\phi_{1}d\phi_{3} \notag \\
&=&\left(\frac{-3f_{yy}}{f_{y}^{3}}+\frac{9a}{f_{y}^{2}}
+\frac{f_{y}}{ff_{y}^{2}}\right) (fd\phi_{1}-f_{2}r_{2}^{2}d\phi_{2}
-f_{3}r_{3}^{2}d\phi_{3})^{2}  \notag \\
&&+2\frac{f}{f_{y}^{3}}(f_{y}d\phi _{1}-f_{2y}r_{2}^{2}d\phi
_{2}-f_{3y}r_{3}^{2}d\phi _{3})^{2} \notag \\
&&+2\frac{1}{f_{y}^{2}}f_{2}f_{2y}r_{2}^{4}d\phi_{2}^{2}
+2\frac{1}{f_{y}^{2}}f_{3}f_{3y}r_{3}^{4}d\phi_{3}^{2}
+2\frac{1}{f_{y}^{2}}(f_{2}f_{3y}+f_{3}f_{2y})r_{2}^{2}r_{3}^{2}
d\phi_{2}d\phi_{3}  \notag \\
&&-\frac{1}{f_{y}^{2}}\left(\frac{f_{y}}{f}-\frac{f_{yy}}{f_{y}}\right)
(f_{2}^{2}r_{2}^{4}d\phi_{2}^{2}+f_{3}^{2}r_{3}^{4}d\phi_{3}{}^{2}
+2f_{2}f_{3}r_{2}^{2}r_{3}^{2}d\phi_{2}d\phi_{3}) \notag \\
&&-2\frac{f}{f_{y}^{3}}(f_{2y}^{2}r_{2}^{4}d\phi_{2}^{2}
+f_{3y}^{2}r_{3}^{4}d\phi_{3}^{2}+2f_{2y}f_{3y}r_{2}^{2}r_{3}^{2}
d\phi_{2}d\phi_{3}),
\end{eqnarray}
where the $d\phi_{2}^{2}$, $d\phi_{3}^{2}$ and $d\phi_{2}d\phi_{3}$ terms
in the last three lines will be combined into $d\widetilde{\Sigma}_{4}^2$.

Thus the ten dimensional metric near $y=0$ goes like
\begin{eqnarray}
ds_{10}^{2} &=&-\frac{2}{f_{y}}dt(fd\phi_{1}-f_{2}r_{2}^{2}d\phi_{2}
-f_{3}r_{3}^{2}d\phi_{3})\bigg|_{y=0}  \notag \\
&&+\left(\frac{-3f_{yy}}{f_{y}^{3}}+\frac{9a}{f_{y}^{2}}
+\frac{f_{y}}{ff_{y}^{2}}\right) (fd\phi_{1}-f_{2}r_{2}^{2}d\phi_{2}
-f_{3}r_{3}^{2}d\phi_{3})^{2}\bigg|_{y=0}  \notag \\
&&+2\frac{f}{f_{y}^{3}}(f_{y}d\phi _{1}-f_{2y}r_{2}^{2}d\phi_{2}
-f_{3y}r_{3}^{2}d\phi_{3})^{2}\bigg|_{y=0}+d\widetilde{\Sigma}_{4}{}^{2}
(r_{2},\phi_{2},r_{3},\phi_{3})  \notag \\
&&+d\mathbb R_{4}{}^{2}.
\end{eqnarray}
We can also rewrite it in the form
\begin{eqnarray}
ds_{10}^{2} &=&-g_{tt}\bigg|_{y=0}dt^{2}+\left(\frac{f}{f_{y}}
+\frac{9af^{2}}{f_{y}^{2}}-\frac{3f^{2}f_{yy}}{f_{y}^{3}}\right)
(d\phi_{1}-\frac{f_{2}r_{2}^{2}}{f}d\phi_{2}-\frac{f_{3}r_{3}^{2}}{f}d\phi_{3}
-w_{t}dt)^{2}\bigg|_{y=0}  \notag \\
&&+\frac{2f}{f_{y}}\left(d\phi_{1}-\frac{f_{2y}r_{2}^{2}}{f_{y}}d\phi_{2}
-\frac{f_{3y}r_{3}^{2}}{f_{y}}d\phi_{3}\right)^{2}\bigg|_{y=0}
+d\widetilde{\Sigma}_{4}{}^{2}(r_{2},\phi_{2},r_{3},\phi_{3})  \notag \\
&&+d\mathbb R_{4}{}^{2},
\end{eqnarray}
where
\begin{eqnarray}
d\widetilde{\Sigma}_{4}{}^{2}(r_{2},\phi_{2},r_{3},\phi_{3})&=&
d\Sigma_{4}^{2}+2\frac{1}{f_{y}^{2}}f_{2}f_{2y}r_{2}^{4}d\phi_{2}^{2}
+2\frac{1}{f_{y}^{2}}f_{3}f_{3y}r_{3}^{4}d\phi_{3}^{2}+2\frac{1}{f_{y}^{2}}
(f_{2}f_{3y}+f_{3}f_{2y})r_{2}^{2}r_{3}^{2}d\phi_{2}d\phi_{3}  \notag \\
&&-\frac{1}{f_{y}^{2}}\left(\frac{f_{y}}{f}-\frac{f_{yy}}{f_{y}}\right)
(f_{2}^{2}r_{2}^{4}d\phi_{2}^{2}+f_{3}^{2}r_{3}^{4}d\phi_{3}^{2}
+2f_{2}f_{3}r_{2}^{2}r_{3}^{2}d\phi_{2}d\phi_{3})  \notag \\
&&-2\frac{f}{f_{y}^{3}}(f_{2y}^{2}r_{2}^{4}d\phi_{2}^{2}
+f_{3y}^{2}r_{3}^{4}d\phi_{3}^{2}+2f_{2y}f_{3y}r_{2}^{2}r_{3}^{2}
d\phi_{2}d\phi_{3}) \notag\\
&=&dr_{2}^{2}\;\frac{1}{f_{y}}\bigg(-f_{2}-f_{22}r_{2}^{2}
+\frac{f_{2}^{2}r_{2}^{2}}{f}\bigg)+dr_{3}^{2}\;\frac{1}{f_{y}}
\bigg(-f_{3}-f_{33}r_{3}^{2}+\frac{f_{3}^{2}r_{3}^{2}}{f}\bigg) \notag \\
&&+2dr_{2}dr_{3}\;\frac{1}{f_{y}}r_{2}r_{3}
\bigg(-f_{23}+\frac{f_{2}f_{3}}{f}\bigg)  \notag \\
&&+d\phi_{2}^{2}\;\frac{r_{2}^{4}}{f_{y}^{2}}\bigg(4f_{2y}f_{2}
-\frac{f_{2}f_{y}}{r_{2}^{2}}-f_{y}f_{22}-\frac{f_{y}f_{2}^{2}}{f}
-\frac{2ff_{2y}^{2}}{f_{y}}\bigg)  \notag \\
&&+d\phi_{3}^{2}\;\frac{r_{3}^{4}}{f_{y}^{2}}\bigg(4f_{3y}f_{3}
-\frac{f_{3}f_{y}}{r_{3}^{2}}-f_{y}f_{33}-\frac{f_{y}f_{3}^{2}}{f}
-\frac{2ff_{3y}^{2}}{f_{y}}\bigg)  \notag \\
&&+2d\phi_{2}d\phi_{3}\;\frac{r_{2}^{2}r_{3}^{2}}{f_{y}}
\bigg(-f_{23}+f_{2y}f_{3}+f_{3y}f_{2}-f_{2}f_{3}f_{yy}
+\frac{f_{2}f_{3y}}{f_{y}}+\frac{f_{3}f_{2y}}{f_{y}} \notag \\
&&\kern7.5em-\frac{f_{2}f_{3}}{f}+\frac{f_{2}f_{3}f_{yy}}{f_{y}^{2}}
-\frac{2ff_{2y}f_{3y}}{f_{y}^{2}}\bigg).
\end{eqnarray}
%

%%%%%%%%%%%%%%%%%%%%%%%%%%%%%%%%%%%%%%%%

\end{document}